\documentclass[twocolumn]{jpsj2}
\usepackage{bm}

\def\dsigma{\underline{\underline{\sigma}}}
\def\dI{\underline{\underline{I}}}

\newcommand{\gtsim}{\protect\raisebox{-0.8ex}{$\:\stackrel{\textstyle >}
	{\sim}\:$}}

\def\runauthor{Online-Journal Subcommittee}

\title{%
Gaussian-Basis Monte Carlo Method for Numerical Study \\
on Ground States of Itinerant and Strongly Correlated Electron Systems
}

\author{%
Takeshi Aimi$^1$ and Masatoshi Imada$^2$
}

\inst{%
Department of Physics$^1$ and Department of Applied Physics$^2$, University of Tokyo\\
7-3-1 Hongo, Bunkyo-ku, Tokyo,
Japan

}

\recdate{April 15, 2007}

\abst{%
We examine Gaussian-basis Monte Carlo method (GBMC) introduced by Corney and Drummond. 
This method is based on an expansion of the density-matrix operator 
$\hat{\rho}$ by means of the coherent 
Gaussian-type operator basis $\hat{\Lambda}$ and does not suffer from the minus sign problem. 
The original method, however, often fails in reproducing the 
true ground state and causes systematic errors of calculated physical quantities 
because the samples are often trapped in 
some metastable or symmetry broken states. 
To overcome this difficulty, 
we combine the quantum-number projection scheme proposed by Assaad, Werner, Corboz, Gull and Troyer
in conjunction with the importance sampling of the original GBMC method. 
This improvement allows us to 
carry out the importance sampling in the quantum-number-projected phase-space. 
Some comparisons with the previous quantum-number projection scheme 
indicate that, in our method, 
the convergence with the ground state is accelerated, 
which makes it possible to extend the applicability and widen the range of 
tractable parameters in the GBMC method. The present scheme 
offers an efficient practical way of computation for 
strongly correlated electron systems beyond the range of system sizes, interaction strengths
and lattice structures tractable by other computational methods such as the quantum Monte Carlo method. 
}

\kword{%
Monte Carlo method, strongly correlated electron systems, Hubbard model, quantum-number projection
}

\begin{document}
\maketitle

\section{Introduction}
Ground state properties of strongly correlated 
electron systems 
are challenging subjects 
in condensed matter physics. 
From the numerical point of view, there exist many numerical algorithms, 
such as the exact diagonalization method, 
the auxiliary-field quantum Monte Carlo (AFQMC) method~\cite{Blankenbecler,Sorella0,Hatsugai,Furukawa}, 
the density matrix renormalization group (DMRG) method~\cite{White} and 
the path-integral renormalization group (PIRG) method~\cite{Kashima0,Kashima1,Kashima2,Morita,Watanabe,Mizusaki}. 
Although the exact diagonalization of the Hamiltonian matrix  
gives accurate results,
the tractable system size is severely limited. 
The AFQMC method can treat larger systems and has been applied to various correlated 
systems. 
In some systems such as doped Mott insulators and 
the Mott insulators with geometrical frustration effects, however, 
the AFQMC method often suffers from the negative sign problem which causes 
large statistical errors by the cancellation of positive and negative 
Monte Carlo samples. 
The DMRG method offers practically exact results without 
suffering from the negative sign problem. However, 
tractable lattice systems are restricted to one-dimensional configurations 
owing to the spatial renormalization process. 
The PIRG method is a powerful sign-free numerical technique for correlated electron systems 
and has been applied to various systems 
beyond the tractable range of the above numerical methods.
A main practical limitation in the PIRG method comes from the extrapolation 
procedure to reach the results for the full Hilbert space.
The truncation error depends on the system size 
as well as on the interaction strength. 

Gaussian-basis Monte Carlo (GBMC) method has been proposed as an alternative 
quantum Monte Carlo method which does not involve any sign problem~\cite{Corney1,Corney2}. 
This method is based on a representation of the density-matrix operator $\hat{\rho}$
by making use of the non-Hermitian Gaussian-type operator basis $\hat{\Lambda}$. 
The Gaussian representation is a natural generalization of the 
positive-$P$ phase-space method 
which is often used in the area of quantum optics~\cite{Drummond1,Drummond2}. 
As well as classical phase-space variables like $(\bm{x},\bm{p})$, 
the Gaussian-basis representation utilizes the one-particle ``Green's function" 
$n_{ij}={\rm Tr}\left[\hat{c}_{i}^{\dag}\hat{c}_{j}\hat{\Lambda}\right]$,
$m_{ij}={\rm Tr}\left[\hat{c}_{i}\hat{c}_{j}\hat{\Lambda}\right]$,
$m_{ij}^{+}={\rm Tr}\left[\hat{c}_{i}^{\dag}\hat{c}_{j}^{\dag}\hat{\Lambda}\right]$ 
and the stochastic weight $\Omega$ as the phase-space variables, 
where $\hat{c}_{i}^{\dag}$ ($\hat{c}_{i}$) 
is a Fermion creation (annihilation) operator of the $i$-th mode. 
In this method, we solve a Fokker-Planck equation with respect to the phase-space 
variables $\underline{\lambda}=(\Omega,\bm{n},\bm{m},\bm{m}^{+})$, which is constructed 
by a mapping from an operator Liouville equation of $\hat{\rho}$. 
One of the phase-space variables $\Omega$ works as a weight of the importance sampling 
and for any two-body Hamiltonian, $\Omega$ remains positive definite. 
Thus there exists no explicit manifestation of the negative sign problem. 
However, in many parameter regions, especially in the low-temperature region, 
the numerical results obtained by the GBMC method often show systematic errors~\cite{Assaad,Assaad2}. 
It will turn out below that this deviation is concerned with the ``spontaneous symmetry breaking
".

Assaad {\it et al.} have used a quantum-number projection scheme to overcome the 
deviations in the low-temperature region~\cite{Assaad,Assaad2}. 
They have proposed to project the density matrix onto given 
quantum numbers of the ground state after the sampling is completed and have reproduced accurate ground states in some 
parameter regions. 
In their method, however, the convergence with the ground state becomes slower 
with the increase of the interaction strength, which determines the practical 
limitations. 

In this study, we combine the quantum-number projection scheme concurrently with the 
importance sampling of the original GBMC method to 
reflect the amount of the overlap with the projected sector in the sampling weight. 
This allows us to perform the importance sampling with respect to the projected distribution. 
The crucial point is that the efficiency of the importance sampling is improved 
because the sampling weight reflects not only the energy but also the overlap 
with the projected state, {\it {\it i.e.}}, 
the overlap with the state which retains the same quantum numbers with the ground state. 
Thanks to the efficient sampling, the tractable parameter region becomes wider than 
the previous method reported by Assaad {\it et al}~\cite{Assaad,Assaad2}. 
Moreover, our method allows us to analyze the projected distribution directly. 
By using this advantage, 
the relation between the numerical convergence and 
the behavior of the projected distribution is also reported. 
We show benchmark analysis up to $10 \times 10$ lattices on the square lattice 
as well as up to the relative interaction strength $U/t=15$ for the Hubbard model,
which indicate the applicability and efficiency of the present method.

The organization of this paper is as follows. 
Section 2 gives an introduction of the GBMC method 
for the general Fermion systems by following the formulation 
in Refs.~\citen{Corney1,Corney2}, which we supply for the self-contained description. 
In \S 3, we explain implementations of the Monte Carlo procedure 
for the Hubbard model. 
The improvements of the GBMC method by quantum-number projections 
are shown in \S 4. In the last part of \S 4, we discuss 
the practical limitation in the applicability of the GBMC method. 
Section 5 is devoted to summary and discussions.

\section{Gaussian-Basis Monte Carlo Method}
Gaussian-basis Monte Carlo (GBMC) method is a numerical method which makes use of 
the mapping between operator equations of motion and stochastic evolution equations of 
generalized phase-space~\cite{Corney1,Corney2}. 
In order to make an exact mapping, we introduce a complete set of Gaussian-type 
operators $\hat{\Lambda}$, which is typically non-Hermitian. 
This basis set allows us to expand 
any physical density-matrix operator $\hat{\rho}(\tau)$ in terms of the phase-space 
variables $\underline{\lambda}$ as 
\begin{align}
\hat{\rho}(\tau)=\int d\underline{\lambda}P(\underline{\lambda},\tau)\hat{\Lambda}(\underline{\lambda}) , \label{eq:expansion}
\end{align}
where $\tau$ is real or imaginary time,  
$P(\underline{\lambda},\tau)$ is the expansion coefficient, 
and $d\underline{\lambda}$ is the integration measure. 
Here, $P(\underline{\lambda},\tau)$ can always be chosen positive as in Appendix B and hence can be regarded as 
a probability distribution. 
In the GBMC method, it is the distribution $P(\underline{\lambda},\tau)$ that is sampled stochastically. 

\subsection{General Gaussian basis}
\subsubsection{Notation}
Before defining the Gaussian basis, we summarize the notation which will be used. 
Consider a $M$-mode Fermionic system characterized by the creation and annihilation 
operators $\hat{c}_{j}^{\dag}$ and $\hat{c}_{j}$, with anticommutation relations 
\begin{align}
[\hat{c}_{i},\hat{c}_{j}^{\dag}]_{+}=\delta_{i,j},\quad
[\hat{c}_{i},\hat{c}_{j}]_{+}=[\hat{c}_{i}^{\dag},\hat{c}_{j}^{\dag}]_{+}=0 , 
\end{align}
where $i,j=1,\ldots,M$. We define a $M$-mode column vector of 
the annihilation operators and its Hermitian conjugate row vector as 
$\hat{\bm{c}}$ and $\hat{\bm{c}}^{\dag}$, respectively. 
In order to define the 
general Gaussian basis in a compact form, we introduce an extended-vector notation 
\begin{align}
\underline{\hat{c}}=\left(
 \begin{array}{c}
 \hat{\bm{c}}  \\
 \hat{\bm{c}}^{\dag T}  \\
\end{array}
 \right),\quad
\underline{\hat{c}}^{\dag}=\left(\hat{\bm{c}}^{\dag},\hat{\bm{c}}^{T}\right) . 
\end{align}
Throughout the paper, we use the bold-type notation for $M$-mode vectors or matrices and 
the underline notation for $2M$-mode extended vectors or matrices. 

For products of operators, we define a normal and an antinormal ordering operators 
denoted by $:\quad:$ and $\{\quad\}$, respectively. 
The normal ordering operator $:\quad:$ reorders so that 
all the creation operators are put to the left of the annihilation operators, 
{\it e.g.}, $:\hat{c}_{i}\hat{c}_{j}^{\dag}:=-\hat{c}_{j}^{\dag}\hat{c}_{i}$. 
Similarly, the antinormal ordering operator $\{\quad\}$ reorders so that
all the annihilation operators are put to the left of the creation operators, 
{\it e.g.}, $\{\hat{c}_{j}^{\dag}\hat{c}_{i}\}=-\hat{c}_{i}\hat{c}_{j}^{\dag}$. 
More generally, in the case of a nested product, the outer ordering operator 
does not reorder the inner one, {\it e.g.}, 
$\{:\hat{c}_{k}\hat{c}_{j}^{\dag}:\hat{c}_{i}\}=\hat{c}_{i}:\hat{c}_{k}\hat{c}_{j}^{\dag}:=-\hat{c}_{i}\hat{c}_{j}^{\dag}\hat{c}_{k}$. 
The sign changes are necessary because of the anticommuting nature 
of the Fermion operators. 

\subsubsection{Definition of the Gaussian basis}
Using the above notation, a general Gaussian operator is defined as : 
\begin{align}
\hat{\Lambda}(\underline{\lambda})=\Omega{\rm Pf}[\dsigma_{A}]:\exp\left[\underline{\hat{c}}^{\dag}(\dI-\dsigma^{-1}/2)\underline{\hat{c}}\right]: ,
\end{align}
where $\dI$ is an extended unit matrix:
\begin{align}
\dI=
\left[
 \begin{array}{cc}
 -\bm{I}    & \bm{0} \\
 \bm{0} & \bm{I}     \\
 \end{array}
 \right] ,
\end{align}
$\dsigma$ is an extended covariance:
\begin{align}
\dsigma=
\left[
 \begin{array}{cc}
 \bm{n}^{{\rm T}}-\bm{I} & \bm{m}     \\
 \bm{m}^{+}        & \bm{I}-\bm{n} \\
 \end{array}
 \right] ,
\end{align}
and the vector parameter $\underline{\lambda}$ is defined as: 
\begin{align}
\underline{\lambda}=(\Omega,\bm{n},\bm{m},\bm{m}^{+}). 
\end{align}
Here, $\bm{n}$ is a $M\times M$ matrix which corresponds to normal Green's 
function, while $\bm{m}$ and $\bm{m}^{+}$ are two independent $M\times M$ 
antisymmetric matrices which correspond to anomalous Green's functions. 
Pfaffian of the antisymmetrized covariance 
$\dsigma_{A}$ appears so as to satisfy the normalization condition 
\begin{align}
{\rm Tr}[\hat{\Lambda}(\underline{\lambda})]=\Omega .
\end{align}
Here, $\dsigma_{A}$ is constructed by 
moving each row in the lower half rows every after the row with the same indices in the upper half, 
and moving each column in the right half columns every before the 
column with the same indices in the left half, {\it i.e.},
\begin{align}
\begin{bmatrix}
{\bf a} & {\bf b} \\
{\bf c} & {\bf d}
\end{bmatrix}_{A}=
\begin{bmatrix}
b_{11} & a_{11} & \cdots & b_{1M} & a_{1M} \\
d_{11} & c_{11} & \cdots & d_{1M} & c_{1M} \\
\vdots & \vdots & \ddots & \vdots & \vdots \\
b_{M1} & a_{M1} & \cdots & b_{MM} & a_{MM} \\
d_{M1} & c_{M1} & \cdots & d_{MM} & c_{MM}
\end{bmatrix} .
\end{align}

The general Gaussian operator $\hat{\Lambda}(\underline{\lambda})$ 
itself may correspond to a density-matrix operator 
under the conditions that $\bm{n}^{\dag}=\bm{n}$, $\bm{m}^{+}=\bm{m}^{\dag}$ 
and the eigenvalues of the matrix $\bm{n}$ lie in the interval $[0,1]$. 
However, we do not restrict the Gaussian operator to such conditions and it 
allows the Gaussian basis to be an overcomplete set which can expand 
any physical density-matrix operator. 

\subsubsection{Properties of Gaussian basis}
From the above definition, the general Gaussian operators satisfy some important 
identities~\cite{Corney1,Corney2}.
First, the Gaussian operator operated by the Fermion operator $\hat{c}$ and $\hat{c}^{\dag}$ 
can be associated with differentiations of the Gaussian operators with respect to their parameters: 
\begin{align}
\hat{\Lambda}&=\Omega\frac{\partial}{\partial \Omega}\hat{\Lambda} , \label{eq:diff1} \\
:\underline{\hat{c}}\,\underline{\hat{c}}^{\dag}\hat{\Lambda}:&=\dsigma\hat{\Lambda}-\dsigma\frac{\partial\hat{\Lambda}}{\partial\dsigma}\dsigma , \label{eq:diff2} \\
\left\{\underline{\hat{c}}:\underline{\hat{c}}^{\dag}\hat{\Lambda}:\right\}&=-\dsigma\hat{\Lambda}+(\dsigma-\dI)\frac{\partial\hat{\Lambda}}{\partial\dsigma}\dsigma . \label{eq:diff3}
\end{align}
Second, the traces of ladder operators with respect to the Gaussian operator 
can be analytically taken by using Grassmann coherent states: 
\begin{align}
{\rm Tr}\left[\hat{c}_{i}^{\dag}\hat{c}_{j}\hat{\Lambda}\right]&=\Omega n_{ij} , \\
{\rm Tr}\left[\hat{c}_{i}\hat{c}_{j}\hat{\Lambda}\right]&=\Omega m_{ij} , \quad
{\rm Tr}\left[\hat{c}_{i}^{\dag}\hat{c}_{j}^{\dag}\hat{\Lambda}\right]=\Omega m_{ij}^{+} .
\end{align}

\subsection{Time evolution}
The time evolution of a density-matrix operator is determined by the Liouville equation 
\begin{align}
\frac{d\hat{\rho}(t)}{dt}=\hat{L}\hat{\rho}(t) , \label{eq:Liu}
\end{align}
where $\hat{L}$ is a Liouville superoperator. 
For a real time evolution, the superoperator is given by the commutator with 
the Hamiltonian: 
\begin{align}
\hat{L}\hat{\rho}(t)\equiv -\frac{i}{\hbar}\left[\hat{H},\hat{\rho}(t)\right] .
\end{align}
In the case of calculating the equilibrium state at $\tau=1/k_{B}T$, 
the imaginary time evolution of a density-matrix operator is determined by the equation 
\begin{align}
\frac{d\hat{\rho}(\tau)}{d\tau}=-\frac{1}{2}\left[\hat{H},\hat{\rho}(\tau)\right]_{+} . 
\end{align}
Therefore the superoperator is given by the anticommutator with the Hamiltonian: 
\begin{align}
\hat{L}\hat{\rho}(\tau)\equiv -\frac{1}{2}\left[\hat{H},\hat{\rho}(\tau)\right]_{+}
\end{align}

To construct a mapping, we first substitute the expansion in Eq.~(\ref{eq:expansion}) into the 
Liouville equation (\ref{eq:Liu}) to get 
\begin{align}
\int\frac{dP(\underline{\lambda},t)}{dt}\hat{\Lambda}(\underline{\lambda})d\underline{\lambda}=\int P(\underline{\lambda},t)\hat{L}\hat{\Lambda}(\underline{\lambda})d\underline{\lambda} . 
\end{align}
Second, using the differential properties in Eqs.(\ref{eq:diff1}-\ref{eq:diff3}), 
one can transform the 
superoperator $\hat{L}\hat{\Lambda}(\underline{\lambda})$ 
into a differential operator ${\cal L}\hat{\Lambda}(\underline{\lambda})$. 
We next apply partial integration to get, provided that boundary terms vanish, 
\begin{align}
\int\hat{\Lambda}(\underline{\lambda})\frac{dP(\underline{\lambda},t)}{dt}d\underline{\lambda}=\int\hat{\Lambda}(\underline{\lambda}){\cal L}'P(\underline{\lambda},t)d\underline{\lambda} , \label{eq:aa}
\end{align}
where ${\cal L}'$ is reordered form of ${\cal L}$. 
Note that from Eq.~(\ref{eq:diff1}-\ref{eq:diff2}) 
${\cal L}'$ contains derivatives only up to the second order for any two-body Hamiltonian. 
As a sufficient solution for Eq.~(\ref{eq:aa}), 
a Fokker-Planck equation of Ito type is obtained:
\begin{align}
\frac{d}{dt}P(\underline{\lambda},t)&={\cal L}'P(\underline{\lambda},t) \notag \\
&=\left[-\sum_{i}\frac{\partial}{\partial\lambda_{i}}A_{i}(\underline{\lambda})\right. \notag \\
&\qquad\quad\left.+\frac{1}{2}\sum_{i,j}\frac{\partial}{\partial\lambda_{i}}\frac{\partial}{\partial\lambda_{j}}D_{ij}(\underline{\lambda})\right]P(\underline{\lambda},t) . \label{eq:Fok}
\end{align}
The imaginary-time evolution equation of the density-matrix operator then boils down to 
the Fokker-Planck equation, which is in practice solved by integrating numerically the corresponding 
stochastic differential equations (SDE). 

\section{Gaussian Representation for Hubbard Model}
\subsection{Mapping}
We consider the following Hubbard Hamiltonian.
\begin{align}
\hat{H}&=\sum_{i,j,\sigma}^{N}t_{ij}\hat{c}_{i\sigma}^{\dag}\hat{c}_{j\sigma}+U\sum_{i}^{N}\hat{c}_{i\uparrow}^{\dag}\hat{c}_{i\uparrow}\hat{c}_{i\downarrow}^{\dag}\hat{c}_{i\downarrow} , \label{eq:standard_H} \\
t_{ij}&=\left\{
 \begin{array}{cl}
 -\mu & {\rm for}\ i=j , \\
 -t   & {\rm for\ }(i,j)\ {\rm being\ a\ nearest}{\text -}{\rm neighbor\ pair} ,    \\
  0   & {\rm otherwise} , \\
 \end{array}
 \right.
\end{align}
where $i$ and $j$ represent the lattice points, $\hat{c}_{i\sigma}^{\dag}(\hat{c}_{i\sigma})$ 
the creation (annihilation) operator of an electron with spin $\sigma$ on the $i$-th site, 
$t_{ij}$ the transfer integral between the $i$-th site and the $j$-th site, $U$ the on-site Coulomb interaction, 
$\mu$ the chemical potential and $N$ the number of the lattice sites. 
Although we treat only this simplest Hubbard model, 
the formation can easily be extended to a more general form including transfers for 
further-site pairs and/or intersite Coulomb interactions.

Although Eq.~(\ref{eq:standard_H}) is a standard representation of the Hubbard Hamiltonian, 
it is necessary that the sign of the interaction term is negative 
so that the diffusion matrix $D(\underline{\lambda})$ 
in Eq.~(\ref{eq:Fok}) being positive definite~\cite{Corney1,Corney2}. 
Thus, we transform the Hamiltonian as follows~\cite{Assaad,Assaad2}: 
\begin{align}
\hat{H}&=\hat{\bm{c}}^{\dag}\mbox{\boldmath $T$}\hat{\bm{c}}-\frac{U}{2}\sum_{i}:(\hat{\bm{c}}_{i}^{\dag}\sigma^{z}\hat{\bm{c}}_{i})^{2}:\\
&=\sum_{x,y=1}^{2N}T_{xy}\hat{n}_{xy}+\frac{U}{2}\sum_{i=1}^{N}\sum_{\substack{\eta,\eta'\\ \sigma,\sigma'}}\left[\delta_{\sigma\eta'}\hat{n}_{(i\eta),(i\sigma')}\right. \notag \\ 
&\qquad\qquad\left. -\hat{n}_{(i\eta),(i\eta')}\hat{n}_{(i\sigma),(i\sigma')}\right]\sigma_{\eta\eta'}^{z}\sigma_{\sigma\sigma'}^{z} ,
\end{align}
where $\hat{n}_{xy}=\hat{c}_{x}^{\dag}\hat{c}_{y}$ and 
suffices $x$ and $y$ denote both the coordinates of site and spin, {\it i.e.}, $x=(i,\sigma)$. 
The vector operators $\hat{\bm{c}}^{\dag}$ and $\hat{\bm{c}}_{i}^{\dag}$ are defined as 
\begin{align}
\hat{\bm{c}}^{\dag}&=(\hat{c}_{1\uparrow}^{\dag},\hat{c}_{2\uparrow}^{\dag},\cdots,\hat{c}_{N\uparrow}^{\dag},\hat{c}_{1\downarrow}^{\dag},\cdots,\hat{c}_{N\downarrow}^{\dag}) , \\
\hat{\bm{c}}_{i}^{\dag}&=(\hat{c}_{i\uparrow}^{\dagger},\hat{c}_{i\downarrow}^{\dag}) .
\end{align}
The $2N\!\times\! 2N$ extended hopping matrix $\bm{T}$ is defined as 
$T_{i,j}=T_{i+N,j+N}=t_{i,j}$ and $T_{i,j+N}=T_{i+N,j}=0$, where $i,j=1,\cdots,N$. 
The matrix $\sigma^{z}$ denotes the $z$ component of the Pauli matrix. 
Since the Hubbard model conserves the total particle number, 
we use the number-conserving subset of the general Gaussian operator to expand 
the density-matrix operator: 
\begin{align}
\hat{\Lambda}(\Omega,\bm{n})=\Omega\det(\bm{I}-\bm{n}):e^{-\hat{\bm{c}}^{\dag}[2\bm{I}+(\bm{n}^{T}-\bm{I})^{-1}]\hat{\bm{c}}}: ,
\end{align}
where $\bm{n}$ is a $2N\times 2N$ matrix. 

The Gaussian operator consists of an overcomplete set and 
it can expand any physical density-matrix operator 
with positive coefficients. 
In the following sections, we express the parameters of the Gaussian 
operator as $\underline{\lambda}=(\Omega,\bm{n})$. 
Similarly to the case of the general Gaussian operators in 
Eqs.(\ref{eq:diff1}-\ref{eq:diff3}), the number-conserving Gaussian satisfies the 
differential identities: 
\begin{align}
\hat{\Lambda}&=\Omega\frac{\partial}{\partial\Omega}\hat{\Lambda} \label{eq:op1}, \\
\hat{n}_{xy}\hat{\Lambda}&=n_{xy}\hat{\Lambda}+(\delta_{xw}-n_{xw})n_{zy}\frac{\partial\hat{\Lambda}}{\partial n_{zw}} , \label{eq:op2} \\
\hat{\Lambda}\hat{n}_{xy}&=n_{xy}\hat{\Lambda}+n_{xw}(\delta_{zy}-n_{zy})\frac{\partial\hat{\Lambda}}{\partial n_{zw}} . \label{eq:op3} 
\end{align}
The trace of the Gaussian operator itself is ${\rm Tr}[\hat{\Lambda}]=\Omega$ 
and the trace of any ladder operators can be calculated by Wick's theorem. 
For instance, we obtain 
\begin{align}
{\rm Tr}[\hat{\Lambda}\hat{c}_{x}^{\dagger}\hat{c}_{y}]&=\Omega n_{xy} , \label{eq:trace1} \\
{\rm Tr}[\hat{\Lambda}\hat{c}_{x}^{\dagger}\hat{c}_{y}\hat{c}_{w}^{\dagger}\hat{c}_{z}]&=\Omega\left[n_{xy}n_{wz}+n_{xz}(\delta_{wy}-n_{wy})\right] . \label{eq:trace2}
\end{align}

To obtain the ground state of the system, one may consider the 
imaginary-time evolution of the density-matrix operator 
\begin{align}
\frac{\partial\hat{\rho}}{\partial\tau}=-\frac{1}{2}\left[\hat{H},\hat{\rho}\right]_{+},\qquad \tau=1/k_{B}T . 
\end{align}
In the GBMC method, instead of solving the Liouville equation above, 
one solves generalized Langevin equations by making use of the 
mapping between the Liouville equation 
and the stochastic equations. To this end, we expand the density-matrix operator as
$\hat{\rho}=\int d\underline{\lambda}P(\underline{\lambda},\tau)\hat{\Lambda}(\underline{\lambda})$. 
Then the Liouville equation becomes
\begin{align}
\int d\underline{\lambda}\hat{\Lambda}(\underline{\lambda})\frac{\partial P(\underline{\lambda},\tau)}{\partial\tau}
=\int d\underline{\lambda}P(\underline{\lambda},\tau)\left\{-\frac{1}{2}\left[\hat{H},\hat{\Lambda}(\underline{\lambda})\right]_{+}\right\} . \label{eq:part}
\end{align}
The differential identities of the Gaussian operator enable us to transform 
$-\frac{1}{2}\left[\hat{H},\hat{\Lambda}(\underline{\lambda})\right]_{+}$ 
into a differential form:
\begin{align}
-\frac{1}{2}\left[\hat{H},\hat{\Lambda}\right]_{+}=&\left[-\Omega H(\bm{n})\frac{\partial}{\partial\Omega}-\sum_{x,y}A_{xy}\frac{\partial}{\partial n_{xy}} \right. \notag \\
&\left. +\frac{1}{2}\sum_{i}\sum_{\substack{x,y\\ w,z}}\left(B_{xy}^{(i)}B_{wz}^{(i)}\frac{\partial^{2}}{\partial n_{xy}\partial n_{wz}}\right.\right. \notag \\
&\quad\qquad\left.\left. +C_{xy}^{(i)}C_{wz}^{(i)}\frac{\partial^{2}}{\partial n_{xy}\partial n_{wz}}\right)\right]\hat{\Lambda} ,
\end{align}
where 
\begin{align}
H(\bm{n})&={\rm Tr}[\hat{\Lambda}(\bm{n})\hat{H}]/{\rm Tr}[\hat{\Lambda}(\bm{n})] , \\
\bm{A}&=\frac{1}{2}\bm{n}(\bm{T}-U\bm{M})(\bm{I}-\bm{n}) \notag \\
&\qquad\qquad\qquad+\frac{1}{2}(\bm{I}-\bm{n})(\bm{T}-U\bm{M})\bm{n} , \label{eq:extension1} \\
B_{xy}^{(i)}&=\sqrt{\frac{U}{2}}\sum_{\sigma,\sigma'}\sigma_{\sigma\sigma'}^{z}n_{x,(i\sigma')}(\delta_{(i\sigma),y}-n_{(i\sigma),y}) , \\
C_{xy}^{(i)}&=\sqrt{\frac{U}{2}}\sum_{\sigma,\sigma'}\sigma_{\sigma\sigma'}^{z}(\delta_{x,(i\sigma')}-n_{x,(i\sigma')})n_{(i\sigma),y} , \label{eq:extension2} \\
M_{(i\sigma),(j\sigma')}&=\delta_{ij}\sum_{\eta,\eta'}n_{(i\eta),(i\eta')}(\sigma_{\sigma\sigma'}^{z}\sigma_{\eta\eta'}^{z}-\sigma_{\sigma\eta'}^{z}\sigma_{\eta\sigma'}^{z}) . 
\end{align}
Partial integration, under the assumption that boundary terms vanish, yields the Fokker-Planck equation for the 
probability distribution $P(\underline{\lambda},\tau)$:
\begin{align}
\frac{\partial P(\underline{\lambda},\tau)}{\partial\tau}=&\left[\frac{\partial}{\partial\Omega}\Omega H(\bm{n})+\sum_{x,y}\frac{\partial}{\partial n_{xy}}A_{xy}\right. \notag \\
&\left. +\frac{1}{2}\sum_{i}\sum_{\substack{x,y\\ w,z}}\left(\frac{\partial^{2}}{\partial n_{xy}\partial n_{wz}}B_{xy}^{(i)}B_{wz}^{(i)}\right.\right. \notag \\
&\qquad\left.\left. +\frac{\partial^{2}}{\partial n_{xy}\partial n_{wz}}C_{xy}^{(i)}C_{wz}^{(i)}\right)\right]P(\underline{\lambda},\tau) . \label{eq:Fokker}
\end{align}
In the actual calculation, instead of solving this equation directly, 
we solve the Ito-type Langevin equations with respect to 
the parameters of the Fokker-Planck equation which 
reproduce the distribution of $P(\underline{\lambda},\tau)$~\cite{Gardiner}:
\begin{align}
d\Omega&=-\Omega H(\bm{n})d\tau , \label{eq:langevin1} \\
d\bm{n}&=-\bm{A}d\tau+\sum_{i}\bm{B}^{(i)}dW_{i}+\sum_{i}\bm{C}^{(i)}dW_{i}' , \label{eq:langevin2}
\end{align}
where $dW$ and $dW'$ are Wiener increments which satisfy 
$\langle dW_{i}\rangle=\langle dW_{i}'\rangle=\langle dW_{i}dW_{j}'\rangle=0$ and 
$\langle dW_{i}dW_{j}\rangle=\langle dW_{i}'dW_{j}'\rangle=\delta_{ij}d\tau$ .

Any expectation values of physical observables are evaluated by using the trace 
properties in Eqs.(\ref{eq:trace1}) and (\ref{eq:trace2}). 
Let $\hat{O}$ be a general observable consisting of the ladder operators, 
then the expectation 
value of $\hat{O}$ becomes 
\begin{align}
\langle\hat{O}\rangle=\frac{{\rm Tr}[\hat{\rho}\hat{O}]}{{\rm Tr}[\hat{\rho}]}
&=\frac{\int d\underline{\lambda}P(\underline{\lambda},\tau){\rm Tr}\left[\hat{\Lambda(\underline{\lambda})}\hat{O}\right]}{\int d\underline{\lambda}P(\underline{\lambda},\tau){\rm Tr}\left[\hat{\Lambda(\underline{\lambda})}\right]} \notag \\
&=\frac{\int d\underline{\lambda}P(\underline{\lambda},\tau)\Omega O(\bm{n})}{\int d\underline{\lambda}P(\underline{\lambda},\tau)\Omega} . 
\end{align}
In the GBMC method, the integration with the weight $P(\underline{\lambda},\tau)$ 
is achieved alternatively by summing up over all the 
walkers of the Langevin equations (\ref{eq:langevin1}) and (\ref{eq:langevin2}), {\it i.e.},
\begin{align}
\langle\hat{O}\rangle=\frac{\sum_{i}\Omega_{i} O(\bm{n}_{i})}{\sum_{i}\Omega_{i}} . \label{eq:weight}
\end{align} 
We now regard $\Omega$ as the weight of the importance sampling in the Monte Carlo procedure. 
Note that from Eq.~(\ref{eq:langevin1}), the formal solution of the weight $\Omega$ becomes 
\begin{align}
\Omega(\tau)=\exp\left[-\int_{0}^{\tau}d\tau' H(\bm{n}(\tau'))\right] . \label{eq:omew}
\end{align}
Since ``Green's function" $\bm{n}(\tau)$ and $H(\bm{n})$ are always real, 
the weight $\Omega$ remains positive. 
Hence the negative sign problem does not appear. 

\subsection{Numerical integration}
When integrating the Langevin equations, one has to be careful about the type of 
the SDEs. Since Eq.~(\ref{eq:langevin2}) is Ito-type SDE, 
the numerical integration must be done by Ito integration~\cite{Gardiner}.
Here, we introduce two schemes of the numerical integration. The simplest one is 
Euler-Maruyama scheme~\cite{Kloeden}:
\begin{align}
x_{i+1}=x_{i}+A(x_{i})\Delta\tau+B(x_{i})\Delta W_{i} . \label{eq:Euler-Maruyama}
\end{align}
This scheme is faster than any other scheme but is not stable in general. 
For a more stable integration, we use a semi-implicit iterative scheme~\cite{Drummond-SDE} : 
\begin{align}
x_{i+1}=x_{i}+A(x_{i+1})\Delta\tau+B(x_{i})\Delta W_{i} . \label{eq:semi-implicit}
\end{align}
To solve the SDE, we make a first guess $\tilde{x}_{i+1}$ by Euler-Maruyama scheme (\ref{eq:Euler-Maruyama}). 
Then, $\tilde{x}_{i+1}$ is substituted into the drift term of (\ref{eq:semi-implicit}) 
iteratively until a self-consistent solution is found. 
For the parameter values of the Hubbard model we have chosen 
the time step $\Delta\tau=0.001$, 
then only a few iterations are needed because the initial guess from the Euler-Maruyama scheme 
is already close to the final solution. 

\subsection{Sampling method}
For an efficient calculation, the importance sampling is needed. From Eq.~(\ref{eq:weight}), 
$\Omega$ can be regarded as a weight. Thus we can construct an importance sampling 
method with  respect to $\Omega$. 
Corney and Drummond use the branching method for 
importance sampling~\cite{Corney1,Corney2,Dowling}. 
The branching method works by cloning the 
samples whose weights are large and by killing whose weights are small. 
Assaad {\it et al.} also 
use a similar reconfiguration method but their method keeps 
total population constant~\cite{Assaad,Sorella}. 

\begin{figure}[htb]
\begin{center}
\includegraphics[width=8.5cm]{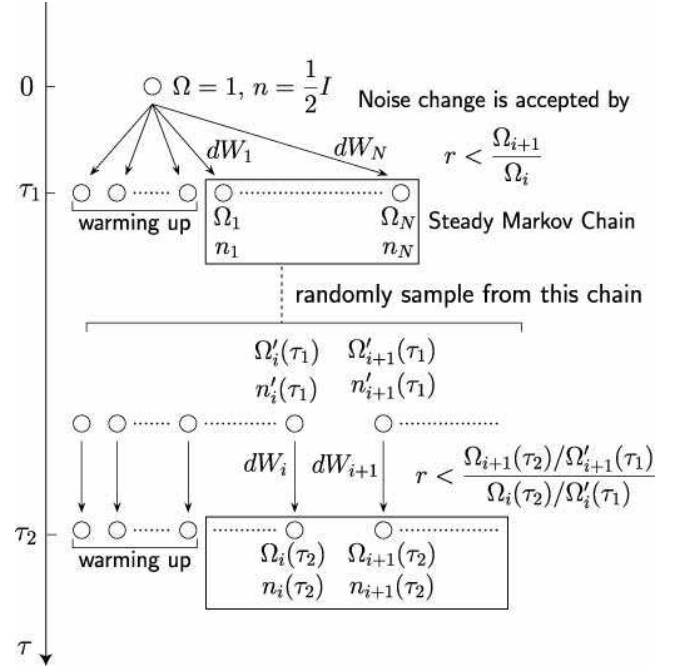}
\caption{sampling method} \label{fig:sampling}
\end{center}
\end{figure}

Here we propose another method which we call ``successive Metropolis method" 
(see Fig.~\ref{fig:sampling}). 
In contrast with the usual Metropolis method, 
this method allows us to evolve $\tau$ successively. 
After a certain number of time steps $N\Delta\tau$, Monte Carlo samples are 
stored by the Metropolis algorithm with the following conditions. 
\begin{align}
\cdot\ &{\rm choose\ one\ starting\ point}\ (\Omega_{i}'(\tau_{1}),\bm{n}_{i}'(\tau_{1}))\ \notag \\
&{\rm and\ Wiener\ increments}\ \vec{dW}_{i} \notag \\
\cdot\ &{\rm evolve}\ N\Delta\tau\ {\rm steps\ to\ get}\ (\Omega_{i}(\tau_{2}),\bm{n}_{i}(\tau_{2})) \notag \\
\cdot\ &{\rm choose\ another\ starting\ point}\ (\Omega_{i+1}^{({\rm try})'}(\tau_{1}),\bm{n}_{i+1}^{({\rm try})'}(\tau_{1})) \notag \\
&{\rm and\ another\ Wiener\ increments}\ \vec{dW}_{i+1}^{({\rm try})} \notag \\
\cdot\ &{\rm evolve}\ N\Delta\tau\ {\rm steps\ to\ get}\ (\Omega_{i+1}^{({\rm try})}(\tau_{2}),\bm{n}_{i+1}^{({\rm try})}(\tau_{2})) \notag \\
\cdot\ &{\rm select\ the\ new\ sample\ with\ the\ condition\ that} \notag \\
&(\Omega_{i+1}(\tau_{2}),\bm{n}_{i+1}(\tau_{2}),\vec{dW}_{i+1}) \notag \\
&\!=\!\left\{
 \begin{array}{l}
 \!\!\! (\Omega_{i+1}^{({\rm try})}(\tau_{2}),\bm{n}_{i+1}^{({\rm try})}(\tau_{2}),\vec{dW}_{i+1}^{({\rm try})})  \\
\qquad\qquad\qquad\qquad\qquad  {\rm if}\ r < \frac{\Omega_{i+1}^{({\rm try})}(\tau_{2})/\Omega_{i+1}^{({\rm try})'}(\tau_{1})}{\Omega_{i}(\tau_{2})/\Omega_{i}'(\tau_{1})},   \\
 \!\!\! (\Omega_{i}(\tau_{2}),\bm{n}_{i}(\tau_{2}),\vec{dW}_{i})\qquad\qquad {\rm otherwise} , \\
 \end{array}
 \right. \notag
\end{align}
where $r$ is a uniform random number distributed in $[0,1)$. 
After a sufficient number of warming-up steps, the stored samples constitute a steady Markov chain 
which can be regarded as the new starting points of further time evolutions.

\subsection{Systematic deviation}\label{sec:deviation}
\begin{figure}[htb]
\begin{center}
\includegraphics[width=7.5cm]{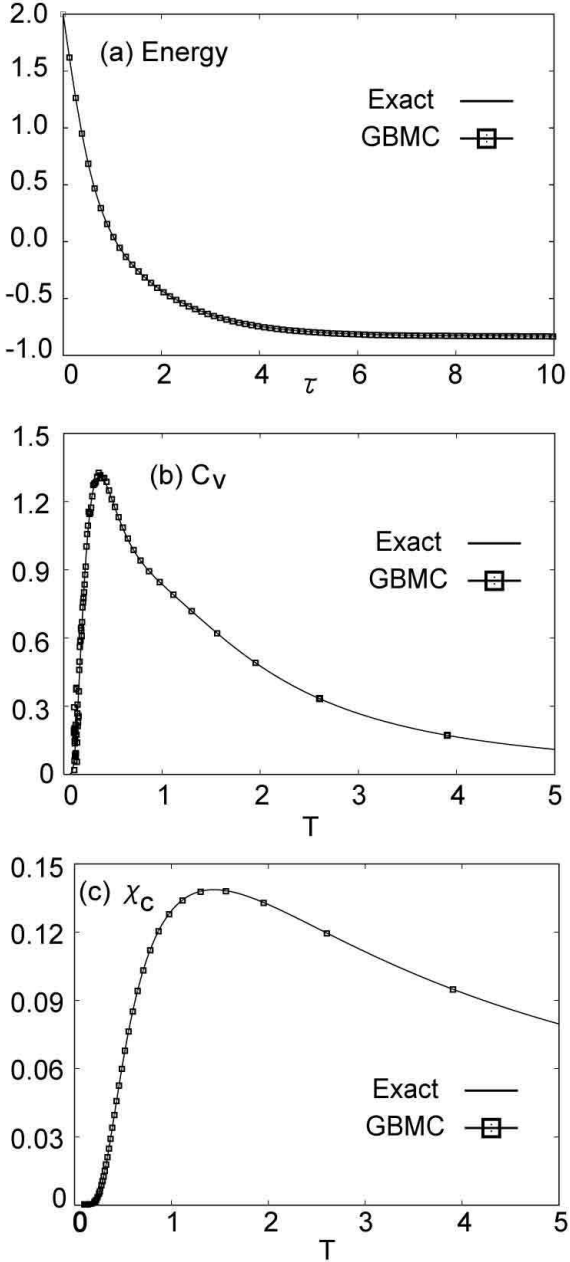}
\caption{(a) Energy of $2\times 1$ lattice with $U/t=4$ and $n=1$ as a function of inverse temperature $\tau$. 
Solid curve is obtained from the exact diagonalization. 
(b) Specific heat as a function of temperature $T$ 
obtained from numerical derivative of (a). 
(c) Charge susceptibility as a function of temperature $T$.} \label{fig:elementary}
\end{center}
\end{figure}

Here, we demonstrate some elementary results in the case of the 
two-site Hubbard model at $U/t=4$ and $n=1$ under the open boundary condition. Figure \ref{fig:elementary} shows the
total energy, the specific heat $C_{v}$ and the charge susceptibility 
%\begin{align}
$\chi_{c}=\frac{1}{NT}(\langle\hat{N}^{2}\rangle-\langle\hat{N}\rangle^{2})$ , 
%\end{align}
where $\hat{N}=\sum_{i\sigma}\hat{c}_{i\sigma}^{\dag}\hat{c}_{i\sigma}$. 
All the numerical results show excellent agreement with the exact diagonalization result. 

However, simulation results deviate from the exact diagonalization results 
if the lattice size or the strength of the on-site interaction $U$ becomes extremely larger. 
Here, as an example we demonstrate the results for the case of the two-site Hubbard model 
at $U/t=100$ and $n=1$. 
\begin{figure}[htb]
\begin{center}
\includegraphics[width=7.5cm]{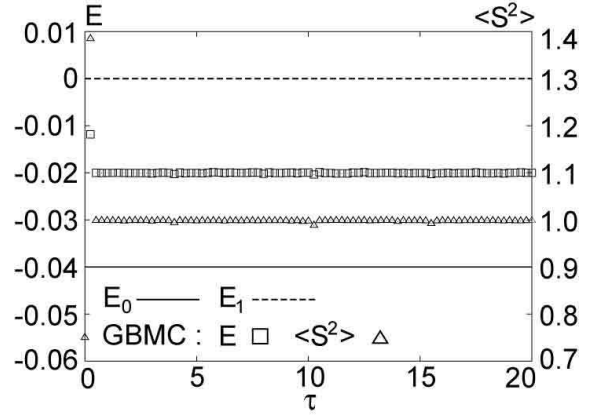}
\caption{Energy and total spin on $2\times 1$ lattice at 
$U/t=100$ and $n=1$ as functions of inverse temperature $\tau$. 
The squares and the triangles represent the GBMC results of the total energy and the total spin, respectively. 
Solid and dashed lines represent the exact value of the energy at 
the ground state $E_{0}$ and the first excited state $E_{1}$, respectively. 
} \label{fig:2x1U100}
\end{center}
\end{figure}
Figure \ref{fig:2x1U100} shows the 
total energy and the total spin $\langle\hat{S}^{2}\rangle$ for $U/t=100$ and $n=1$, 
where the total spin operator $\hat{S}=(\hat{S}^{x},\hat{S}^{y},\hat{S}^{z})$ 
is defined as $\hat{S}^{\alpha}=\frac{1}{2}\sum_{i=1}^{N}\hat{\bm{c}}_{i}^{\dag}\sigma^{\alpha}\hat{\bm{c}}_{i},\ (\alpha=x,y,z)$. 
As is seen from Fig.~\ref{fig:2x1U100}, 
the energy obtained by the GBMC method is located just at the middle 
between the ground state and the triplet first excited states. 
This means that the GBMC method reproduces the state which 
is represented by the superposition of the ground state and the triplet first excited states. 
Indeed, the expectation value of the total spin is $\langle\hat{S}^{2}\rangle=1$, which is 
the middle point between the singlet state and the triplet states. 
Here, the ground state of the two-site Hubbard model at $n=1$ is known to be represented as 
\begin{align}
|\Phi_{0}\rangle&=\frac{1}{\sqrt{2}}\sqrt{1+\frac{1}{\sqrt{1+r^2}}}|\phi_{s1}\rangle \notag \\
&\qquad+\frac{1}{2}\sqrt{1-\frac{1}{\sqrt{1+r^2}}}\left(|\phi_{s2}\rangle+|\phi_{s3}\rangle\right) ,
\end{align}
where $r=4t/U$ and 
\begin{align}
|\phi_{s1}\rangle&=\frac{1}{\sqrt{2}}(|\uparrow,\downarrow\rangle-|\downarrow,\uparrow\rangle) \\ 
|\phi_{s2}\rangle&=|\uparrow\downarrow,0\rangle,\quad |\phi_{s3}\rangle=|0,\uparrow\downarrow\rangle. 
\end{align}
Note that for $r\ll 1$, the ground state can be represented as 
\begin{align}
|\Phi_{0}\rangle&\simeq|\phi_{s1}\rangle=\frac{1}{\sqrt{2}}(|\uparrow,\downarrow\rangle-|\downarrow,\uparrow\rangle) . 
\end{align}
On the other hand, the triplet first excited states are represented as 
\begin{align}
|\phi_{t1}\rangle&=\frac{1}{\sqrt{2}}(|\uparrow,\downarrow\rangle+|\downarrow,\uparrow\rangle) \\ 
|\phi_{t2}\rangle&=|\uparrow,\uparrow\rangle,\quad |\phi_{t3}\rangle=|\downarrow,\downarrow\rangle. 
\end{align}
The GBMC result of $S^{z}$ is nearly zero in the whole range of $\tau$ (not shown), 
and the GBMC method converges with
\begin{align}
|\phi_{1s}\rangle\pm|\phi_{1t}\rangle\propto |\uparrow,\downarrow\rangle\ {\rm or}\ |\downarrow,\uparrow\rangle. 
\end{align}
This means that samples obtained by the GBMC method are trapped in 
a symmetry broken states as $|\uparrow,\downarrow\rangle\ {\rm or}\ |\downarrow,\uparrow\rangle$, 
because the electron hopping is prohibited owing to the energy loss by the 
strong Coulomb repulsion. 

\begin{figure}[htb]
\begin{center}
\includegraphics[width=7.5cm]{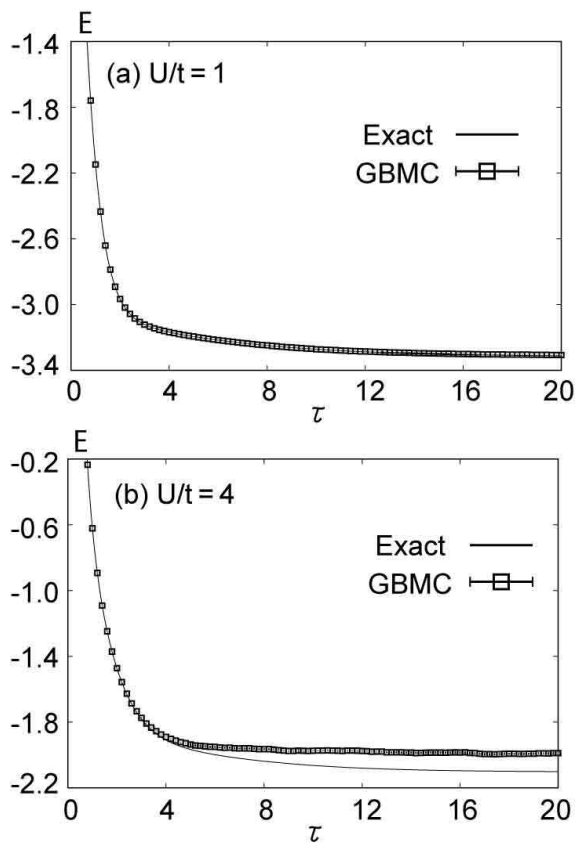}
\caption{(a) Energy of $4\times 1$ lattice with $U/t=1$ and $n=1$ as a function of inverse temperature $\tau$. 
Solid curve is obtained from the exact diagonalization. 
(b) Energy of $4\times 1$ lattice with $U/t=4$ and $n=1$. } \label{fig:elementary2}
\end{center}
\end{figure}

\begin{figure}[htb]
\begin{center}
\includegraphics[width=7.5cm]{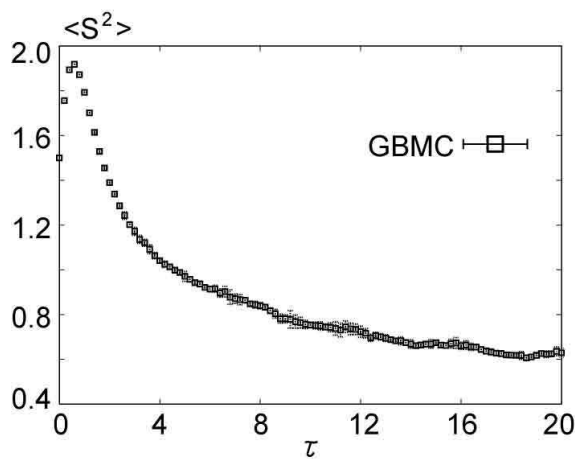}
\caption{Expectation value of total spin as a function of inverse temperature $\tau$ for the same case as Fig.~\ref{fig:elementary2} (b).} \label{fig:U04S}
\end{center}
\end{figure}

When the system size becomes larger, the systematic deviation caused by 
the same reason as the case of the two-site Hubbard model at $U/t=100$ occurs 
at a relatively small $U/t$. 
Here, for example we demonstrate the results for the case of 
the Hubbard model on the $4\times 1$ ring with $n=1$ under the periodic boundary condition in the $x$ direction. 
Hereafter, the $4\times 1$ lattice results are all obtained from the same boundary condition. 
Figure \ref{fig:elementary2} shows the
total energy in the case of $U/t=1$ and $U/t=4$. 
In the case of $U/t=1$, the total energy agrees with the exact diagonalization result, 
while in the case of $U/t=4$, the total energy deviates from the result of 
the exact diagonalization systematically. 
As is shown in the Table \ref{tab:spectra}, 
$4\times 1$ lattice Hubbard model with $U/t=4$ and $n=1$ has 
triplet exicited states with the energy $E=-1.806424$. 
We conclude that the GBMC samples are 
trapped in a symmetry broken state 
constructed from a linear combination of 
the singlet ground state and 
the triplet excited states. 
\begin{table}[h]
\caption{Energy spectra of $4\times 1$ lattice with $U/t=4$ and $n=1$ 
under the periodic boundary condition in the one-dimensional direction 
obtained by the exact diagonalization.} \label{tab:spectra}
\begin{center}
\begin{tabular}{cc}
\hline
    Energy      & Number of Degeneracy \\ \hline
  $-2.102748$ & 1   \\
  $-1.806424$ & 3   \\
  $-1.068140$ & 1   \\
%  $-0.8284271247$ & 6   \\
    $\vdots$    & $\vdots$ \\ \hline
\end{tabular}
\end{center}
\end{table}
To confirm this, we have calculated the expectation 
value of the total spin $\langle\hat{S}^{2}\rangle$. 
As is shown in Fig.~\ref{fig:U04S}, the total spin has a 
nonzero value and these overlaps with excited $S>0$ sectors 
cause the systematic deviation. 

\subsection{Power-law tails}\label{sec:tail}
To investigate the reason of the systematic deviation in detail, 
we calculate the distributions of the parameters of the Fokker-Planck equation (\ref{eq:Fokker}) 
to analyze whether or not the boundary term in the partial integration of Eq.~(\ref{eq:part}) 
exists. 

First, we calculate the distribution of the weight $\Omega$. 
\begin{figure}[htb]
\begin{center}
\includegraphics[width=7.5cm]{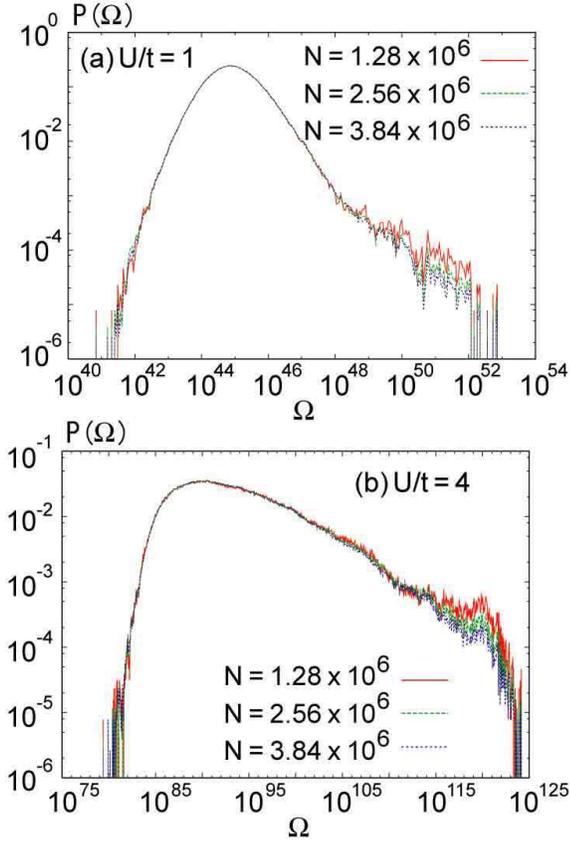}
\caption{(color online): Distribution of weight $\Omega$ at $\tau=20$ for the $4\times 1$ lattice at $n=1$ in case of (a) $U/t=1$ 
and (b) $U/t=4$. 
Red (solid), green (dashed) and blue (dotted) curves represent the distribution obtained by 
$1.28\times 10^{6}$, $2.56\times 10^{6}$ and $3.84\times 10^{6}$ Monte Carlo steps, 
respectively.} \label{fig:omega_comp}
\end{center}
\end{figure}
\begin{figure}[htb]
\begin{center}
\includegraphics[width=7.5cm]{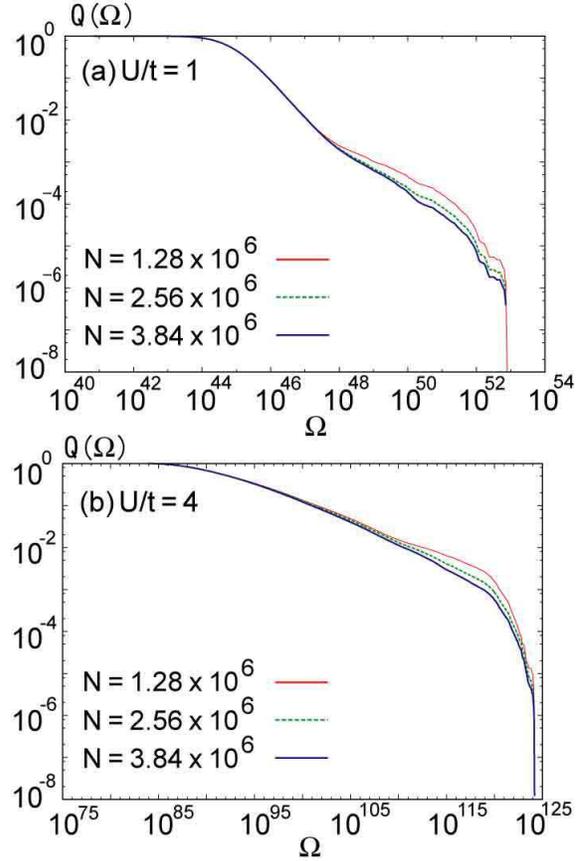}
\caption{(color online): Integrated distribution of weight $\Omega$ at $\tau=20$ for the case of $4\times 1$ lattice at $n=1$ 
in case of (a) $U/t=1$ 
and (b) $U/t=4$. 
Red (light), green (dashed) and blue (dark) curves represent the distribution obtained by 
$1.28\times 10^{6}$, $2.56\times 10^{6}$ and $3.84\times 10^{6}$ Monte Carlo steps, 
respectively.} \label{fig:qomega4x1}
\end{center}
\end{figure}
\begin{figure}[htb]
\begin{center}
\includegraphics[width=7.5cm]{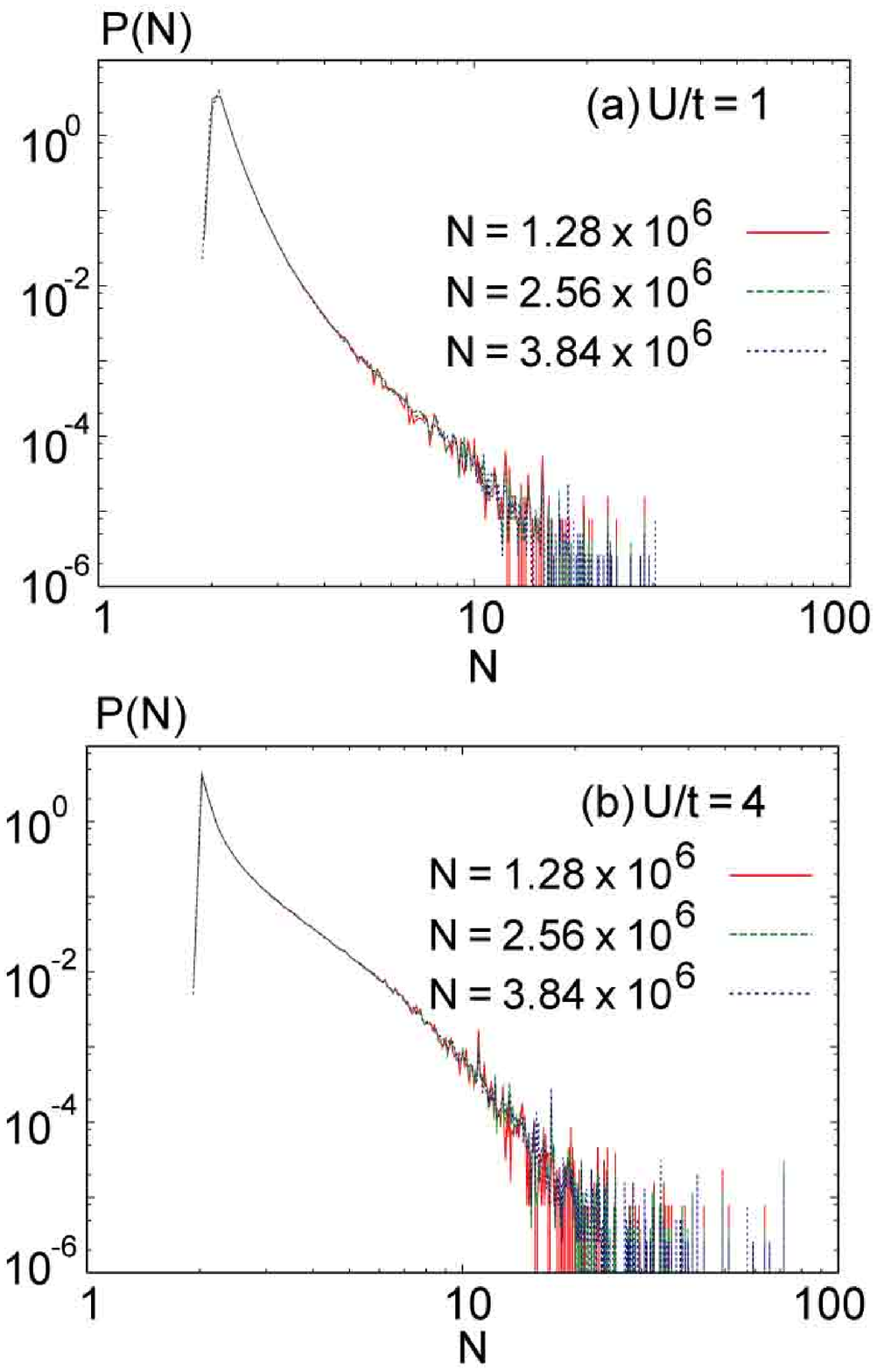}
\caption{(color online): Distribution of Green's function at $\tau=20$ for the $4\times 1$ lattice at $n=1$ in case of (a) $U/t=1$ 
and (b) $U/t=4$. 
Red (solid), green (dashed) and blue (dotted) curves represent the distribution obtained by 
$1.28\times 10^{6}$, $2.56\times 10^{6}$ and $3.84\times 10^{6}$ Monte Carlo steps, 
respectively.} \label{fig:n2_comp}
\end{center}
\end{figure}
As is seen form Fig.~\ref{fig:omega_comp}, 
the Monte Carlo step dependence of $P(\Omega)$ implies that 
the distributions of both $U/t=1$ and $U/t=4$ have upper limits of $\Omega$, 
even though they have 
broad peak structures. 
To analyze this cutoff of $P(\Omega)$ in detail, we also calculate the integrated distribution 
defined by 
\begin{align}
Q(\Omega)=1-\int_{0}^{\Omega}P(\Omega')d\Omega' .
\end{align}
As we see from Fig.~\ref{fig:qomega4x1}, the integrated distributions of both $U/t=1$ and $U/t=4$ 
have cutoffs around $\Omega\simeq 10^{53}$ and $\Omega\simeq 10^{124}$, respectively. 
Thus there exists no boundary term with respect to the weight $\Omega$. 

Next, we calculate the distribution of Green's function $\bm{n}$. 
Figure \ref{fig:n2_comp} shows the distribution of Green's function. 
The abscissa $N$ denotes the square root of the 
sum of each squared Green's function element, {\it i.e.},
\begin{align}
N=\sqrt{\sum_{ij\sigma}n_{(i\sigma),(j\sigma)}^{2}}, 
\end{align}
and the ordinate $P(N)$ denotes its probability distribution. 
Similarly to the case of $P(\Omega)$, 
the distributions $P(N)$ of both $U/t=1$ and $U/t=4$ have the upper limits. 
Thus there is no boundary term with respect to Green's function $\bm{n}$, either. 
However, as is seen from Fig.~\ref{fig:n2_comp}, the distribution tails of both 
$U/t=1$ and $U/t=4$ show power-law-like behaviors, {\it i.e.}, $P(N)\propto N^{-r}$, 
below their cutoffs. 
If the exponent of the power law $r$ is less or equal to $p+1$, 
the $p$-th moment of Green's function diverges, {\it i.e.}, 
\begin{align}
\int N^{p}P(N)dN\propto \int N^{p-r}dN\rightarrow\infty\,,\quad {\rm if\ \ }r\le p+1 . \label{eq:criterion}
\end{align}

\begin{figure}[htb]
\begin{center}
\includegraphics[width=7.5cm]{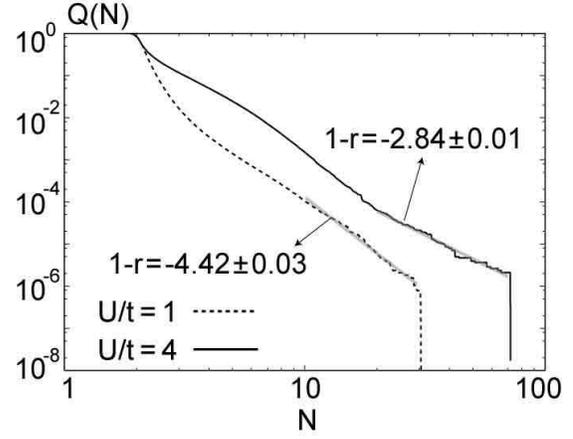}
\caption{Integrated distributions at $\tau=20$ for the $4\times 1$ lattice at $n=1$. 
Dashed curve represents the distribution of $U/t=1$, whereas solid curve is that of $U/t=4$.
The logarithmic fit to the tails leads to the exponent $r=5.42\pm 0.03$ for $U/t=1$ and 
$r=3.84\pm 0.01$ for $U/t=4$. 
Both data are obtained from $3.84\times 10^6$ Monte Carlo steps.} \label{fig:Qn2}
\end{center}
\end{figure}

To estimate the exponent of the power law in detail, 
we have also calculated the integrated distribution defined as
\begin{align}
Q(N)=1-\int_{0}^{N}P(N')dN' .
\end{align}
Figure \ref{fig:Qn2} shows the integrated distribution of both $U/t=1$ and $U/t=4$. 
From the tails of $Q(N)$, we obtain the power-law exponent of $P(N)$,
$r=5.42\pm 0.03$ for $U/t=1$ and $r=3.84\pm 0.01$ for $U/t=4$, respectively. 
To make the energy be well defined, the power-law exponent $r$ must be larger than three, 
since the  energy is the second-order moment of Green's function (see Eq.~(\ref{eq:criterion})). 
In the case of $U/t=4$, the exponent $r$ is larger than three, which also supports the absence 
of the boundary terms in the partial integration of Eq.~(\ref{eq:part}). 
From the analysis of the distribution $P(\Omega)$ and $P(N)$, we conclude that 
the systematic deviations observed in the results of the original GBMC method 
are caused not from the boundary terms but from the trap to the quasi-stable states. 
In the next section, we introduce the 
quantum-number projection method which can remove the systematic deviation. 

\section{Quantum-Number Projection}
Generally speaking, quantum many-body systems 
have several symmetries inherent in a Hamiltonian such as 
translational symmetry, U(1) symmetry, SU(2) symmetry, 
point group symmetry of the lattice, {\it etc}.
Although these symmetries are sometimes broken in the thermodynamic limit, 
they must be preserved in finite size systems. 
In actual numerical calculations, however, these symmetries 
are not always preserved in restricted Hilbert space and the numerical calculation often 
suffers from systematic errors. 

One of the most promising device to 
restore these symmetries is the quantum-number projection which 
has been used successfully in the framework of the path-integral 
renormalization group method~\cite{Mizusaki}. Also in the framework of the GBMC method, 
Assaad {\it et al.} have used the quantum-number projection~\cite{Assaad,Assaad2}. 
They proposed to project the density 
matrix onto given quantum numbers of the ground state 
after the ordinary GBMC sampling is performed and 
reproduced accurate ground states 
in some parameter regions. In this paper, we call their method 
post-projected GBMC (GBMC-PS) method.

In this section, we first review the mathematical framework of 
the quantum-number projection method, then
introduce 
an alternative method for performing the quantum-number projection. 
In this method, we combine the quantum-number projection scheme 
in conjunction with the importance sampling of the original GBMC method. 
This allows us to perform the importance sampling with respect to the quantum-number-projected distribution, 
which makes it possible to treat wider parameter region than the previous studies~\cite{Assaad,Assaad2}. 
In this paper, we call this method pre-projected GBMC (PR-GBMC) method. 

\subsection{Unitary transformation of a Gaussian operator}
Before introducing the quantum-number projectors, we define a unitary transformation 
of a Gaussian operator which will be used in all the projectors. 
For any Hermitian matrix $\bm{h}(\bm{x})$, a Gaussian operator is transformed as~\cite{Assaad,Assaad2} 
\begin{align}
e^{i\hat{\bm{c}}^{\dag}\bm{h}(\bm{x})\hat{\bm{c}}}\hat{\Lambda}(\Omega,\bm{n})
=\hat{\Lambda}(\Omega(\bm{x}),\bm{n}(\bm{x})) , \label{eq:unitary}
\end{align}
where
\begin{align}
\Omega(\bm{x})&=\Omega\det\left[(e^{i\bm{h}(\bm{x})}-\bm{I})\bm{n}^{T}+\bm{I}\right] , \label{eq:4.2} \\
\bm{I}-\bm{n}(\bm{x})&=\left\{\left[(e^{i\bm{h}(\bm{x})}-\bm{I})\bm{n}^{T}+\bm{I}\right]^{-1}\right\}^{T}(\bm{I}-\bm{n}) . \label{eq:4.3}
\end{align}
To prove Eqs.(\ref{eq:4.2}) and (\ref{eq:4.3}), 
we first introduce several identities of the Grassmann algebra~\cite{Corney2} : 
\begin{align}
\langle\bm{\xi}|\bm{\xi}'\rangle &= \exp\left[\overline{\bm{\xi}}\bm{\xi}'-\frac{1}{2}\overline{\bm{\xi}}\bm{\xi}-\frac{1}{2}\overline{\bm{\xi}'}\bm{\xi}'\right] , \\
\langle\bm{\xi}|:A(\hat{\bm{c}}^{\dag},\hat{\bm{c}}):|\bm{\xi}'\rangle &= A(\overline{\bm{\xi}},\bm{\xi}') \notag \\
&\quad\times\exp\left[\overline{\bm{\xi}}\bm{\xi}'-\frac{1}{2}\overline{\bm{\xi}}\bm{\xi}-\frac{1}{2}\overline{\bm{\xi}'}\bm{\xi}'\right] , \\
\bm{1} &= \int\underbrace{\prod_{x}[d\overline{\xi}_{x}d\xi_{x}]}_{\cal{D}\bm{\xi}}|\bm{\xi}\rangle\langle\bm{\xi}| , 
\end{align}
where $\bm{\xi}$ are Grassmann vectors and $|\bm{\xi}\rangle$ are Fermi coherent states. 

In order to use the above identities, it is necessary to transform 
$e^{i\hat{\bm{c}}^{\dag}\bm{h}\hat{\bm{c}}}$ into a normal ordered form. 
Because $\bm{h}$ is Hermitian, it can be diagonalized 
by the unitary transformation such that 
$\bm{h} = \bm{U}\bm{D}\bm{U}^{\dag}$, 
where $\bm{U}$ is a unitary matrix and $\bm{D}$ is a diagonal one. 
With the canonical transformation $\hat{\bm{\gamma}}^{\dag}=\hat{\bm{c}}^{\dag}\bm{U}$, 
$e^{i\hat{\bm{c}}^{\dag}\bm{h}\hat{\bm{c}}}$ becomes 
\begin{align}
e^{i\hat{\bm{c}}^{\dag}\bm{h}\hat{\bm{c}}} &= \prod_{i}e^{i\hat{\gamma}_{i}^{\dag}\hat{\gamma}_{i}D_{i}}
 = \prod_{i}\left[1+\sum_{n=1}^{\infty}\frac{(iD_{i})^{n}}{n!}\hat{\gamma}_{i}^{\dag}\hat{\gamma}_{i}\right] \notag \\
 &= \prod_{i}\left[1+(e^{iD_{i}}-1)\hat{\gamma}_{i}^{\dag}\hat{\gamma}_{i}\right] \notag \\
 &= \prod_{i}:e^{(e^{iD_{i}}-1)\hat{\gamma}_{i}^{\dag}\hat{\gamma}_{i}}:=:e^{\hat{\bm{c}}^{\dag}(e^{i\bm{h}}-\bm{I})\hat{\bm{c}}}: , 
\end{align}
where $(\hat{\gamma}_{i}^{\dag}\hat{\gamma}_{i})^{n}=\hat{\gamma}_{i}^{\dag}\hat{\gamma}_{i}$ is used.
Then for any matrix $\bm{B}$, we obtain
\begin{align}
&e^{i\hat{\bm{c}}^{\dag}\bm{h}\hat{\bm{c}}}:e^{\hat{\bm{c}}^{\dag}\bm{B}\hat{\bm{c}}}: \notag \\
&= :e^{\hat{\bm{c}}^{\dag}(e^{i\bm{h}}-\bm{I})\hat{\bm{c}}}::e^{\hat{\bm{c}}^{\dag}\bm{B}\hat{\bm{c}}}: \notag \\
&=\! \int\!{\cal D}\bm{\xi}{\cal D}\bm{\eta}{\cal D}\bm{\gamma} |\bm{\xi}\rangle\langle\bm{\xi}|:e^{\hat{\bm{c}}^{\dag}(e^{i\bm{h}}-\bm{I})\hat{\bm{c}}}:|\bm{\eta}\rangle\langle\bm{\eta}|:e^{\hat{\bm{c}}^{\dag}\bm{B}\hat{\bm{c}}}:|\bm{\gamma}\rangle\langle\bm{\gamma}| \notag \\
&=\! \int\!{\cal D}\bm{\xi}{\cal D}\bm{\eta}{\cal D}\bm{\gamma} |\bm{\xi}\rangle \notag \\
&\qquad\times \exp[\overline{\bm{\xi}}(e^{i\bm{h}}-\bm{I})\bm{\eta}]\exp\left[\overline{\bm{\xi}}\bm{\eta}-\frac{1}{2}\overline{\bm{\xi}}\bm{\xi}-\frac{1}{2}\overline{\bm{\eta}}\bm{\eta}\right] \notag \\
&\qquad\times \exp[\overline{\bm{\eta}}\bm{B}\bm{\gamma}]\exp\left[\overline{\bm{\eta}}\bm{\gamma}-\frac{1}{2}\overline{\bm{\eta}}\bm{\eta}-\frac{1}{2}\overline{\bm{\gamma}}\bm{\gamma}\right]\langle\bm{\gamma}| \notag \\
&=\! \int\!{\cal D}\bm{\xi}{\cal D}\bm{\epsilon}{\cal D}\bm{\gamma} |\bm{\xi}\rangle\exp\left[\overline{\bm{\xi}}\bm{\epsilon}-\frac{1}{2}\overline{\bm{\xi}}\bm{\xi}-\frac{1}{2}\overline{\bm{\epsilon}}\bm{\epsilon}\right] \notag \\
&\qquad\times \exp\left[\overline{\bm{\epsilon}}\{e^{i\bm{h}}(\bm{B}+\bm{I})-\bm{I}\}\bm{\gamma}\right] \notag \\
&\qquad\times \exp\left[\overline{\bm{\epsilon}}\bm{\gamma}-\frac{1}{2}\overline{\bm{\epsilon}}\bm{\epsilon}-\frac{1}{2}\overline{\bm{\gamma}}\bm{\gamma}\right]\langle\bm{\gamma}| \notag \\
&= :e^{\hat{\bm{c}}^{\dag}[e^{i\bm{h}}(\bm{B}+\bm{I})-\bm{I}]\hat{\bm{c}}}: , 
\end{align}
where $\bm{\epsilon}=e^{i\bm{h}}\bm{\eta}$ and 
\begin{align}
{\cal D}\bm{\epsilon}=\prod_{x}d\overline{\epsilon}_{x}d\epsilon_{x}=\prod_{x}d\overline{\eta}_{x}e^{-i\bm{h}}e^{i\bm{h}}d\eta_{x}={\cal D}\bm{\eta}. 
\end{align}
Thus Eq.~(\ref{eq:unitary}) is proven by taking 
$\bm{B}=-[2\bm{I}+(\bm{n}^{T}-\bm{I})^{-1}]$. 

\subsection{Examples of quantum-number projector}
\subsubsection{Particle-number projector}
Since the GBMC method is a grand canonical approach, 
particle-number projection is needed when treating the canonical ensemble. 
The projector onto a state with a given particle-number $N$ is defined as 
\begin{align}
\hat{P}_{{\cal N}}(N)&=\frac{1}{2\pi}\int_{0}^{2\pi}d\phi\langle N|\hat{T}_{{\cal N}}(\phi)|N\rangle^{\dagger}\hat{T}_{{\cal N}}(\phi) \notag \\
&=\frac{1}{2\pi}\int_{0}^{2\pi}d\phi g_{{\cal N}}(\phi,N)e^{i\hat{\bm{c}}^{\dag}\bm{h}_{{\cal N}}(\phi)\hat{\bm{c}}} , 
\end{align}
where $g_{{\cal N}}(\phi,N)=e^{-i\phi N}$ and 
\begin{align}
\hat{T}_{{\cal N}}(\phi)=e^{i\phi\sum_{i}\hat{\bm{c}}_{i}^{\dag}\hat{\bm{c}}_{i}} ,\quad
e^{i\bm{h}_{{\cal N}}(\phi)}=e^{i\phi}\bm{I} . 
\end{align}

Similarly, the projection onto the state which has 
$N_{\sigma}$ electrons with spin $\sigma$ is defined as 
\begin{align}
\hat{P}_{{\cal N}_{\sigma}}(N_{\sigma})&=
\frac{1}{2\pi}\int_{0}^{2\pi}d\phi\langle N_{\sigma}|\hat{T}_{{\cal N}_{\sigma}}(\phi)|N_{\sigma}\rangle^{\dagger}\hat{T}_{{\cal N}_{\sigma}}(\phi) \notag \\
&=\frac{1}{2\pi}\int_{0}^{2\pi}d\phi g_{{\cal N}_{\sigma}}(\phi,N_{\sigma})e^{i\hat{\bm{c}}^{\dag}\bf{h}_{{\cal N}_{\sigma}}(\phi,\sigma)\hat{\bm{c}}} , 
\end{align}
where $g_{{\cal N}_{\sigma}}(\phi,N_{\sigma})=e^{-i\phi N_{\sigma}}$ and 
\begin{align}
\hat{T}_{{\cal N}_{\sigma}}(\phi)&=e^{i\phi\sum_{i}\hat{c}_{i\sigma}^{\dagger}\hat{c}_{i\sigma}} ,\quad
e^{i\bm{h}_{{\cal N}_{\sigma}}(\phi,\sigma)}=e^{i\phi}\bm{I}_{\sigma} ,  \\
\bm{I}_{\sigma}&=\delta_{\sigma,\uparrow}\begin{bmatrix} \bm{I} & {\bf 0} \\ {\bf 0} & {\bf 0} \end{bmatrix}
+\delta_{\sigma,\downarrow}\begin{bmatrix} {\bf 0} & {\bf 0} \\ {\bf 0} & \bm{I} \end{bmatrix} . 
\end{align}

\subsubsection{Total-spin projector}
The SU(2) symmetry of the total spin is recovered by summing up all the Euler angles 
in the spin space~\cite{Mizusaki,Assaad,Assaad2}. 
Thus, projection onto a given total-spin state is defined as 
(here we restrict ourselves to the case of $S_{z}=0$)
\begin{align}
\hat{P}_{\cal{S}}(S)&=\frac{2S+1}{\int d\bm{\omega}}\int d\bm{\omega}\langle S,0|\hat{T}_{\cal{S}}(\bm{\omega})|S,0\rangle^{\dagger}\hat{T}_{\cal{S}}(\bm{\omega}) \notag \\
&=\frac{2S+1}{8\pi^{2}}\int_{0}^{2\pi}d\alpha\int_{0}^{\pi}d\beta\sin\beta\int_{0}^{2\pi}d\gamma \notag \\
&\qquad \times P_{S}(\cos\beta)e^{i\hat{\bm{c}}^{\dag}\bm{h}_{S_z}(\alpha)\hat{\bm{c}}}e^{i\hat{\bm{c}}^{\dag}\bm{h}_{S_y}(\beta)\hat{\bm{c}}}e^{i\hat{\bm{c}}^{\dag}\bm{h}_{S_z}(\gamma)\hat{\bm{c}}} , 
\end{align}
where $P_{S}(\cos\beta)$ is the $S^{\rm{th}}$ Legendre polynomial and 
\begin{align}
\hat{T}_{\cal{S}}(\bm{\omega})&=e^{i\alpha\hat{S}^{z}}e^{i\beta\hat{S}^{y}}e^{i\gamma\hat{S}^{z}} , \\
\bm{h}_{S_z}(\alpha)&=\frac{\alpha}{2}\begin{bmatrix} \bm{I} & {\bf 0} \\ {\bf 0} & -\bm{I} \end{bmatrix} ,\ 
\bm{h}_{S_y}(\beta)=\frac{\beta}{2}\begin{bmatrix} {\bf 0} & -i\bm{I} \\ i\bm{I} & {\bf 0} \end{bmatrix} . 
\end{align}
Here, $\hat{S}^{\alpha},(\alpha=x,y,z)$ corresponds to the total $\alpha$-component 
of spin : 
\begin{align}
\hat{S}^{\alpha}=\frac{1}{2}\sum_{i}\hat{\bm{c}}_{i}^{\dag}\bm{\sigma}^{\alpha}\hat{\bm{c}}_{i} .
\end{align}

Since the total-spin projection involves triple integrals, 
the computational cost is rather high. However, if one takes 
$N_{\uparrow}$ projection and $N_{\downarrow}$ projection before 
the total-spin projection, the integrations about Euler angles 
$\alpha$ and $\gamma$ can be done analytically. 
Thus, the total-spin projection is reduced to 
\begin{align}
\hat{P}_{\cal{S}}(S)=\frac{2S+1}{2}\int_{0}^{\pi}d\beta g_{\cal{S}}(\beta,S)e^{i\hat{\bm{c}}^{\dag}\bm{h}_{S_y}(\beta)\hat{\bm{c}}} , 
\end{align}
where $g_{\cal{S}}(\beta,S)=\sin\beta\, P_{S}(\cos\beta)$.

\subsubsection{Lattice-symmetry projector}
\begin{figure}[htb]
\begin{center}
\includegraphics[width=6cm]{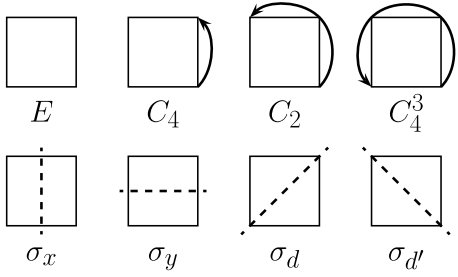}
\caption{Each element of $C_{4v}$ group.} \label{fig:c4v}
\end{center}
\end{figure}

When the Hamiltonian is invariant under certain 
geometrical transformations, 
such geometrical symmetry is recovered by summing up all the transformations. 
For instance, when treating square lattice systems, 
they have $C_{4v}$ lattice symmetry (see Fig.~\ref{fig:c4v}). 
By assuming that the square lattice lies in the $x$-$y$ plane, 
the $\pi/2$ rotations around the $z$-axis is achieved by 
the $z$-component of the angular momentum $\hat{L}_{z}$. 
Let $\hat{T}_{C_{4}}=e^{i\frac{\pi}{2}\hat{L}_{z}}$ such that 
\begin{align}
\hat{T}_{C_{4}}=e^{i\frac{\pi}{2}\hat{L}_{z}}&=e^{i\hat{\bm{c}}^{\dag}\bm{h}_{C_{4}}\hat{\bm{c}}} \\
\hat{T}_{C_{4}}\hat{\bm{c}}_{i\sigma}^{\dag}\hat{T}_{C_{4}}^{-1}&=\hat{\bm{c}}_{C_{4}(i)\sigma}^{\dag} , 
\end{align}
where $C_{4}(i)$ denotes a $\pi/2$ rotation around $z$-axis. 
From the above, $e^{i\hat{\bm{c}}^{\dag}\bm{h}_{C_{4}}\hat{\bm{c}}}$ is represented as 
\begin{align}
\left[e^{i\hat{\bm{c}}^{\dag}\bm{h}_{C_{4}}\hat{\bm{c}}}\right]_{(j\sigma'),(i\sigma)}=\delta_{\sigma\sigma'}\delta_{jC_{4}(i)} .
\end{align}
Similarly, the $x$-$y$ mirror transformation $T_{\sigma_{d}}=e^{i\hat{\bm{c}}^{\dag}\bm{h}_{\sigma_{d}}\hat{\bm{c}}}$ 
is defined as 
\begin{align}
\hat{T}_{\sigma_{d}}\hat{\bm{c}}_{i\sigma}^{\dag}\hat{T}_{\sigma_{d}}^{-1}=\hat{\bm{c}}_{\sigma_{d}(i)\sigma}^{\dag} ,
\end{align}
where $\sigma_{d}(i)$ denotes a $x$-$y$ mirror transformation of $i$-th site, {\it i.e.}, 
$\sigma_{d}(x,y)=(y,x)$. 
Then, the representation of $\sigma_{d}$ becomes 
\begin{align}
\left[e^{i\hat{\bm{c}}^{\dag}\bm{h}_{\sigma_{d}}\hat{\bm{c}}}\right]_{(j\sigma'),(i\sigma)}=\delta_{\sigma\sigma'}\delta_{j\sigma_{d}(i)} .
\end{align}

Although $C_{4v}$ group has other elements as in Fig.~\ref{fig:c4v}, 
all can be generated by $C_{4}$ and $\sigma_{d}$ 
(see Table.\ref{tab:c4v_tab}). For example, $\sigma_{x}=C_{4}\cdot\sigma_{d}$, {\it i.e.}, 
\begin{align}
C_{4}\cdot\sigma_{d}\left(
  \begin{array}{c}
     x \\
     y \\
  \end{array}
\right)&=
C_{4}\left(
  \begin{array}{c}
     y \\
     x \\
  \end{array}
\right)=
\left(
  \begin{array}{cc}
     0 & -1 \\
     1 & 0 \\
  \end{array}
\right)
\left(
  \begin{array}{c}
     y \\
     x \\
  \end{array}
\right) \notag \\
&=
\left(
  \begin{array}{c}
     -x \\
     y \\
  \end{array}
\right)=
\sigma_{x}\left(
  \begin{array}{c}
     x \\
     y \\
  \end{array}
\right)
\end{align}
Thus, the projection onto the $C_{4v}$ symmetry sector reads
\begin{align}
\hat{P}_{C_{4v}}=\frac{1}{8}\sum_{\alpha}g_{C_{4v}}(\alpha)\hat{T}_{C_{4v}}(\alpha) , 
\end{align}
where $\alpha=\{E,C_{4},C_{2},C_{4}^{3},\sigma_{x},\sigma_{y},\sigma_{d},\sigma_{d'}\}$ and
\begin{align}
&{\rm for}\ s\ {\rm wave} \notag \\
&\qquad g_{C_{4v}}(\alpha)=1\ (\,\forall \ \alpha\,), \\
&{\rm for}\ d_{x^{2}-y^{2}}\ {\rm wave} \notag \\
&\qquad g_{C_{4v}}(\alpha)=
\left\{
  \begin{array}{cl}
     1 & (\alpha=E,C_{2},\sigma_{x},\sigma_{y}) \\
    -1 & (\alpha=C_{4},C_{4}^{3},\sigma_{d},\sigma_{d'}) \\
  \end{array}
\right. , \\
&{\rm for}\ d_{xy}\ {\rm wave} \notag \\
&\qquad g_{C_{4v}}(\alpha)=
\left\{
  \begin{array}{cl}
     1 & (\alpha=E,C_{2},\sigma_{d},\sigma_{d'}) \\
    -1 & (\alpha=C_{4},C_{4}^{3},\sigma_{x},\sigma_{y}) \\
  \end{array}
\right. . 
\end{align}

For a simpler case, 
$C_{4}$-symmetry projector is obtained by 
restricting $\alpha$ to the subgroup of $C_{4v}$, {\it i.e.},
\begin{align}
\hat{P}_{C_{4}}&=\frac{1}{4}\sum_{\alpha}g_{C_4}(\alpha)\hat{T}_{C_4}(\alpha) , \\
\alpha&=\{E,C_{4},C_{2},C_{4}^{3}\} ,
\end{align}
where the definition of $g_{C_4}(\alpha)$ is same as $g_{C_{4v}}(\alpha)$ within $\alpha=\{E,C_{4},C_{2},C_{4}^{3}\}$. 
\begin{table}[h]
\begin{center}
\caption{The multiplication table of $C_{4v}$}\label{tab:c4v_tab}
\begin{tabular}{c|cccccccc} \hline
              & $E$           & $C_{4}$       & $C_{2}$       & $C_{4}^{3}$   & $\sigma_{x}$  & $\sigma_{y}$  & $\sigma_{d}$  & $\sigma_{d'}$ \\ \hline
$E$           & $E$           & $C_{4}$       & $C_{2}$       & $C_{4}^{3}$   & $\sigma_{x}$  & $\sigma_{y}$  & $\sigma_{d}$  & $\sigma_{d'}$ \\ 
$C_{4}$       & $C_{4}$       & $C_{2}$       & $C_{4}^{3}$   & $E$           & $\sigma_{d'}$ & $\sigma_{d}$  & $\sigma_{x}$  & $\sigma_{y}$  \\ 
$C_{2}$       & $C_{2}$       & $C_{4}^{3}$   & $E$           & $C_{4}$       & $\sigma_{y}$  & $\sigma_{x}$  & $\sigma_{d'}$ & $\sigma_{d}$  \\ 
$C_{4}^{3}$   & $C_{4}^{3}$   & $E$           & $C_{4}$       & $C_{2}$       & $\sigma_{d}$  & $\sigma_{d'}$ & $\sigma_{y}$  & $\sigma_{x}$  \\ 
$\sigma_{x}$  & $\sigma_{x}$  & $\sigma_{d}$  & $\sigma_{y}$  & $\sigma_{d'}$ & $E$           & $C_{2}$       & $C_{4}$       & $C_{4}^{3}$   \\ 
$\sigma_{y}$  & $\sigma_{y}$  & $\sigma_{d'}$ & $\sigma_{x}$  & $\sigma_{d}$  & $C_{2}$       & $E$           & $C_{4}^{3}$   & $C_{4}$       \\ 
$\sigma_{d}$  & $\sigma_{d}$  & $\sigma_{y}$  & $\sigma_{d'}$ & $\sigma_{x}$  & $C_{4}^{3}$   & $C_{4}$       & $E$           & $C_{2}$       \\ 
$\sigma_{d'}$ & $\sigma_{d'}$ & $\sigma_{x}$  & $\sigma_{d}$  & $\sigma_{y}$  & $C_{4}$       & $C_{4}^{3}$   & $C_{2}$       & $E$           \\ \hline
\end{tabular}
\end{center}
\end{table}

\subsubsection{Total-momentum projector}
When the Hamiltonian has the translational symmetry, the total momentum 
must be conserved. A translation by a lattice vector $\bm{R}$ is 
achieved by the operator $\hat{T}(\bm{R})=e^{i\bm{R}\cdot\bm{k}_{\rm{tot}}}$. Here, 
$\bm{k}_{\rm{tot}}$ is defined by the Fourier transformation of the creation and 
the annihilation operators as 
\begin{align}
\bm{k}_{\rm{tot}}=\sum_{\bm{k},\sigma}\bm{k}\hat{c}_{\bm{k}\sigma}^{\dag}\hat{c}_{\bm{k}\sigma} ,\quad
\hat{c}_{\bm{k}\sigma}^{\dag}=\frac{1}{\sqrt{N}}\sum_{\bm{i}}e^{i\bm{k}\cdot\bm{i}}\hat{c}_{\bm{i}\sigma}^{\dag} , 
\end{align}
where $\bm{i}$ denotes a vector to the $i$-th site and $N$ is the number of sites. 
The sum over $\bm{k}$ goes over all the points in the first Brillouin zone. 
The projection onto the Hilbert space with the total momentum $\bm{K}_{0}$ then reads: 
\begin{align}
\hat{P}_{\bm{{\cal K}}}(\bm{K}_{0})&=\frac{1}{N}\sum_{\bm{R}}\langle\bm{K}_{0}|\hat{T}_{\bm{{\cal K}}}(\bm{R})|\bm{K}_{0}\rangle^{\dag}\hat{T}_{\bm{{\cal K}}}(\bm{R}) \notag \\
&=\frac{1}{N}\sum_{\bm{R}}g_{\bm{{\cal K}}}(\bm{R},\bm{K}_{0})e^{i\hat{\bm{c}}^{\dag}\bm{h}_{\bm{{\cal K}}}(\bm{R})\hat{\bm{c}}} , 
\end{align}
where $g_{\bm{{\cal K}}}(\bm{R},\bm{K}_{0})=e^{-i\bm{R}\cdot\bm{K}_{0}}$ and the sum over $\bm{R}$ 
goes over all the lattice sites and $\bm{h}_{\bm{{\cal K}}}(\bm{R})$ is 
\begin{align}
(\bm{h}_{\bm{{\cal K}}}(\bm{R}))_{ij}=\frac{1}{N}\bm{R}\cdot\sum_{\bm{k}}\bm{k}e^{i\bm{k}\cdot(\bm{i}-\bm{j})} .
\end{align}
Then the matrix representation of $e^{i\hat{\bm{c}}^{\dag}\bm{h}_{\bm{{\cal K}}}(\bm{R})\hat{\bm{c}}}$ reads 
\begin{align}
\left[e^{i\hat{\bm{c}}^{\dag}\bm{h}_{\bm{{\cal K}}}(\vec{R})\hat{\bm{c}}}\right]_{(\bm{i}\sigma),(\bm{j}\sigma')}
&=\delta_{\sigma\sigma'}\langle\bm{i}|e^{i\bm{R}\cdot\bm{k}_{{\rm tot}}}|\bm{j}\rangle \notag \\
&=\delta_{\sigma\sigma'}\sum_{\bm{k}}\langle\bm{i}|\bm{k}\rangle e^{i\bm{R}\cdot\bm{k}}\langle \bm{k}|\bm{j}\rangle \notag \\
&=\delta_{\sigma\sigma'}\sum_{\bm{k}}e^{i\bm{k}\cdot(\bm{R}+\bm{i}-\bm{j})} \notag \\
&=\delta_{\sigma\sigma'}\delta_{\bm{i}+\bm{R},\bm{j}}\, . 
\end{align}

\subsection{Projected expectation values of observables}
From the previous subsection, one can define a general quantum-number projector as 
\begin{align}
\hat{P}=\int d\bm{x} g(\bm{x})\hat{T}(\bm{x}) , 
\end{align}
where $\hat{T}(\bm{x})$ is unitary and thus $\hat{P}^{\dag}=\hat{P}$. 
For simplicity, we assume that the physical observable $\hat{O}$ commutes with 
$\hat{P}$, {\it i.e.}, $\left[\hat{P},\hat{O}\right]_{-}=0$ . 
Then, the projected expectation value of the observable $\hat{O}$ becomes 
\begin{align}
\langle\hat{O}\rangle_{\hat{P}}=\frac{{\rm Tr}\left[\hat{P}\hat{\rho}\hat{P}\hat{O}\right]}{{\rm Tr}\left[\hat{P}\hat{\rho}\hat{P}\right]}=\frac{{\rm Tr}\left[\hat{P}\hat{\rho}\hat{O}\right]}{{\rm Tr}\left[\hat{P}\hat{\rho}\right]} . 
\end{align}
Here we use the projection property $\hat{P}^{2}=\hat{P}$. Replacing the density-matrix operator 
by the sum over all the walkers 
$\rho=\int d\underline{\lambda}P(\underline{\lambda})\hat{\Lambda}(\underline{\lambda})\simeq\sum_{i}\hat{\Lambda}(\underline{\lambda}_{i})$ yields
\begin{align}
\langle\hat{O}\rangle_{\hat{P}}&=\frac{\sum_{i}\int d\bm{x}{\rm Tr}\left[\hat{T}(\bm{x})\hat{\Lambda}(\underline{\lambda}_{i})\hat{O}\right]}{\sum_{i}\int d\bm{x}{\rm Tr}\left[\hat{T}(\bm{x})\hat{\Lambda}(\underline{\lambda}_{i})\right]} \notag \\
&=\frac{\sum_{i}\int d\bm{x} g(\bm{x})\Omega_{i}(\bm{x})O(\bm{n}_{i}(\bm{x}))}{\sum_{i}\int d\bm{x} g(\bm{x})\Omega_{i}(\bm{x})} , \label{eq:expectation}
\end{align}
where $\hat{T}(\bm{x})\hat{\Lambda}(\underline{\lambda})=\hat{\Lambda}(\underline{\lambda}(\bm{x}))$. 

\subsection{Post-projected sampling method}
\begin{figure}[htb]
\begin{center}
\includegraphics[width=7.5cm]{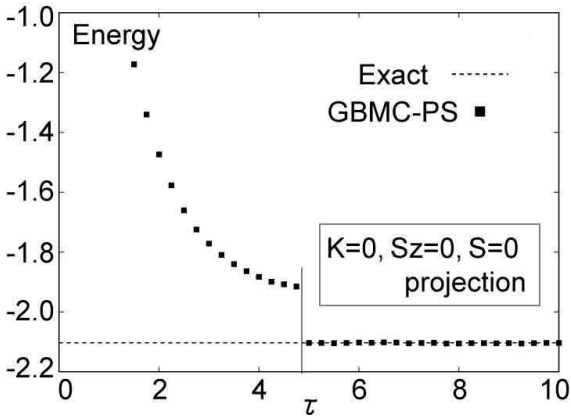}
\caption{Energy of the $4\times 1$-site Hubbard model with $U/t=4$ and $n=1$. 
Here, $\bm{K}=0,S_{z}=0$ and $S=0$ projections are performed for $\tau>5$. 
The integrals of $S_{z}$ and $S$ are evaluated 
by the Gauss-Legendre quadrature~\cite{Numerical} 
with the summation over 10 points mesh each.} \label{fig:Ene4x1_postsym}
\end{center}
\end{figure}

In this section, we demonstrate the results of the GBMC-PS method. 
As we see in \S \ref{sec:deviation}, the original GBMC method fails in reproducing 
the ground state of $4\times 1$ lattice Hubbard model at 
$U/t=4$ and $n=1$. 
Here, we assess the accuracy obtained from the 
quantum-number projection 
to the density matrix obtained by 
the original GBMC method by following the idea of Assaad {\it et al}~\cite{Assaad,Assaad2}. 
Figure \ref{fig:Ene4x1_postsym} shows the total energy compared with 
the exact diagonalization result. 
Here, we have projected onto the state which has the total momentum $\bm{K}=0$, 
the total $z$-component of spin $S_{z}=0$ and the total spin $S=0$. 
In this case, the exact ground state energy is $-2.102748$ whereas our data is
$-2.1037\pm 0.0010$. The error is obtained by averaging the data over the imaginary time after the convergence.
This result shows that 
the quantum-number projection method well 
reproduces the ground state energy, which is not obtained in the framework 
of the original GBMC method. 
This result is consistent with the result of Assaad {\it et al}~\cite{Assaad,Assaad2}.

\begin{figure}[htb]
\begin{center}
\includegraphics[width=7.5cm]{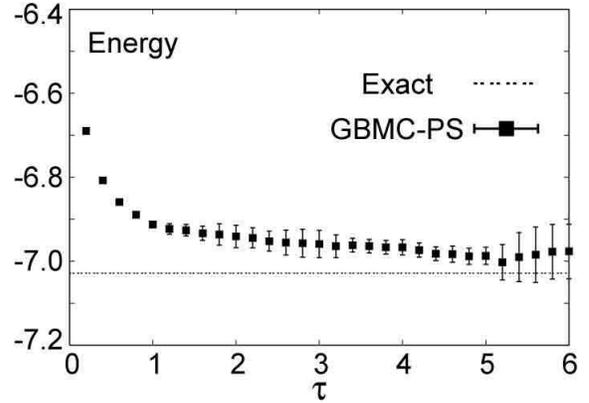}
\caption{Energy as function of $\tau$ at $U/t=10$ and $n=1$ on the $4\times 4$ lattice
under the periodic boundary condition. 
Dashed line is the exact ground state energy obtained from the exact diagonalization. 
Here, for a fast convergence, we use Green's function obtained by the Hartree-Fock calculation as an 
alternative starting point. This scheme for the fast convergence will be discussed in \S \ref{sec:AOC} 
in detail. 
} \label{fig:4x4U10post}
\end{center}
\end{figure}

However, the GBMC-PS method suffers from a slow convergence 
when the interaction strength $U/t$ becomes larger. 
As is illustrated in Fig.~\ref{fig:4x4U10post}, the energy in the case of $U/t=10$ and $n=1$ 
on $4\times 4$ lattice under the full periodic boundary condition
obtained by the GBMC-PS method is not yet converged with the ground state at $\tau=6$. 
Since the strong on-site repulsion $U/t$ prevents the state from updating, 
the efficiency of the importance sampling of the original GBMC becomes worse with 
the increase of the interaction strength $U/t$. 
Thus, in the framework of the GBMC-PS method, 
it is difficult to store the samples 
which has a large overlap with the ground state. 
In the next subsection, we introduce 
the PR-GBMC method 
to overcome this slow convergence in the strong interaction regions. 

\subsection{Pre-projected sampling method}
In this subsection, we introduce an alternative method for performing the quantum-number projection 
which is based on the importance sampling in combination with the quantum-number projection. 
This allows us to perform the sampling with the projected weight, which is more efficient than 
performing the sampling with the original weight. 
We call this pre-projection method the PR-GBMC method 
in contrast with the GBMC-PS method 
introduced by Assaad {\it et al}~\cite{Assaad,Assaad2}. 

For the pre-projected sampling, we rewrite Eq.~(\ref{eq:expectation}) as 
\begin{align}
\langle\hat{O}\rangle_{\hat{P}}=\sum_{i}\tilde{\Omega}_{i}\frac{\int d\bm{x} g(\bm{x})\Omega_{i}(\bm{x})O(\bm{n}_{i}(\bm{x}))}{\tilde{\Omega}_{i}}/\sum_{i}\tilde{\Omega}_{i} , 
\end{align}
where $\tilde{\Omega}_{i}=\int d\bm{x} g(\bm{x})\Omega_{i}(\bm{x})$. 
Estimating $\langle\hat{O}\rangle_{\hat{P}}$ is now reduced to the calculation of 
the weighted average of 
$\int d\bm{x} g(\bm{x})\Omega_{i}(\bm{x})O(\bm{n}_{i}(\bm{x}))/\tilde{\Omega}_{i}$ 
with respect to the projected weight $\tilde{\Omega}_{i}$. 

The projected weight $\tilde{\Omega}_{i}$ stems from 
the projected density-matrix operator, {\it i.e.},
\begin{align}
{\rm Tr}[\hat{P}\hat{\rho}\hat{P}^{\dag}]
&=\sum_{i}\int d\bm{x}g(\bm{x}){\rm Tr}\left[\hat{T}(\bm{x})\hat{\Lambda}(\Omega_{i},\bm{n}_{i})\right] \notag \\
%&=\sum_{i}\int d\bm{x}g(\bm{x})\Omega_{i}\det\left[(e^{i\bm{h}(\bm{x})}-\bm{I})\bm{n}_{i}^{T}+\bm{I}\right] \notag \\
&=\sum_{i}\int d\bm{x}g(\bm{x})\Omega_{i}(\bm{x})=\sum_{i}\tilde{\Omega}_{i}, 
\end{align}
where $\hat{T}(\bm{x})=e^{i\hat{\bm{c}}^{\dag}\bm{h}(\bm{x})\hat{\bm{c}}}$ and 
$\Omega_{i}(\bm{x})=\Omega_{i}\det\left[(e^{i\bm{h}(\bm{x})}-\bm{I})\bm{n}_{i}^{T}+\bm{I}\right]$. 
If the original samples have no overlap with the projected sector, 
\begin{align}
{\rm Tr}[\hat{P}\hat{\rho}\hat{P}^{\dag}]&=\sum_{i}\tilde{\Omega}_{i} \notag \\
&=\sum_{i}\Omega_{i}\int d\bm{x}g(\bm{x})\det\left[(e^{i\bm{h}(\bm{x})}-\bm{I})\bm{n}_{i}^{T}+\bm{I}\right] \label{eq:neg} 
\end{align}
becomes zero. 
However, from Eq.~(\ref{eq:omew}) the unprojected weight $\Omega_{i}$ is always positive. 
It suggests that the factor $\int d\bm{x}g(\bm{x})\det\left[(e^{i\bm{h}(\bm{x})}-\bm{I})\bm{n}_{i}^{T}+\bm{I}\right]$ 
in Eq.~(\ref{eq:neg}) causes the reduction of $\Omega_{i}$ to $\tilde{\Omega}_{i}$ 
when the original samples have small overlap with the projected sector. 
Empirically, we find that this reduction is realized by the cancellation of positive $\tilde{\Omega}_{i}$ and 
negative $\tilde{\Omega}_{i}$, thus $\tilde{\Omega}_{i}$ is not positive definite. 
This is the source of the negative sign problem in the PR-GBMC method. 
In this case, we introduce the sign variable $S_{i}=\pm 1$ by 
$\tilde{\Omega}_{i}=S_{i}|\tilde{\Omega}_{i}|$, and the importance 
sampling is performed with the 
absolute value of $\tilde{\Omega}_{i}$, namely we calculate 
\begin{align}
\langle\hat{O}\rangle_{\hat{P}}=\sum_{i}|\tilde{\Omega}_{i}|S_{i}\frac{\int d\bm{x}g(\bm{x})\Omega_{i}(\bm{x})O(\bm{n}_{i}(\bm{x}))}{\tilde{\Omega}_{i}}/\sum_{i}|\tilde{\Omega}_{i}|S_{i} .
\end{align}
Appearance of the negative sign is evaluated by calculating the expectation value 
of the sign defined as 
\begin{align}
\langle S\rangle=\frac{\sum_{i}|\tilde{\Omega}_{i}|S_{i}}{\sum_{i}|\tilde{\Omega}_{i}|} .
\end{align}

\begin{figure}[htb]
\begin{center}
\includegraphics[width=7.5cm]{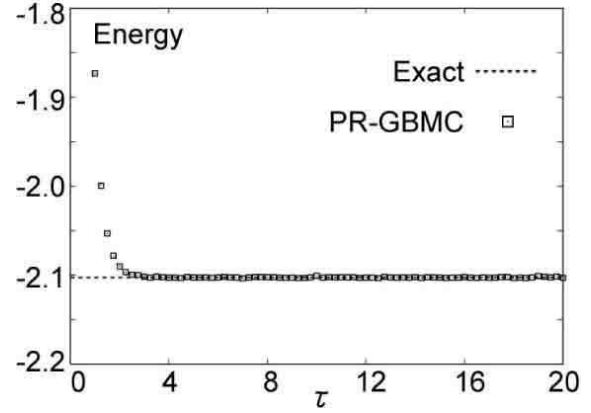}
\caption{Energy of the $4\times 1$-site lattice Hubbard model with $U/t=4$ and $n=1$. 
Here, $\bm{K}=0,S_{z}=0$ and $S=0$ projections are performed. 
} \label{fig:4x1pre}
\end{center}
\end{figure}

Below we show results of the PR-GBMC method with the quantum-number projection $\bm{K}=0$, $S_{z}=0$ and $S=0$. 
Figure \ref{fig:4x1pre} shows the energy of $4\times 1$ lattice Hubbard model 
with $U/t=4$ and $n=1$ calculated by the PR-GBMC method. 
Averaging the data over imaginary time gives the energy $-2.1026\pm 0.0002$ 
in agreement with the exact diagonalization result $-2.102748$. 
During the simulation, the average sign $\langle S\rangle$ 
is kept unity, {\it i.e.}, there is no negative sign. 
We will discuss later the negative sign problem in more detail.

One of the main advantages in the PR-GBMC method is that this method allows us 
to analyze directly the change in the probability distributions caused by 
the quantum-number projection: 
\begin{align}
\frac{\partial P(\Omega,\bm{n})}{\partial \tau}&={\cal L}P(\Omega,\bm{n}) \notag \\
&\xrightarrow{{\rm projection}}
\frac{\partial \tilde{P}(\tilde{\Omega},\tilde{\bm{n}})}{\partial \tau}=\tilde{{\cal L}}\tilde{P}(\tilde{\Omega},\tilde{\bm{n}}) ,
\end{align}
where the projected variables are denoted by tilde. 
The reweighted distribution $\tilde{P}$ cannot be calculated 
in the framework of the GBMC-PS method, because 
in the GBMC-PS method, the quantum-number projection is performed by 
reweighting the importance sampling 
of the original GBMC method, {\it i.e.},
\begin{align}
\langle\hat{O}\rangle_{\hat{P}}&=\frac{\sum_{i}\int d\bm{x}g(\bm{x})\Omega_{i}(\bm{x})O(\bm{n}_{i}(\bm{x}))}{\sum_{i}\int d\bm{x}g(\bm{x})\Omega_{i}(\bm{x})} \notag \\
&=\frac{\sum_{i}\Omega_{i}\frac{\int d\bm{x}g(\bm{x})\Omega_{i}(\bm{x})O(\bm{n}_{i}(\bm{x}))}{\Omega_{i}}}{\sum_{i}\Omega_{i}\frac{\int d\bm{x}g(\bm{x})\Omega_{i}(\bm{x})}{\Omega_{i}}} \notag \\
&\xrightarrow{{\rm sampling\ by\ } \Omega_{i}}
\frac{\frac{1}{N_{\rm mcs}}\sum_{i}\frac{\int d\bm{x}g(\bm{x})\Omega_{i}(\bm{x})O(\bm{n}_{i}(\bm{x}))}{\Omega_{i}}}{\frac{1}{N_{\rm mcs}}\sum_{i}\frac{\int d\bm{x}g(\bm{x})\Omega_{i}(\bm{x})}{\Omega_{i}}}, \label{eq:po}
\end{align}
where $N_{\rm mcs}$ is the number of Monte Carlo samples.
As is seen from Eq.~(\ref{eq:po}), what one can obtain by the GBMC-PS method is the distribution of 
$\int d\bm{x}g(\bm{x})\Omega_{i}(\bm{x})O(\bm{n}_{i}(\bm{x}))/\Omega_{i}$ and 
$\int d\bm{x}g(\bm{x})\Omega_{i}(\bm{x})/\Omega_{i}$. 
Thus, the distribution of $\langle\hat{O}\rangle_{\hat{P}}$ itself 
can not be calculated by GBMC-PS method. 

\begin{figure}[htb]
\begin{center}
\includegraphics[width=7.5cm]{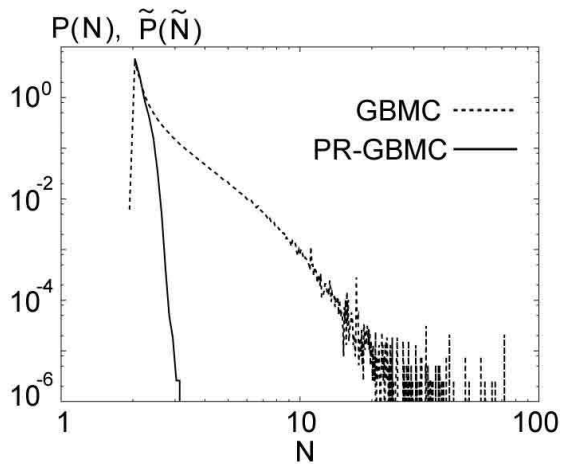}
\caption{Distribution of Green's function. 
The solid curve shows the distribution obtained by the PR-GBMC and 
the dashed curve shows the distribution obtained by the GBMC which is the same data as in Fig.~\ref{fig:n2_comp}(b).
The data are obtained for the $4\times 1$-site Hubbard model with $U/t=4$ and $n=1$ 
at $\tau=20$ with $3.84\times 10^6$ Monte Carlo steps.} \label{fig:4x1prepost}
\end{center}
\end{figure}

We now calculate the projected probability distribution $\tilde{P}(\tilde{N})$ to 
compare with the unprojected distribution $P(N)$ obtained by the original GBMC method. 
Figure \ref{fig:4x1prepost} shows the distribution of Green's function. 
The abscissa $N$ represents $\sqrt{\sum_{ij\sigma}n_{(i\sigma),(j\sigma)}^{2}}$ for the 
GBMC method and $\sqrt{\sum_{ij\sigma\sigma'}\tilde{n}_{(i\sigma),(j\sigma')}^{2}}$ 
for the PR-GBMC method, where 
\begin{align}
\tilde{\bm{n}}=\int d\bm{x}g(\bm{x})\Omega(\bm{x})\bm{n}(\bm{x})/|\tilde{\Omega}| .
\end{align}
As is illustrated in Fig.~\ref{fig:4x1prepost}, the projected distribution $\tilde{P}(\tilde{N})$ 
decays exponentially. This means that the quantum-number projection actually reduces 
the phase-space. In the PR-GBMC method, 
the importance sampling is performed with respect to this reduced phase-space. 
Therefore, 
the convergence to the ground state is faster than the GBMC-PS method which 
is based on the importance sampling with the original weight. 

\subsection{Comparison between GBMC-PS and PR-GBMC}
In the previous subsection, we have confirmed that the quantum-number projection 
changes the probability distribution and removes errors arising in the 
original GBMC procedure. Here, we make a comparison between the GBMC-PS and
PR-GBMC methods to discuss their merits and demerits. 

\begin{figure}[htb]
\begin{center}
\includegraphics[width=7.5cm]{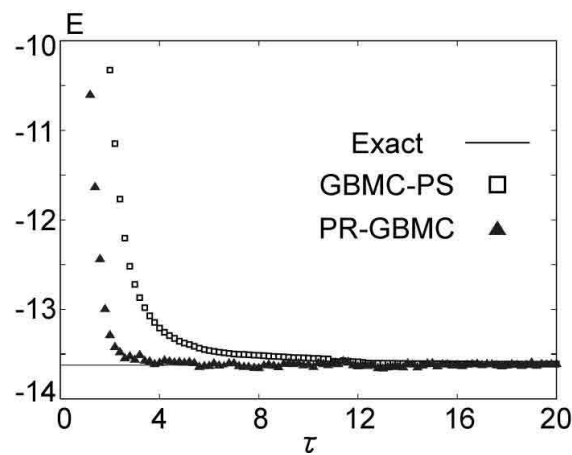}
\caption{Energy of the $4\times 4$ lattice with $U/t=4$ and $n=1$ 
as functions of $\tau$.
The squares represent the result of the GBMC-PS method and the triangles represent 
the result of the PR-GBMC method.} \label{fig:4x4prepost}
\end{center}
\end{figure}

Figure \ref{fig:4x4prepost} shows the energy of the $4\times 4$-site
Hubbard model with $U/t=4$ and $n=1$ under the periodic boundary condition. 
As is seen from Fig.~\ref{fig:4x4prepost}, the energy 
obtained by the PR-GBMC method converges with the ground state faster 
than that obtained by the GBMC-PS method. 
This comes from the difference of the sampling 
procedure. In GBMC-PS method importance sampling is performed with 
respect to the unprojected weight $\Omega=e^{-\int_{0}^{\tau}H(\bm{n})d\tau'}$. 
Thus the sampling depends only on the energy. On the other hand, PR-GBMC method makes use of 
the projected weight $\tilde{\Omega}=\Omega\int d\bm{x}g(\bm{x})\det[(e^{i\bm{h}}-\bm{I})\bm{n}^{T}+\bm{I}]$, 
which reflects not only the energy but also the overlap with the projected sector. 
Therefore, the PR-GBMC method allows the convergence to the ground state at smaller $\tau$ than the GBMC-PS method.
However, in the PR-GBMC method we have to perform the projection for every sample, 
while the GBMC-PS method requires the projection only for accepted samples, 
which makes the computation time shorter. 
Empirically, we find that the energy obtained by both the GBMC-PS and the PR-GBMC 
methods converges with 
the ground state when the 
on-site interaction $U/t$ is not too large, 
whereas the PR-GBMC method is more efficient. 
Actually the PR-GBMC method offers a better convergence at larger $U/t$. 
This is because the original GBMC sampling fails in making samples which have enough 
overlap with the ground state at relatively large $U/t$. 
This possibly causes a serious minus sign problem. 
Thus when treating large $U/t$ systems (typically $U/t\gtsim 4$), 
the PR-GBMC method has to be employed. 

\subsection{Negative sign problem}\label{sec:negative}
\begin{figure}[htb]
\begin{center}
\includegraphics[width=7.5cm]{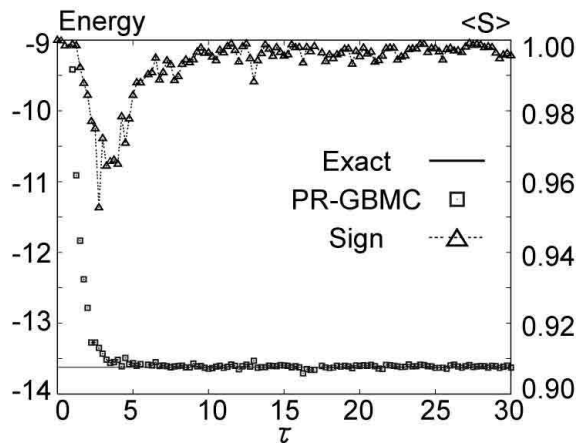}
\caption{Energy and average sign of PR-GBMC results as functions of $\tau$. 
The squares represent the energy of $4\times 4$ 
lattices under the periodic boundary condition with $U/t=4$ and $n=1$ and the triangles represent the average sign. The initial condition at $\tau=0$ is Green's functions for the infinite temperature.} \label{fig:4x4U4negaE}
\end{center}
\end{figure}

Since the PR-GBMC method is based on samplings with respect to the projected weights $\tilde{\Omega}$, 
there appears minus sign if a sample has small overlap with the projected sector. 
Figure \ref{fig:4x4U4negaE} shows 
PR-GBMC results of the 
total energy and the expectation value of the sign 
%\begin{align}
$\langle S\rangle=\sum_{i}|\tilde{\Omega}_{i}|S_{i}/\sum_{i}|\tilde{\Omega}_{i}|$ 
%\end{align}
on the $4\times 4$ lattice at 
$U/t=4$ and $n=1$. 
As we see in Fig.~\ref{fig:4x4U4negaE}, the average
sign decreases first, 
but it recovers along with the convergence of the energy. 
This means that if the overlap with the projected sector is small, 
the expectation value of the sign becomes small,  
while $\langle S\rangle$ recovers when the samples gain a large overlap 
with the quantum-number-projected state. 
We note that the $\tau$ dependence of $\langle S\rangle$ is completely different 
from that in the conventional AFQMC method 
as well as from that in other methods, where $\langle S\rangle$ exponentially decreases 
to zero with increasing $\tau$. 

\subsection{Acceleration of convergence}\label{sec:AOC}
\begin{figure}[htb]
\begin{center}
\includegraphics[width=7.5cm]{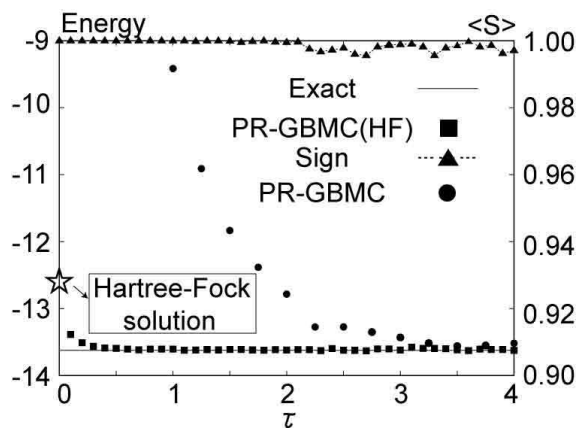}
\caption{Energy and average sign of PR-GBMC results as functions of $\tau$. 
The squares represent the energy of $4\times 4$ 
lattices under the periodic boundary condition 
with $U/t=4$ and $n=1$ started from the Hartree-Fock solution (open star) 
and the triangles represent the 
average sign. 
For comparison, the PR-GBMC result obtained from the infinite temperature at $\tau=0$ 
is plotted (circles). 
%Here $\tau$ does not have a meaning of the inverse temperature any more because $\tau=0$ is 
%the Hartree-Fock solution.
} \label{fig:4x4U4negaEHF}
\end{center}
\end{figure}

In the previous subsections, we have proposed and studied 
the PR-GBMC method and examined 
the negative sign problem. 
In the PR-GBMC method, the negative sign appears when the overlap 
with the projected sector is small. To avoid this problem, instead of employing  
$\bm{n}_{0}=\frac{1}{2}\bm{I}$ as the initial condition, it is better to use 
Green's function obtained by the Hartree-Fock calculation 
as an alternative starting point. 
Since the initial state is already that of the Hartree-Fock solution, 
the overlap with the ground state is expected to be relatively 
larger than $\bm{n}_{0}=\frac{1}{2}\bm{I}$. 
Therefore, the convergence to the ground state becomes faster and 
the expectation value of the average sign becomes stable. 
Figure \ref{fig:4x4U4negaEHF} shows the total energy of $4\times 4$ lattices 
under the periodic boundary condition with 
$U/t=4$ and $n=1$ together with its average sign. 
As is seen from Fig.~\ref{fig:4x4U4negaEHF}, the total energy converges already 
at $\tau=0.5$ and the average sign is nearly unity 
in the whole range of $\tau$. 

\subsection{Applicability of PR-GBMC method}
\subsubsection{$U/t$ dependence of convergence}
In the PR-GBMC method, the distributions of the phase-space variables $P(\Omega)$ and $P(N)$ 
are transformed to the projected distributions $\tilde{P}(\tilde{\Omega})$ and $\tilde{P}(\tilde{N})$ 
which decay faster than the original distributions. 
However, the decay of the projected distributions becomes slower with the 
increase of $U/t$. In this subsection, 
by comparing the $U/t$ dependence of the energy convergence with 
that of the projected distribution, 
we discuss the applicable range of the PR-GBMC method. 

\begin{figure}[htb]
\begin{center}
\begin{tabular}{c}
\includegraphics[width=7.5cm]{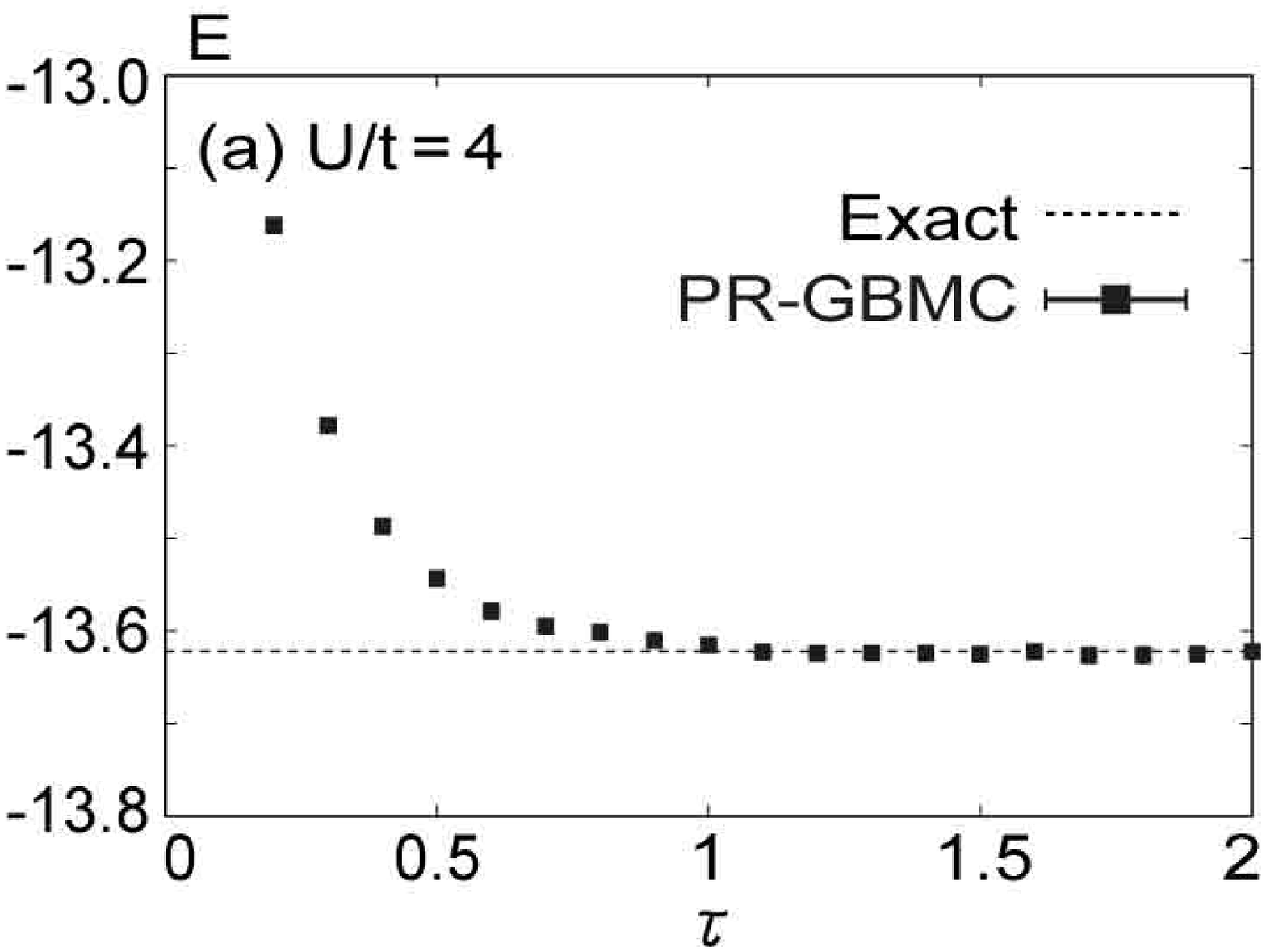}  \\
\includegraphics[width=7.5cm]{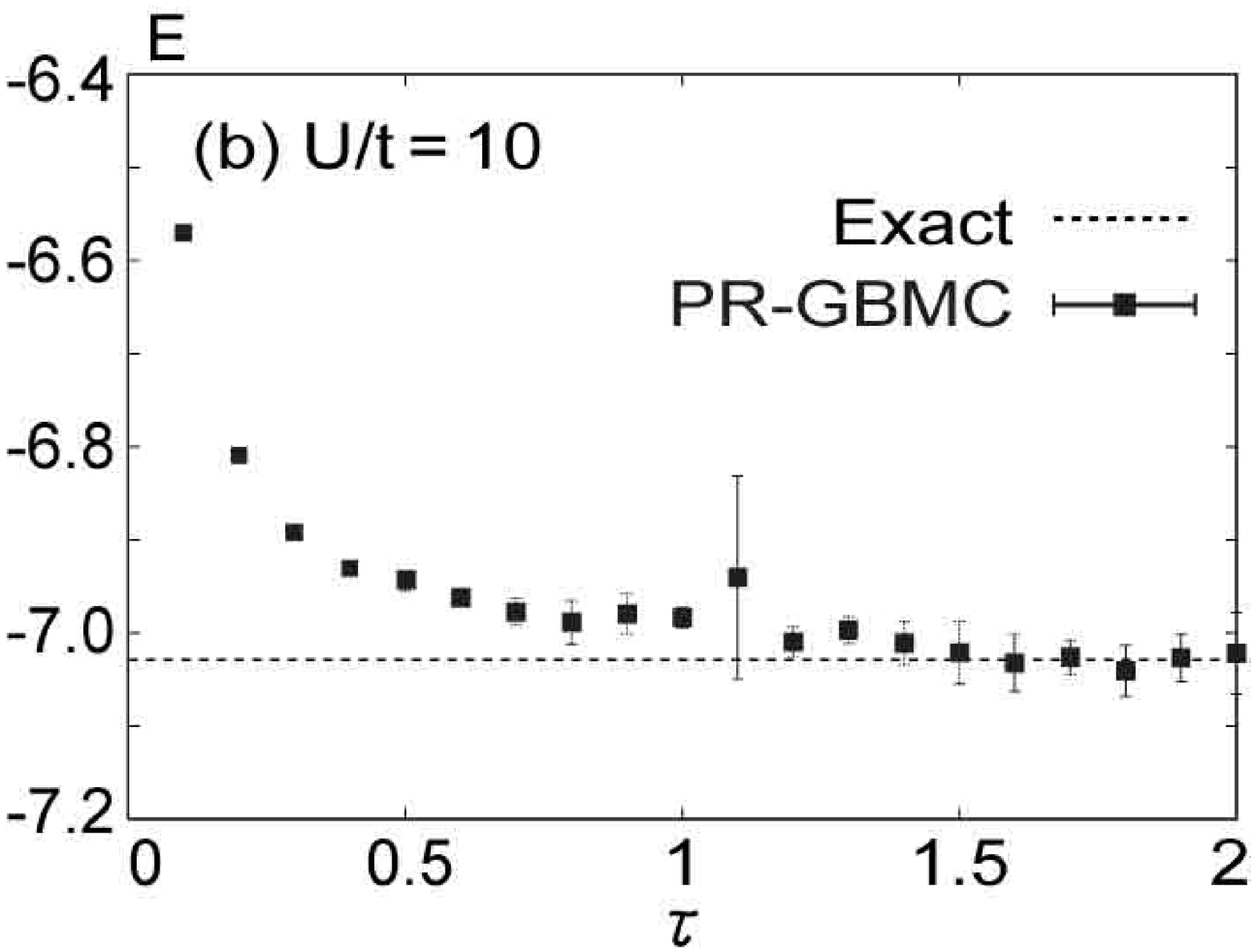} \\
\includegraphics[width=7.5cm]{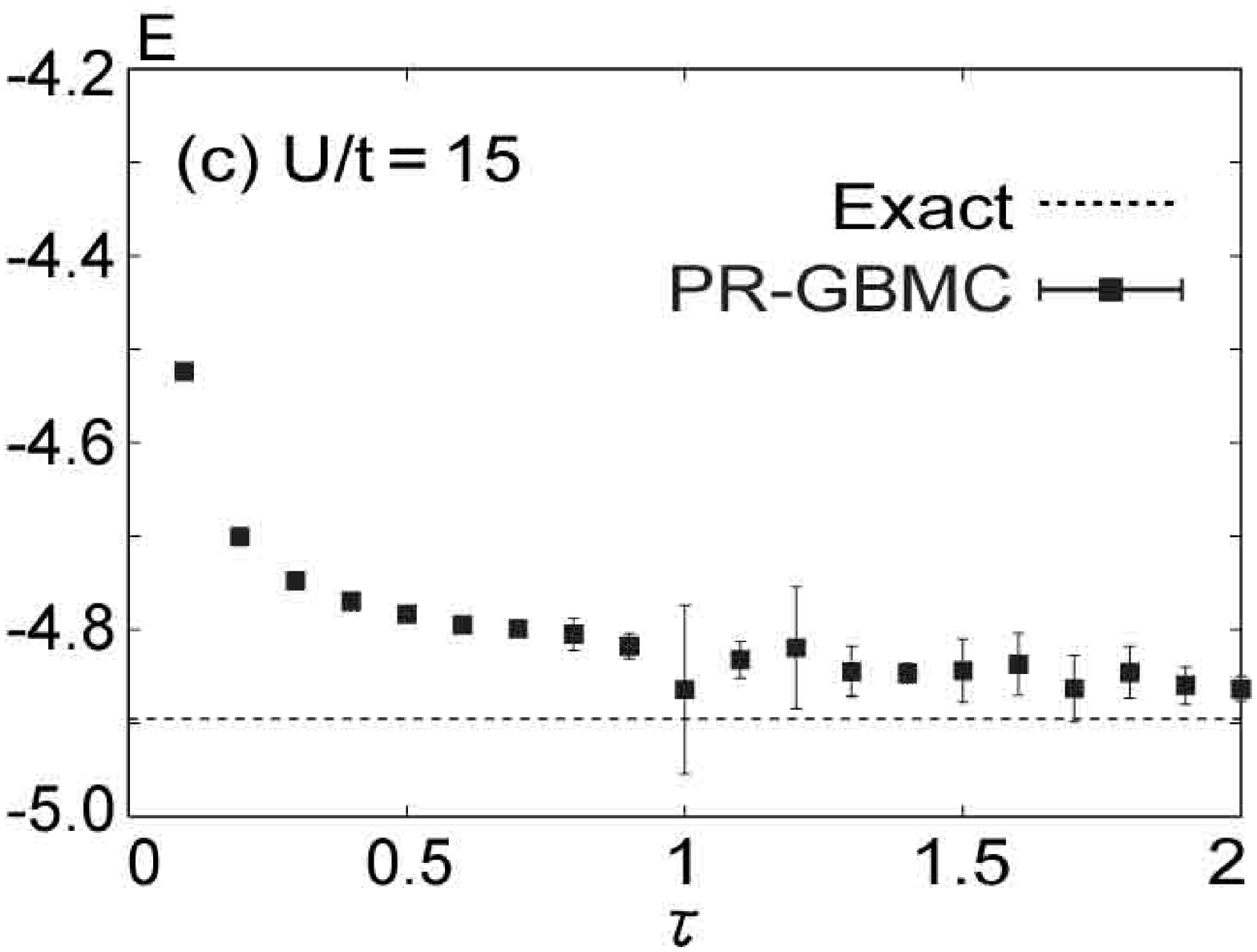} \\
\end{tabular}
\caption{Energy as functions of $\tau$ at 
(a) $U/t=4$, (b) $U/t=10$ and (c) $U/t=15$ for the $4\times 4$-site Hubbard model under the periodic boundary condition . 
Dashed line is the exact ground state energy obtained from the exact diagonalization.} \label{fig:4x4enes}
\end{center}
\end{figure}

Figure \ref{fig:4x4enes} shows the energy of $4\times 4$ lattice 
under the periodic boundary condition with $n=1$ at $U/t=4$, 10 and 15. All the results are obtained by $1.6\times 10^6$ 
Monte Carlo steps with the pre-projection at $\bm{K}=0$, $S_{z}=0$ and $S=0$. 
We employ the data at $\tau=0$ from the solutions of the Hartree-Fock calculation with 
$U_{{\rm int}}=U/2$. 
Measurements are divided into 5 bins and the error bars are estimated by the 
variance among the 5 bins. 
As is seen from Fig.~\ref{fig:4x4enes}, 
the convergence with the ground state becomes slower 
with the increase of $U/t$ and at $U/t=15$, the energy obtained by the PR-GBMC method 
does not yet converge with the ground state at $\tau=2$ because of 
the slow convergence. 
At $U/t=15$, since the energy seems to show further decrease 
beyond $\tau=2$, further time evolution is required whereas 
the statistical error becomes large. 
This large statistical error comes from the fast diffusion of 
the probability distribution of the sampling weight $\Omega$. 
Since the definition of the weight is 
given by $\Omega=e^{-\int_{0}^{\tau}H(\bm{n})d\tau'}$, 
the weight $\Omega$ grows exponentially with $\tau$. 
Thus, the phase-space to be sampled becomes larger if the convergence with the 
ground state becomes slower because of the strong interaction. 
This requires much more computation time, which determines the practical limitation 
of the PR-GBMC method. 

\begin{figure}[h!]
\begin{center}
\includegraphics[width=7.5cm]{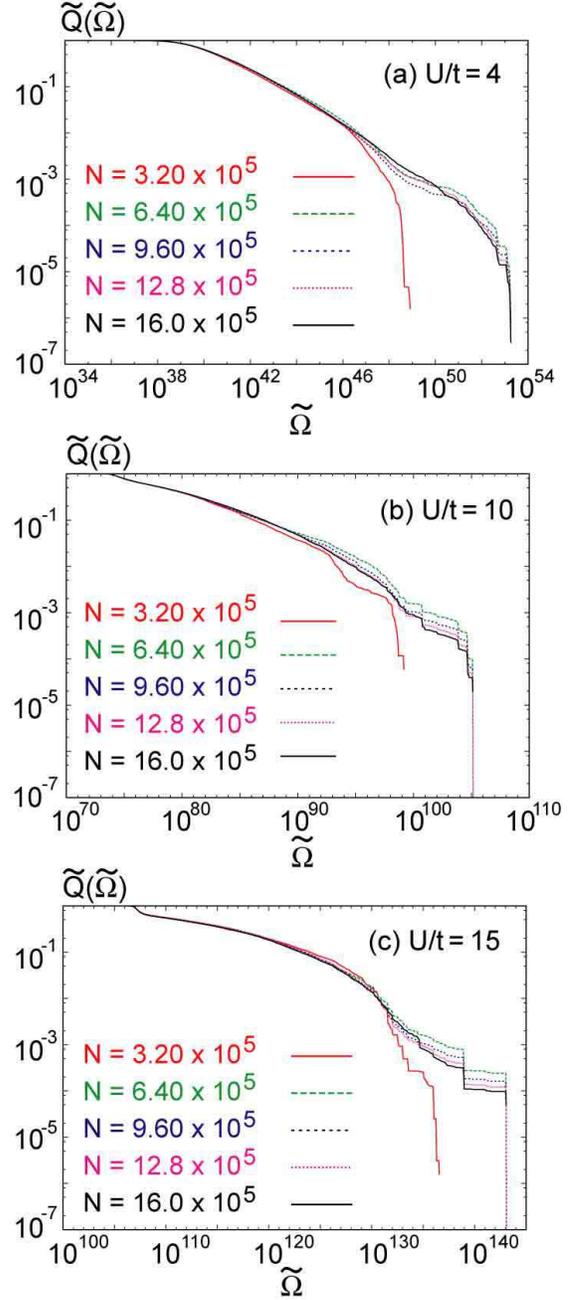}
\caption{(color online) Integrated distribution of projected weight $\tilde{\Omega}$ at 
$\tau=2$ for the $4\times 4$-site square lattice under the periodic boundary 
condition with (a) $U/t=4$, (b) $U/t=10$ and (c) $U/t=15$. 
In all the panels, 
red (light), green (long dashed), blue (dashed), pink (dotted) and black (dark) curves represent 
the distribution obtained by 
$3.20\times 10^{5}$, $6.40\times 10^{5}$, $9.60\times 10^{5}$, $12.8\times 10^{5}$ and 
$16.0\times 10^{5}$ Monte Carlo steps, 
respectively.} \label{fig:4x4omegas}
\end{center}
\end{figure}

\begin{figure}[h!]
\begin{center}
\includegraphics[width=7.5cm]{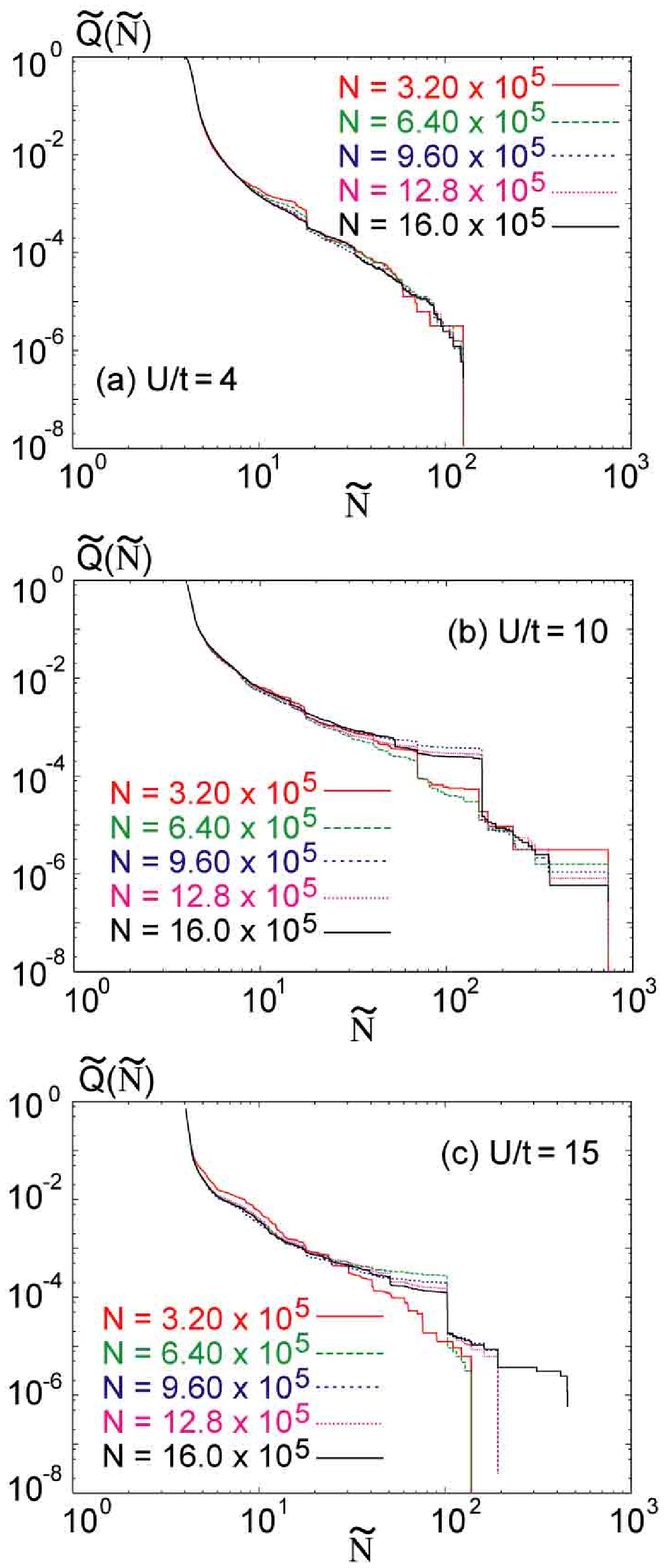}
\caption{(color online): Integrated distribution of projected Green's function $\tilde{N}$ at 
$\tau=2$ for the $4\times 4$-site square lattice under the periodic boundary 
condition with (a) $U/t=4$, (b) $U/t=10$ and (c) $U/t=15$. 
In all the panels, 
red (light), green (long dashed), blue (dashed), pink (dotted) and black (dark) curves represent 
the distributions obtained from 
$3.20\times 10^{5}$, $6.40\times 10^{5}$, $9.60\times 10^{5}$, $12.8\times 10^{5}$ and 
$16.0\times 10^{5}$ Monte Carlo steps, 
respectively.} \label{fig:4x4ns}
\end{center}
\end{figure}

\begin{figure}[h1]
\begin{center}
\includegraphics[width=7cm]{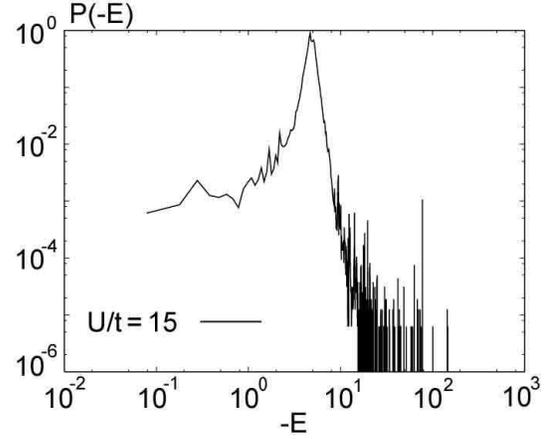}
\caption{Distribution of energy at $U/t=15$ for $4\times 4$-site Hubbard model under the periodic boundary 
condition at $n=1$. 
The data are obtained from $1.6\times 10^6$ Monte Carlo steps at $\tau=2$} \label{fig:hist_eneU15}
\end{center}
\end{figure}

To confirm this, 
we calculate the distributions of the projected phase-space variables 
$\tilde{\Omega}$ and $\tilde{N}=\sqrt{\sum_{ij\sigma\sigma'}\tilde{n}_{(i\sigma),(j\sigma')}}$. 
Figures \ref{fig:4x4omegas} and \ref{fig:4x4ns} show the integrated distributions $\tilde{Q}(\tilde{\Omega})$ 
and $\tilde{Q}(\tilde{N})$ defined by 
\begin{align}
\tilde{Q}(\tilde{\Omega})&=1-\int_{0}^{\tilde{\Omega}}\tilde{P}(\tilde{\Omega}')d\tilde{\Omega}' , \\
\tilde{Q}(\tilde{N})&=1-\int_{0}^{\tilde{N}}\tilde{P}(\tilde{N}')d\tilde{N}' .
\end{align}
As we see in Fig.~\ref{fig:4x4omegas}, the Monte Carlo step dependence of 
$\tilde{Q}(\tilde{\Omega})$ implies the existence of the cutoffs in the distributions 
of $\tilde{\Omega}$ for all $U/t$. 
Thus, the boundary terms 
with respect to the projected weight $\tilde{\Omega}$ 
do not appear to exist for all $U/t$. 
However, the phase-space of the projected weight $\tilde{\Omega}$ becomes 
larger with $U/t$ and at $U/t=15$, and a distinct plateau structure with steps caused 
by the lack of the large $\tilde{\Omega}$ samples is seen in 
the tail of the distribution 
(see Fig.~\ref{fig:4x4omegas}.c). 
This means that in the case of $U/t=15$, $1.6\times 10^{6}$ Monte Carlo steps are not 
enough for the accurate sampling of events at 
large weight $\Omega$. 
Since the samples with large weight seldom appear, 
but contribute to lowering the energy, 
these samples cause the spike structure in the distribution of the energy 
(see Fig.~\ref{fig:hist_eneU15}). 
Thus, the statistical error of the energy becomes larger. 

The slow convergence of the distribution due to the strong interaction 
is also observed in 
the distribution of the projected Green's function $\tilde{Q}(\tilde{N})$. 
As is seen from Fig.~\ref{fig:4x4ns}, 
the convergence of the distribution tail becomes slower as $U/t$ increases. 
Although the energy converges with the ground state at $U/t=10$ as in Fig.~\ref{fig:4x4enes}.b, 
the plateau structure in $\tilde{Q} (\tilde{N})$ visible at $U/t=10$ (Fig.~\ref{fig:4x4ns}.b) 
signals the slow convergence 
in the PR-GBMC method. 
However, the convergence of the energy at $U/t=10$ shows that 
a small plateau structure arising in the tail part of $\tilde{Q}(\tilde{N})$ 
does not yet cause a bad effect on the convergence of the energy. 
At $U/t=15$, the distribution tail of $\tilde{Q}(\tilde{N})$ 
shows no indication of the existence of the cutoff at least up to $1.6\times 10^6$ Monte Carlo steps, 
which indicates that the number of Monte Carlo steps $1.6\times 10^6$ is 
not enough at $U/t=15$. 

These slow convergences of the 
distributions at large $U/t$ come from the 
expansion of the phase-space to be sampled. 
From Eqs.(\ref{eq:extension1}-\ref{eq:extension2}), 
the drift term and the diffusion terms in Langevin equation (\ref{eq:langevin2}) 
are proportional to $U$ and $\sqrt{U}$, respectively. 
Thus the diffusion speed of the distributions becomes faster with 
the increase of $U/t$, which means that the phase-space to be sampled becomes 
larger. This larger sampling space requires larger Monte Carlo steps and causes 
an insufficient sampling at large $U/t$. 
Therefore, it is advisable to monitor the convergence of the distributions 
$\tilde{Q}(\tilde{\Omega})$ and $\tilde{Q}(\tilde{N})$ to ensure that the number of  
Monte Carlo steps is sufficiently large. 

\subsubsection{Size dependence of convergence}
\begin{figure}[h!]
\begin{center}
\includegraphics[width=7cm]{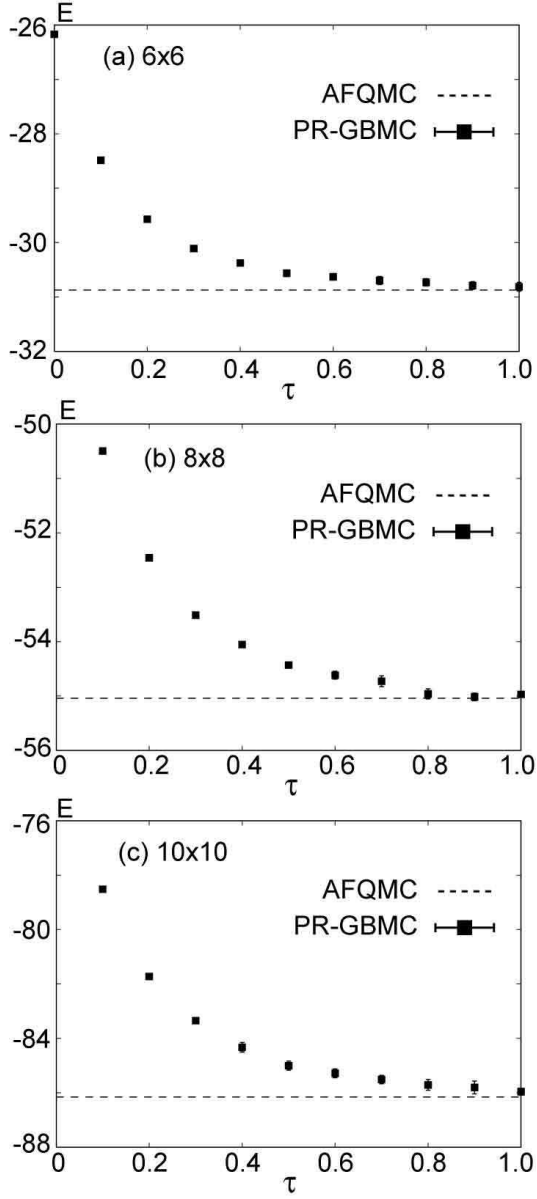}
\caption{Energy as functions of $\tau$ at $U/t=4$ and $n=1$ for 
(a) $6\times 6$, (b) $8\times 8$ and (c) $10\times 10$ lattices under the
periodic boundary condition. 
Dashed lines are the ground state energy obtained by the AFQMC method.
Error bars are as large as the symbol size.} \label{fig:sizeE}
\end{center}
\end{figure}

\begin{figure}[h!]
\begin{center}
\includegraphics[width=7.5cm]{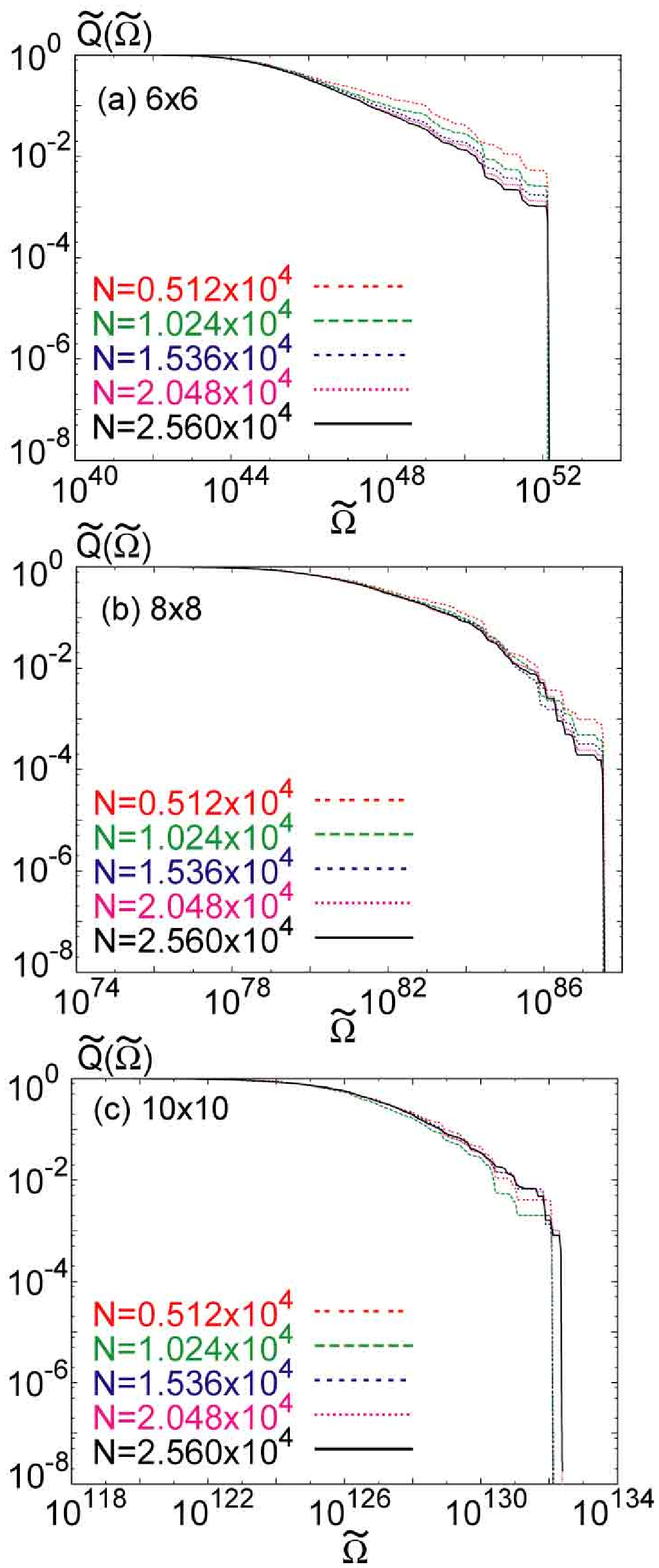}
\caption{(color online): Integrated distribution of projected weight $\tilde{\Omega}$ 
obtained at $U/t=4$ $n=1$ and $\tau=1$
on (a) $6\times 6$,  (b) $8\times 8$ and (c) $10\times 10$ lattices
under the periodic boundary condition. 
In all the panels, 
red (light), green (long dashed), blue (dashed), pink (dotted) and black (dark) curves represent 
the distribution obtained by 
$0.512\times 10^{4}$, $1.024\times 10^{4}$, $1.536\times 10^{4}$, $2.048\times 10^{4}$ and 
$2.560\times 10^{4}$ Monte Carlo steps, 
respectively.
} \label{fig:sizeomega}
\end{center}
\end{figure}

\begin{figure}[h!]
\begin{center}
\includegraphics[width=7.5cm]{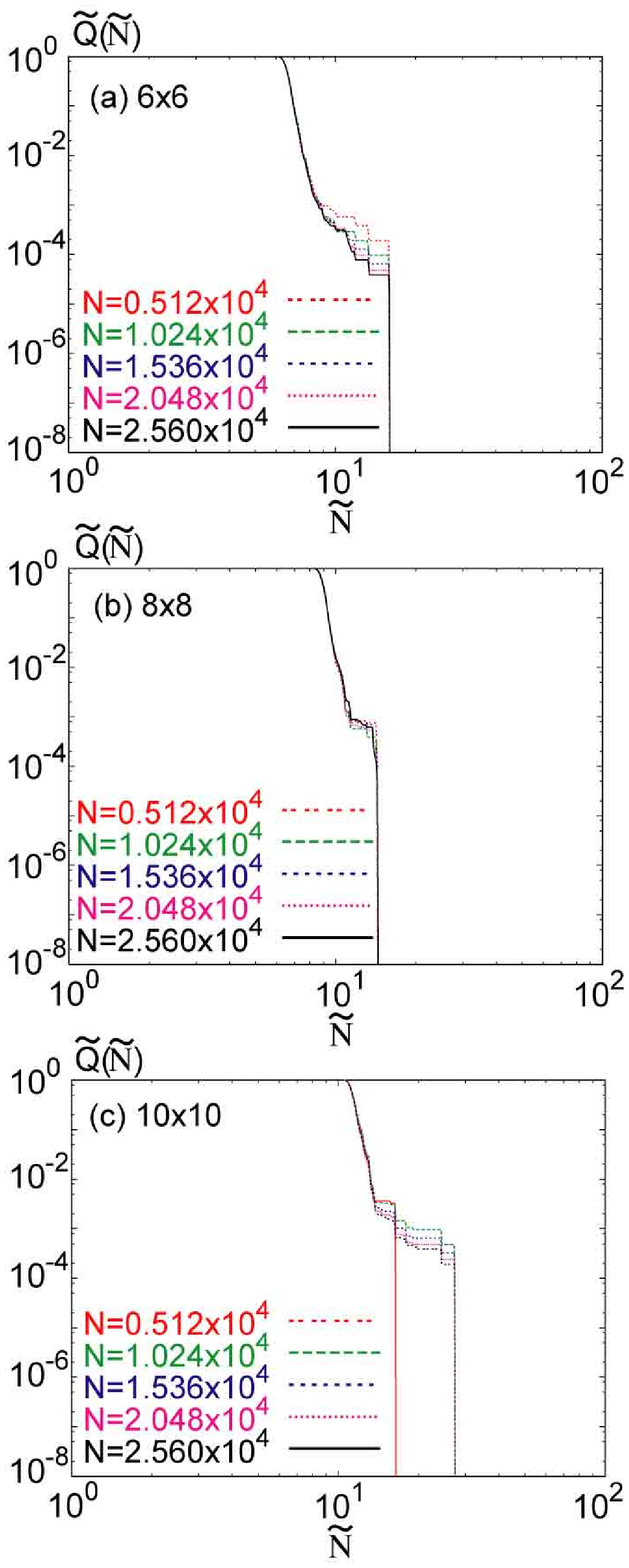}
\caption{(color online): Integrated distribution of Green's function $\tilde{N}$ obtained at $U/t=4$ and 
$\tau=1$ on (a) $6\times 6$, (b) $8\times 8$ and (c) $10\times 10$ lattices
under the periodic boundary condition. 
In all the panels, 
red (light), green (long dashed), blue (dashed), pink (dotted) and black (dark) curves represent 
the distribution obtained by 
$0.512\times 10^{4}$, $1.024\times 10^{4}$, $1.536\times 10^{4}$, $2.048\times 10^{4}$ and 
$2.560\times 10^{4}$ Monte Carlo steps, 
respectively.
} \label{fig:sizeN}
\end{center}
\end{figure}

In this subsection, we analyze the size dependence of the convergence 
by employing the PR-GBMC method with $\bm{K}=0$, $S_{z}=0$ and $S=0$ projections. 
Figure \ref{fig:sizeE} shows the energy of $6\times 6$, $8\times 8$ and $10\times 10$ lattices 
at $U/t=4$ and $n=1$. 
As is seen from Fig.~\ref{fig:sizeE}, all the simulation results 
converge with the ground state energy 
obtained by the AFQMC method with the Trotter discretization of $\Delta\tau=0.025$. 
As we see in Fig.~\ref{fig:sizeomega}, 
in accordance with the convergence of the energy, 
the integrated distributions of the projected weight $\tilde{Q}(\tilde{\Omega})$ 
at any size show 
no distinct signal of insufficiency in sampling which is observed in 
the tail part of $\tilde{Q}(\tilde{\Omega})$ at $U/t=15$ on $4\times 4$ lattice 
as illustrated in Fig.~\ref{fig:4x4omegas}.c. 
The existence of the cutoffs in the integrated distributions of 
Green's function $\tilde{Q}(\tilde{N})$ also supports the fact that 
the number of Monte Carlo steps 
$2.56\times 10^{4}$ is sufficient in these systems (see Fig.~\ref{fig:sizeN}). 

\begin{figure}[h!]
\begin{center}
\begin{tabular}{c}
\includegraphics[width=7.5cm]{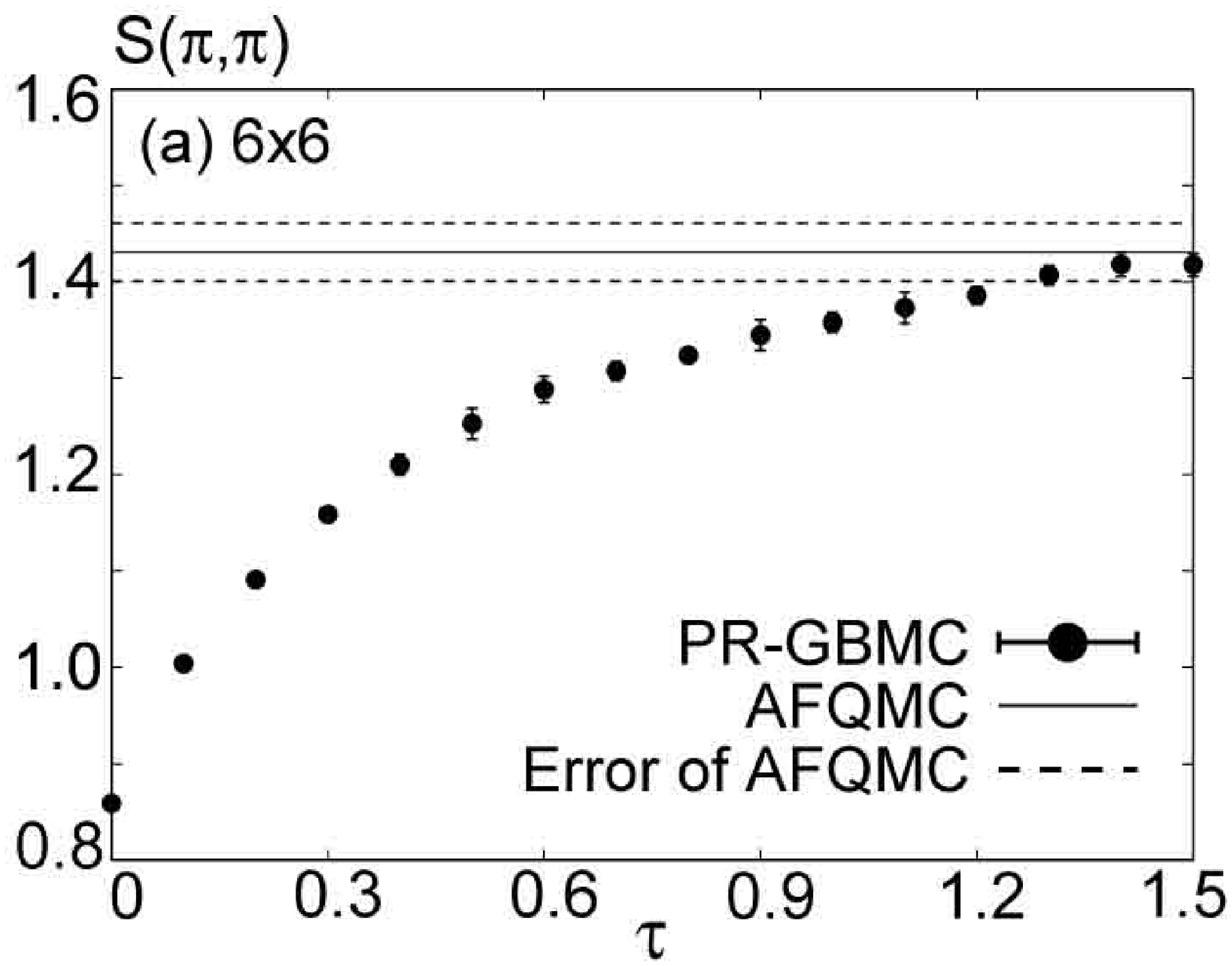}  \\
\includegraphics[width=7.5cm]{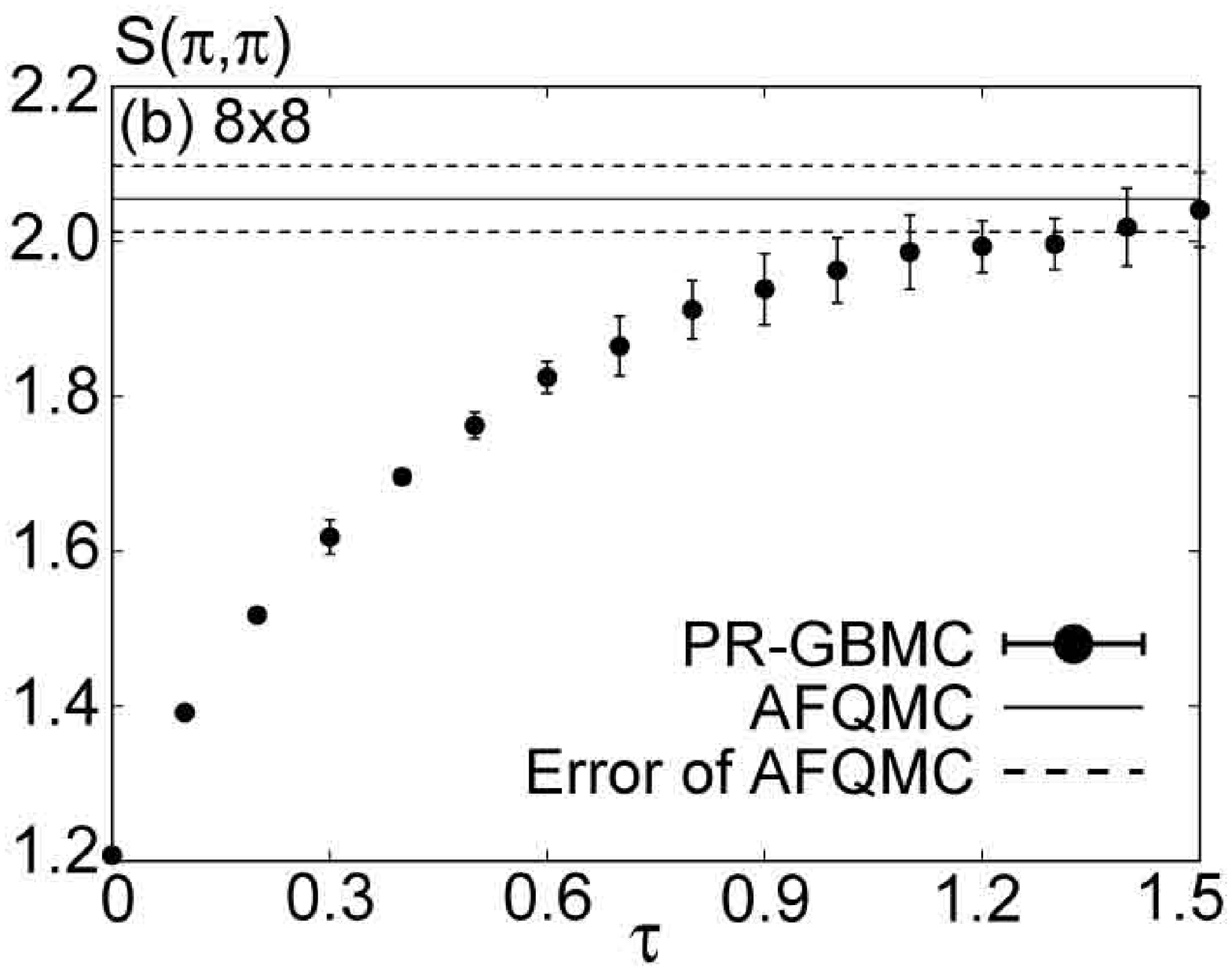} \\
\includegraphics[width=7.5cm]{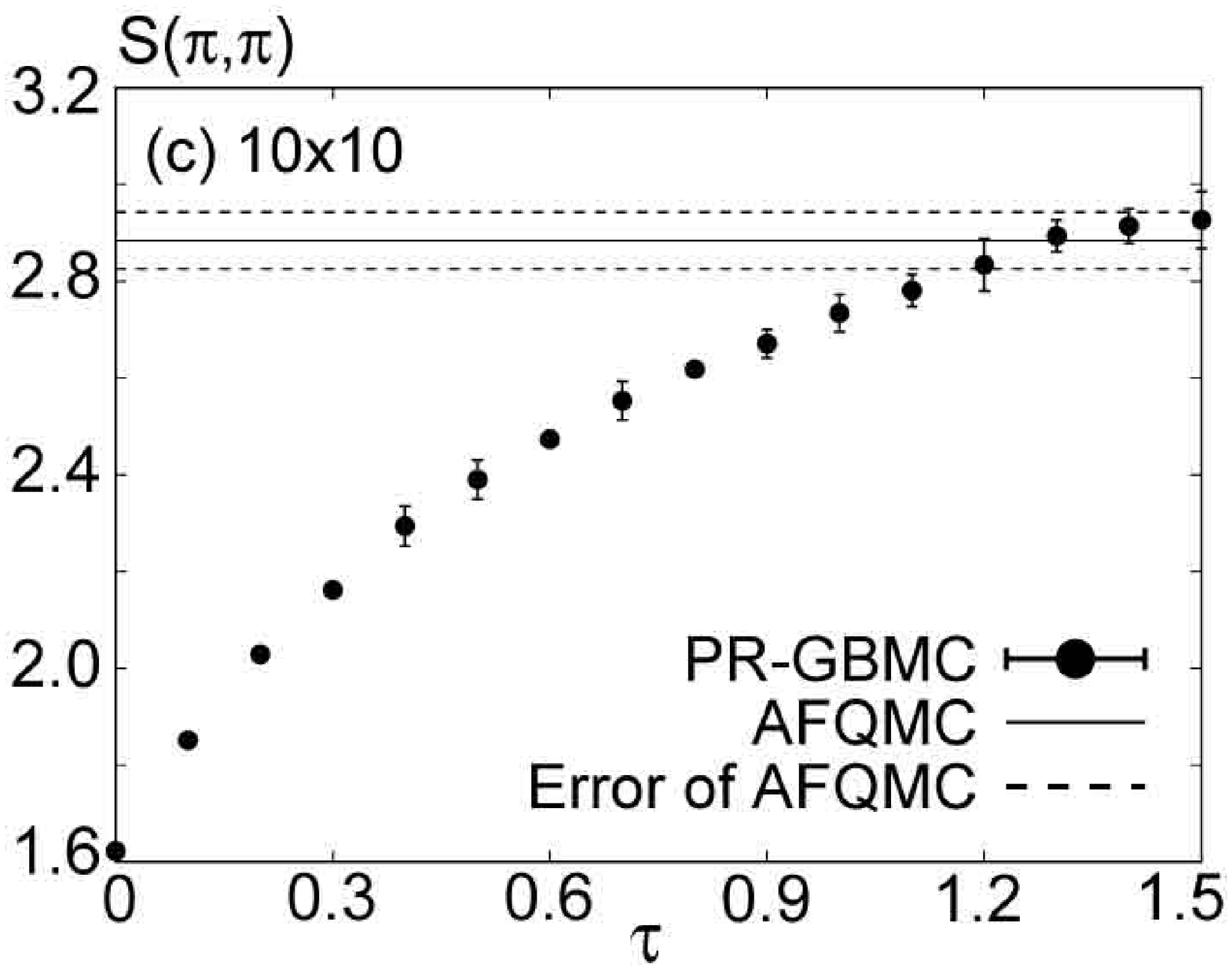} \\
\end{tabular}
\caption{
Peak values of spin correlation $S(\pi,\pi)$ on 
(a) $6\times 6$, (b) $8\times 8$ and (c) $10\times 10$ lattices
under the periodic boundary condition at $U/t=4$ and $n=1$. 
In all the panels, solid and dashed lines represent the 
results obtained by the AFQMC method and their error ranges, respectively. 
} \label{fig:sq}
\end{center}
\end{figure}

Here, to confirm the convergence, 
we have calculated not only the ground state energy but also 
several physical quantities. 
First, we evaluate the equal-time spin and charge correlations defined by 
\begin{align}
S(\bm{k})&=\frac{1}{3N}\sum_{i,j}e^{i\bm{k}\cdot(\bm{r}_{i}-\bm{r}_{j})}\langle\hat{\bm{S}}_{i}\cdot\hat{\bm{S}}_{j}\rangle , \\
N(\bm{k})&=\frac{1}{N}\sum_{i,j}e^{i\bm{k}\cdot(\bm{r}_{i}-\bm{r}_{j})}\langle(\hat{n}_{i\uparrow}+\hat{n}_{i\downarrow})(\hat{n}_{j\uparrow}+\hat{n}_{j\downarrow})\rangle . 
\end{align}
As we see in Fig.~\ref{fig:sq}, for all the system sizes, 
the peak values of the spin correlation 
$S(\pi,\pi)$ obtained by the PR-GBMC method 
are consistent with the values obtained by the AFQMC method.
The peak values of $S(\pi,\pi)$ as well as the charge correlation $N(\pi,\pi)$ are 
listed in Table \ref{tab:benchmark}. A small discrepancies observed in $N(\pi,\pi)$ may be attributed either to the effect of
finite intervals in the imaginary time step of AFQMC or to the slight insufficiency of $\tau$ taken 
for the ground-state average in the PR-GBMC calculation. 

Next, we show the superconducting correlation defined by 
\begin{align}
S_{\alpha}=\frac{1}{4}\sum_{r}P_{\alpha}(r), 
\end{align}
where $P_{\alpha}(r)$ denotes the equal-time pairing correlation 
defined as 
\begin{align}
P_{\alpha}(r)=\frac{1}{2N}\sum_{i=1}^{N}\langle\Delta_{\alpha}^{\dag}(i)\Delta_{\alpha}(i+r)+\Delta_{\alpha}(i)\Delta_{\alpha}^{\dag}(i+r)\rangle , 
\end{align}
where $r$ is the distance from the $i$-th site and $\Delta_{\alpha}$ is the 
superconducting order parameter. The latter is defined as 
\begin{align}
\Delta_{\alpha}(i)=\frac{1}{\sqrt{2}}\sum_{r}f_{\alpha}(r)(\hat{c}_{i\uparrow}\hat{c}_{i+r\downarrow}-\hat{c}_{i\downarrow}\hat{c}_{i+r\uparrow}) , 
\end{align}
where $f_{\alpha}(r)$ is the form factor of the pairing correlation 
defined as 
\begin{align}
f_{1s}(r)&=4\delta_{r_{x},0}\delta_{r_{y},0} ,                                                                          \\
f_{2s}(r)&=\delta_{r_{y},0}(\delta_{r_{x},1}+\delta_{r_{x},-1}) \notag \\
&\qquad\qquad +\delta_{r_{x},0}(\delta_{r_{y},1}+\delta_{r_{y},-1}) ,  \\
f_{2d}(r)&=\delta_{r_{y},0}(\delta_{r_{x},1}+\delta_{r_{x},-1}) \notag \\
&\qquad\qquad -\delta_{r_{x},0}(\delta_{r_{y},1}+\delta_{r_{y},-1}) ,  \\
f_{3s}(r)&=\delta_{r_{x},1}(\delta_{r_{y},1}+\delta_{r_{y},-1}) \notag \\
&\qquad\qquad +\delta_{r_{x},-1}(\delta_{r_{y},-1}+\delta_{r_{y},1}) , \\
f_{3d}(r)&=\delta_{r_{x},1}(\delta_{r_{y},1}-\delta_{r_{y},-1}) \notag \\
&\qquad\qquad +\delta_{r_{x},-1}(\delta_{r_{y},-1}-\delta_{r_{y},1}) ,   
\end{align}
where $\delta_{ij}$ is Cronecker's delta. 
The suffices $\alpha=1s$,\,$2s$,\,$2d$,\,$3s$ and $3d$ represent 
the on-site $s$-wave, the extended $s$-wave directing along the $x$ and $y$ axes, 
the $d_{x^{2}-y^{2}}$-wave, 
the extended $s$-wave along the diagonals and the $d_{xy}$-wave, respectively.
Numerical results for all the quantities defined above 
are listed in Table \ref{tab:benchmark}. Here, for comparison with the other 
method, we demonstrate the numerical results obtained by the AFQMC method 
with the Trotter discretization of $\Delta\tau=0.025$ and 
the number of Monte Carlo steps $N_{{\rm mcs}}=10^4$.
\begin{table}
\begin{center}
\caption{Comparison between PR-GBMC and AFQMC methods.
For all the parameter sets, we employ the quantum-number projections onto
the total momentum $\bm{K}=0$, total $z$-component of the spin $S_{z}=0$ and
the total spin $S=0$. All the simulation results are obtained by
$2.56\times 10^{4}$ Monte Carlo steps under the periodic boundary condition.
}\label{tab:benchmark}
\begin{tabular}{|c|c|c|}
\hline
  $6\times 6,U/t=4,n=1$  & PR-GBMC          & AFQMC            \\ \hline
  Energy       & $-30.87\pm 0.02$ & $-30.87\pm 0.03$ \\
  $S(\pi,\pi)$ &   $1.43\pm 0.01$ &   $1.43\pm 0.03$ \\
  $N(\pi,\pi)$ & $0.409\pm 0.001$ & $0.403\pm 0.001$ \\
  $S_{1s}$     & $1.636\pm 0.005$ & $1.610\pm 0.004$ \\
  $S_{2s}$     & $1.188\pm 0.004$ & $1.184\pm 0.001$ \\
  $S_{2d}$     & $1.101\pm 0.007$ & $1.116\pm 0.018$ \\
  $S_{3s}$     & $0.539\pm 0.004$ & $0.532\pm 0.003$ \\
  $S_{3s}$     & $0.397\pm 0.003$ & $0.399\pm 0.012$ \\\hline\hline
  $8\times 8,U/t=4,n=1$  & PR-GBMC          & AFQMC            \\ \hline
  Energy       & $-55.01\pm 0.03$ & $-55.09\pm 0.06$ \\
  $S(\pi,\pi)$ &   $2.08\pm 0.04$ &   $2.05\pm 0.04$ \\
  $N(\pi,\pi)$ & $0.420\pm 0.003$ & $0.412\pm 0.002$ \\
  $S_{1s}$     & $1.681\pm 0.011$ & $1.650\pm 0.001$ \\
  $S_{2s}$     & $1.191\pm 0.001$ & $1.190\pm 0.001$ \\
  $S_{2d}$     & $1.077\pm 0.004$ & $1.110\pm 0.018$ \\
  $S_{3s}$     & $0.515\pm 0.004$ & $0.505\pm 0.002$ \\
  $S_{3s}$     & $0.457\pm 0.005$ & $0.457\pm 0.007$ \\\hline\hline
  $10\times 10,U/t=4,n=1$& PR-GBMC          & AFQMC            \\ \hline
  Energy       & $-86.25\pm 0.10$ & $-86.11\pm 0.03$ \\
  $S(\pi,\pi)$ &   $2.88\pm 0.03$ &   $2.88\pm 0.06$ \\
  $N(\pi,\pi)$ & $0.424\pm 0.002$ & $0.417\pm 0.001$ \\
  $S_{1s}$     & $1.694\pm 0.006$ & $1.668\pm 0.004$ \\
  $S_{2s}$     & $1.195\pm 0.005$ & $1.192\pm 0.001$ \\
  $S_{2d}$     & $1.076\pm 0.005$ & $1.111\pm 0.018$ \\
  $S_{3s}$     & $0.510\pm 0.003$ & $0.507\pm 0.002$ \\
  $S_{3s}$     & $0.471\pm 0.005$ & $0.459\pm 0.027$ \\\hline
\end{tabular}
\end{center}
\end{table}

We have also calculated the momentum distribution defined by 
\begin{align}
n(\bm{k})&=\frac{1}{2N}\sum_{i,j}e^{i\bm{k}\cdot(\bm{r}_{i}-\bm{r}_{j})}\langle\hat{c}_{i\uparrow}^{\dag}\hat{c}_{j\uparrow}+\hat{c}_{i\downarrow}^{\dag}\hat{c}_{j\downarrow}\rangle . 
\end{align}
In Table \ref{tab:moment_dist}, we show the numerical results for the 
momentum distribution 
along the line from $(0,0)$ to $(\pi,\pi)$ in the momentum space.
\begin{table}
\begin{center}
\caption{Momentum distribution $n(k_{x},k_{y})$ obtained by PR-GBMC and 
AFQMC method. The results have been obtained from the same simulation as those of Table \ref{tab:benchmark}
}\label{tab:moment_dist}
\begin{tabular}{|c|c|c|}
\hline
  \multicolumn{3}{|c|}{$6\times 6$, $U/t=4$, $n=1$}\\ \hline
  $(k_{x},k_{y})$   & PR-GBMC             & AFQMC  \\ \hline
  $(0,0)$           & $0.967\pm 0.001$    & $0.966\pm 0.001$ \\
  $(\pi/3,\pi/3)$   & $0.928\pm 0.001$    & $0.926\pm 0.001$ \\
  $(2\pi/3,2\pi/3)$ & $0.072\pm 0.001$    & $0.074\pm 0.001$ \\
  $(\pi,\pi)$       & $0.033\pm 0.001$    & $0.034\pm 0.001$ \\ \hline\hline
  \multicolumn{3}{|c|}{$8\times 8$, $U/t=4$, $n=1$}\\ \hline
  $(k_{x},k_{y})$   & PR-GBMC & AFQMC  \\ \hline
  $(0,0)$           & $0.967\pm 0.001$  & $0.966\pm 0.001$ \\
  $(\pi/4,\pi/4)$   & $0.951\pm 0.001$  & $0.950\pm 0.001$ \\
  $(\pi/2,\pi/2)$   & $0.500\pm 0.000$  & $0.500\pm 0.000$ \\
  $(3\pi/4,3\pi/4)$ & $0.049\pm 0.001$  & $0.050\pm 0.001$ \\
  $(\pi,\pi)$       & $0.033\pm 0.001$  & $0.034\pm 0.001$ \\ \hline\hline
  \multicolumn{3}{|c|}{$10\times 10$, $U/t=4$, $n=1$}\\ \hline
  $(k_{x},k_{y})$   & PR-GBMC & AFQMC  \\ \hline
  $(0,0)$           & $0.968\pm 0.001$  & $0.966\pm 0.001$ \\
  $(\pi/5,\pi/5)$   & $0.959\pm 0.001$  & $0.957\pm 0.001$ \\
  $(2\pi/5,2\pi/5)$ & $0.883\pm 0.003$  & $0.878\pm 0.001$ \\
  $(3\pi/5,3\pi/5)$ & $0.117\pm 0.003$  & $0.121\pm 0.001$ \\
  $(4\pi/5,4\pi/5)$ & $0.041\pm 0.001$  & $0.043\pm 0.001$ \\
  $(\pi,\pi)$       & $0.032\pm 0.001$  & $0.033\pm 0.001$ \\ \hline
\end{tabular}
\end{center}
\end{table}

In the numerical results shown 
in Tables \ref{tab:benchmark} and \ref{tab:moment_dist}, 
all the numerical data are essentially consistent each other. 
We also show a hard test of the convergence for the square-lattice Hubbard model
with the next nearest neighbor transfer $t'=0.5t$ in one of the diagonal direction.
This is nothing but the anisotropic trianglar lattice.
This geometrically frustrated lattice structure generates a serious difficulty in various
simulations.  The PIRG method offers the only available results~\cite{Kashima2, Morita}.
Here PR-GBMC results for the $6\times 6$ lattice with the periodic boundary condition 
at $U/t=4$ shows the ground state enrgy $E=-32.54\pm 0.05$ and the double occupancy $\langle D \rangle =0.164\pm 0.001$,
which is favorably compared with the PIRG result of $E=-32.6\pm 0.1$ and $\langle D \rangle =0.168\pm 0.006$.
This parameter value is, as is estimated from the systematic studies of the PIRG result, just near the Mott transition point and provides us with a severe numerical benchmark.
From the analysis of the size dependence, 
we conclude that 
the convergence and the applicability of the PR-GBMC method is not restricted by the system size.

\section{Summary and Discussion}
In this paper, we have reexamined the Gaussian-basis Monte Carlo method (GBMC) 
proposed by Corney and Drummond~\cite{Corney1,Corney2}. 
This method does not suffer from the minus sign problem for any 
Hamiltonian composed of up to two-body interactions (see Appendix A and B). 
However, the original method often shows systematic 
errors especially in the low-temperature region. 
We have elucidated how the systematic errors come from 
the slow relaxation caused by the trap in the excited states.

To overcome the systematic error, we have improved the quantum-number 
projection scheme proposed by Assaad {\it et al}.~\cite{Assaad,Assaad2} 
to make it possible to 
combine the projection procedure in conjunction 
with the importance sampling of the original GBMC method. 
This method allows us to 
project out the excited states and improve 
the behavior of the probability distributions, 
which makes it possible to widen the region of 
tractable parameters in the GBMC method. 

We have also discussed the applicability of our algorithm. 
In the PR-GBMC method, 
the convergence with the ground state 
becomes slower with the increase of the interaction strength $U/t$. 
This slow convergence is caused by the 
increasing inefficiency in 
the importance sampling procedure owing to 
the barrier in the phase-space coming from the 
strong on-site repulsion. 
Nevertheless, good convergence at $U/t=10$ indicates 
a better efficiency of the PR-GBMC method compared to other numerical 
methods such as the AFQMC method. 
The system size dependence up to $10\times 10$ lattice 
shows no distinct symptom of the slow convergence 
in physical quantities 
with the increase of the system size. 
Thus the slow convergence does not restrict the applicability of 
this method for large 
system sizes at least up to $10\times 10$ lattices. 

%In the large $U/t$ region, the slow convergence causes the exponential 
%growth of the weight $\Omega$, which 
%results in the inefficient sampling. 
%Especially, the exponential growth of $\Omega$ and the resultant 
%insufficient sampling causes the plateau structure 
%in the tail part of the distribution $\tilde{Q}(\tilde{\Omega})$. 
%This means the lack of 
%the low-energy samples because of the definition of 
%$\Omega=\exp\left[-\int Hd\tau'\right]$. 
In the large $U/t$ region, the slow convergence is tightly associated with 
the exponential broadening of the distribution of the weight $\Omega$, 
which causes an expansion of the 
phase-space to be sampled and results in inefficient importance sampling. 
Especially, the exponential growth of $\Omega$ and corresponding expansion of the 
phase-space often results in the lack of rare event with large $\Omega$. 
The lack of Monte Carlo samples with large $\Omega$ causes a spike structure 
in the tail part of the distribution $\tilde{P}(\tilde{\Omega})$ and 
causes a plateau structure in the tail part of the integrated distribution $\tilde{Q}(\tilde{\Omega})$. 
Despite samples with large $\Omega$ contribute to physical quantities, 
such samples seldom appear at large $U/t$, which causes large statistical errors. 
Thus, with the increase of $U/t$, the convergence of the energy becomes 
slower accompanied by the increase of the statistical errors. 
If that is the case, the computation time required 
for the convergence of the energy with required statistical errors 
goes beyond allowed computation time, which determines the practical limitation of the PR-GBMC method. 
Therefore, it is advisable to monitor the 
convergence of the distribution 
to evaluate the number of Monte Carlo steps needed for the convergence, 
especially when one calculates large 
$U/t$ regions. 

With the inspection of the large $U/t$ systems and the system size dependence, 
the PR-GBMC method offers a powerful 
tool which can be applied to the systems in several cases beyond tractable 
parameters of the conventional 
numerical methods such as the AFQMC and PIRG methods and is at least complementary to the existing methods.

\section*{Acknowledgements}
We would like to thank J. F. Corney for useful discussions, especially on the 
positivity of the distribution function discussed in Appendix B.
The present work is supported by Grant-in-Aids for scientific research 
from Ministry of Education, Culture, Sports, Science and Technology 
under the grant numbers 16340100 and 17064004.
A part of our computation has been done at 
the supercomputer center at the Institute for 
Solid State Physics, University of Tokyo.

\appendix
\section{Langevin Equations for a General Hamiltonian}
In this Appendix, we show how to construct a Gaussian representation for a 
general Hamiltonian. 
Here for simplicity, we treat a general number-conserving Hamiltonian given by 
\begin{align}
\hat{H}&=\sum_{ij\sigma}t_{ij\sigma}\hat{n}_{(i\sigma),(j\sigma)}+\!\!\sum_{ijkl\sigma\sigma'}V_{kl\sigma'}^{ij\sigma}\hat{n}_{(i\sigma),(j\sigma)}\hat{n}_{(k\sigma'),(l\sigma')} \notag \\
&=\sum_{ij\sigma}\hat{H}_{ij\sigma}+\sum_{ijkl\sigma\sigma'}\hat{H}_{ijkl\sigma\sigma'} . \label{eq:Gham}
\end{align}
Using the operator identities in Eqs.(\ref{eq:op1})-(\ref{eq:op3}), 
one-body term $\hat{H}_{ij\sigma}$ gives a contribution 
$A_{ij\sigma}^{\Omega}$ to the equation of $\Omega$ and 
a contribution $A_{ij\sigma}^{n_{xy}}$ to the drift term of the Green's function: 
\begin{align}
d\Omega=\sum_{ij\sigma}A_{ij\sigma}^{\Omega}\Omega d\tau , \quad dn_{xy}=\sum_{ij\sigma}A_{ij\sigma}^{n_{xy}}d\tau, 
\end{align}
where $x$ (and $y$) denotes the site and the spin, i.e, $x=(i,\sigma)$. 
The concrete expressions of $A_{ij\sigma}^{\Omega}$ and $A_{ij\sigma}^{n_{xy}}$ are 
\begin{align}
A_{ij\sigma}^{\Omega}&=-t_{ij\sigma}n_{(i\sigma),(j\sigma)} , \\
A_{ij\sigma}^{n_{xy}}&=-\frac{t_{ij\sigma}}{2}\left[n_{x,(j\sigma)}\{\delta_{(i\sigma),y}-n_{(i\sigma),y}\}\right. \notag \\
&\qquad\qquad\left.+\{\delta_{x,(j\sigma)}-n_{x,(j\sigma)}\}n_{(i\sigma),y}\right] .
\end{align}
Next, to guarantee the positive diffusion, we change the two-body term $\hat{H}_{ijkl\sigma\sigma'}$ by adding the identity 
${\displaystyle \hat{n}_{(i\sigma),(j\sigma)}^{2}\!-\!\delta_{ij}\hat{n}_{(i\sigma),(j\sigma)}\!=\!0}$ as 
\begin{align}
\hat{H}_{ijkl\sigma\sigma'}&=V_{kl\sigma'}^{ij\sigma}\hat{n}_{(i\sigma),(j\sigma)}\hat{n}_{(k\sigma'),(l\sigma')} \notag \\
&=-\frac{|V_{kl\sigma'}^{ij\sigma}|}{2}\left[\hat{n}_{(i\sigma),(j\sigma)}-{\rm sign}(V_{kl\sigma'}^{ij\sigma})\hat{n}_{(k\sigma'),(l\sigma')}\right]^{2} \notag \\ 
&\quad+\frac{|V_{kl\sigma'}^{ij\sigma}|}{2}\left[\delta_{ij}\hat{n}_{(i\sigma),(j\sigma)}+\delta_{kl}\hat{n}_{(k\sigma'),(l\sigma')}\right] , 
\end{align}
The two-body term then gives a contribution $A_{ijkl\sigma\sigma'}^{\Omega}$ to the equation of $\Omega$, 
a contribution $A_{ijkl\sigma\sigma'}^{n_{xy}}$ to the drift term of $n_{xy}$ 
and the contributions $B_{ijkl\sigma\sigma'}$, $C_{ijkl\sigma\sigma'}$ to the diffusion term of $n_{xy}$: 
\begin{align}
d\Omega&=\sum_{ijkl\sigma\sigma'}A_{ijkl\sigma\sigma'}^{\Omega}\Omega d\tau , \\
dn_{xy}&=\sum_{ijkl\sigma\sigma'}\left[A_{ijkl\sigma\sigma'}^{n_{xy}}d\tau+B_{ijkl\sigma\sigma'}^{n_{xy}}dW_{ijkl\sigma\sigma'}^{(1)} \right. \notag \\
&\quad\qquad\qquad\qquad\qquad\left.+C_{ijkl\sigma\sigma'}^{n_{xy}}dW_{ijkl\sigma\sigma'}^{(2)}\right] , 
\end{align}
where the Wiener increment $dW$ satisfies 
\begin{align}
\langle dW_{ijkl\sigma\sigma'}^{(r)}dW_{i'j'k'l'\eta\eta'}^{(r')}\rangle=d\tau\delta_{rr'}\delta_{ii'}\delta_{jj'}\delta_{kk'}\delta_{ll'}\delta_{\sigma\eta}\delta_{\sigma'\eta'} .
\end{align}
The concrete expressions of each term become 
\begin{align}
A_{ijkl\sigma\sigma'}^{\Omega}&=-\frac{V_{kl\sigma'}^{ij\sigma}}{2}\left[2\left\{n_{(i\sigma),(j\sigma)}n_{(k\sigma'),(l\sigma')}\right.\right. \notag \\
&\qquad\qquad\qquad\left.-n_{(i\sigma),(l\sigma')}n_{(k\sigma'),(j\sigma)}\right\} \notag \\
& \left.+n_{(i\sigma),(l\sigma')}\delta_{kj}\delta_{\sigma\sigma'}+n_{(k\sigma'),(j\sigma)}\delta_{il}\delta_{\sigma\sigma'}\right] , \\
A_{ijkl\sigma\sigma'}^{n_{xy}}&=-\frac{V_{kl\sigma'}^{ij\sigma}}{2}
 \left[n_{(i\sigma),(j\sigma)}\left\{n_{x,(l\sigma')}(\delta_{(k\sigma'),y}-n_{(k\sigma'),y})\right.\right. \notag \\
&\qquad\qquad\qquad\qquad\left.+(\delta_{x,(l\sigma')}-n_{x,(l\sigma')})n_{(k\sigma'),y}\right\} \notag \\
&\qquad\qquad+n_{(k\sigma'),(l\sigma')}\left\{n_{x,(j\sigma)}(\delta_{(i\sigma),y}-n_{(i\sigma),y})\right. \notag \\
&\qquad\qquad\qquad\qquad\left.+(\delta_{x,(j\sigma)}-n_{x,(j\sigma)})n_{(i\sigma),y}\right\} \notag \\
&\qquad -\{n_{(i\sigma),(l\sigma')}-\frac{1}{2}\delta_{(i\sigma),(l\sigma')}\} \notag \\
&\qquad\times   \left\{n_{x,(j\sigma)}(\delta_{(k\sigma'),y}-n_{(k\sigma'),y})\right. \notag \\
&\qquad\qquad\qquad\left.+(\delta_{x,(j\sigma)}-n_{x,(j\sigma)})n_{(k\sigma'),y}\right\} \notag \\
&\qquad -\{n_{(k\sigma'),(j\sigma)}-\frac{1}{2}\delta_{(k\sigma'),(j\sigma)}\} \notag \\
&\qquad\times   \left\{n_{x,(l\sigma')}(\delta_{(i\sigma),y}-n_{(i\sigma),y})\right.\notag \\
&\qquad\qquad\qquad\left.\left.+(\delta_{x,(l\sigma')}-n_{x,(l\sigma')})n_{(i\sigma),y}\right\}
\right] , 
\end{align}
and 
\begin{align}
B_{ijkl\sigma\sigma'}^{n_{xy}}&=\sqrt{\frac{|V_{kl\sigma'}^{ij\sigma}|}{2}}\left[n_{x,(j\sigma)}\{\delta_{(i\sigma),y}-n_{(i\sigma),y}\}\right. \notag \\
&\left.-{\rm sign}(V_{kl\sigma'}^{ij\sigma})n_{x,(l\sigma')}\{\delta_{(k\sigma'),y}-n_{(k\sigma'),y}\}\right] , \\
C_{ijkl\sigma\sigma'}^{n_{xy}}&=\sqrt{\frac{|V_{kl\sigma'}^{ij\sigma}|}{2}}\left[\{\delta_{x,(j\sigma)}-n_{x,(j\sigma)}\}n_{(i\sigma),y}\right. \notag \\
&\left.-{\rm sign}(V_{kl\sigma'}^{ij\sigma})\{\delta_{x,(l\sigma')}-n_{x,(l\sigma')}\}n_{(k\sigma'),y}\right] . 
\end{align}
In all, the general Hamiltonian (\ref{eq:Gham}) gives the Langevin equations of Ito-type 
\begin{align}
d\Omega&=\left[\sum_{ij\sigma}A_{ij\sigma}^{\Omega}+\sum_{ijkl\sigma\sigma'}A_{ijkl\sigma\sigma'}^{\Omega}\right]\Omega d\tau , \\
dn_{xy}&=\left[\sum_{ij\sigma}A_{ij\sigma}^{n_{xy}}+\sum_{ijkl\sigma\sigma'}A_{ijkl\sigma\sigma'}^{n_{xy}}\right]d\tau \notag \\
&\quad +\sum_{ijkl\sigma\sigma'}\left[B_{ijkl\sigma\sigma'}^{n_{xy}}dW_{ijkl\sigma\sigma'}^{(1)}+C_{ijkl\sigma\sigma'}^{n_{xy}}dW_{ijkl\sigma\sigma'}^{(2)}\right] .
\end{align}

\section{Completeness and Positivity of the Gaussian Basis}
In order to establish a phase-space representation based on the Gaussian operators, 
one must show that they form a complete basis for the class of density-matrix operators 
which we wish to represent. Furthermore, the expansion of any density-matrix operators 
in terms of the Gaussian operators must involve the positive coefficients. 
For the self-contained description of this paper, we prove 
this `positive completeness' by following the idea of J. F. Corney~\cite{Corney3}. 

To prove this `positive completeness', we will relate the Gaussian operators to 
the number-state projection operators which form a complete basis set and we will show 
that any term which appears in a number-state expansion of a density-matrix operator 
can be written as a sum of Gaussian operators with positive coefficients. 
Here, we prove only the number-conserving case in this Appendix, 
a proof for the most general Gaussian 
can be similarly constructed~\cite{Corney3}. 
To this end, we rewrite the Gaussian operator in terms of $\bm{\mu}$ which is defined as
\begin{align}
\bm{\mu} = 2\bm{I}+(\bm{n}^{T}-\bm{I})^{-1}. \label{eq:lamex}
\end{align}
A number-conserving Gaussian operator $\hat{\Lambda}(\bm{n})$ is then written as 
\begin{align}
\hat{\Lambda}&=\det(\bm{I}-\bm{n}):e^{-\hat{\bm{b}}^{\dag}[2\bm{I}+(\bm{n}^{T}-\bm{I})^{-1}]\hat{\bm{b}}}: \notag \\
&=\frac{1}{\det(2\bm{I}-\bm{\mu})}:e^{-\hat{\bm{b}}^{\dag}\bm{\mu}\hat{\bm{b}}}: \notag \\
&=:\prod_{i,j=1}^{M}(\hat{1}-\hat{b}_{i}^{\dag}\mu_{ij}\hat{b}_{j}):/\det(2\bm{I}-\bm{\mu}).
\end{align}
To avoid singular behavior, the limit of any $n_{jj}\rightarrow 1$ is taken only in the 
normalized form of the Gaussian.

\subsection{Number-state expansion}
Let $\vec{n}$ be a Fermionic occupation number vector 
$\vec{n}=(n_{1},n_{2},\cdots,n_{M})$, where $n_{i}=0,1$, 
then a complete set of Fermionic number-state is 
represented by $\{|\vec{n}\rangle\}$, where $\{|\vec{n}\rangle\}$ runs 
over all the $2^{M}$ permutations. 
This set defines a complete operator basis of dimension $2^{2M}$, 
and it enables us to expand the density-matrix operator as
\begin{align}
\hat{\rho}&=\sum_{\vec{n}}\sum_{\vec{m}}|\vec{n}\rangle\langle\vec{n}|\hat{\rho}|\vec{m}\rangle\langle\vec{m}| \notag \\
&=\sum_{\vec{n}}\sum_{\vec{m}}\rho_{\vec{n}\vec{m}}|\vec{n}\rangle\langle\vec{m}|=\sum_{\vec{n}}\sum_{\vec{m}}\hat{\rho}_{\vec{n}\vec{m}} , 
\end{align}
where we impose a number-conserving condition $\sum_{i}n_{i}=\sum_{i}m_{i}$. 
From the positive definiteness of the density-matrix operator, 
all the diagonal density-matrix elements 
are real and positive: $\rho_{\vec{n}\vec{n}}\ge 0$. Here we require additionally that $\rho_{\vec{n}\vec{n}}\le 1$ for 
the normalization. 

Since a density-matrix operator is Hermitian, it can be always diagonalized. 
Let $|\Psi_{k}\rangle=\sum_{\vec{n}}C_{k\vec{n}}|\vec{n}\rangle$ are the eigenvectors 
and $P_{k}$ the corresponding positive eigenvalues of the density matrix, we can write 
\begin{align}
\hat{\rho}=\sum_{k}P_{k}|\Psi_{k}\rangle\langle\Psi_{k}|=\sum_{k}\sum_{\vec{n},\vec{m}}P_{k}C_{k\vec{n}}C_{k\vec{m}}^{*}|\vec{n}\rangle\langle\vec{m}| . 
\end{align}
Thus the coefficients of the number-state expansion can be represented as 
\begin{align}
\rho_{\vec{n}\vec{m}}=\sum_{k}P_{k}C_{k\vec{n}}C_{k\vec{m}}^{*}=\sum_{k}D_{k\vec{n}}D_{k\vec{m}}^{*} , 
\end{align}
where $D_{k\vec{n}}=\sqrt{P_{k}}C_{k\vec{n}}$. 
By using a Cauchy-Schwartz inequality, the magnitude of these coefficients is given by 
\begin{align}
|\rho_{\vec{n}\vec{m}}|^{2}&=\sum_{k,k'}D_{k,\vec{n}}D_{k\vec{m}}^{*}D_{k'\vec{n}}^{*}D_{k'\vec{m}} \notag \\
&\le\sum_{k,k'}D_{k,\vec{n}}D_{k\vec{n}}^{*}D_{k'\vec{m}}D_{k'\vec{m}}^{*} \notag \\
&=\rho_{\vec{n}\vec{n}}\rho_{\vec{m}\vec{m}}\le 1 .
\end{align}
Thus the magnitude of any off-diagonal element is bounded, 
Conversely, any diagonal element is at least as large as 
the squared magnitude of anything else on the same row or column: 
\begin{align}
|\rho_{\vec{n}\vec{m}}|^{2}&\le\rho_{\vec{n}\vec{n}}\rho_{\vec{m}\vec{m}}\le\rho_{\vec{n}\vec{n}} \\
|\rho_{\vec{m}\vec{n}}|^{2}&\le\rho_{\vec{m}\vec{m}}\rho_{\vec{n}\vec{n}}\le\rho_{\vec{n}\vec{n}} .
\end{align}
Thus we obtain the lower limit of the diagonal elements: 
\begin{align}
\rho_{\vec{n}\vec{n}}&\ge\max\left(|\rho_{\vec{n}\vec{m}}|^2,|\rho_{\vec{m}\vec{n}}|^2\right) \notag \\
&\ge\frac{1}{2(M-1)}\sum_{\vec{m}\ne\vec{n}}\left(|\rho_{\vec{n}\vec{m}}|^2+|\rho_{\vec{m}\vec{n}}|^2\right) \notag \\
&\ge\frac{1}{2(2^{M}-1)}\sum_{\vec{m}\ne\vec{n}}\left(|\rho_{\vec{n}\vec{m}}|^2+|\rho_{\vec{m}\vec{n}}|^2\right) . 
\end{align}
The number-state expansion of the density-matrix operator can then be written as 
\begin{align}
\hat{\rho}=\sum_{\vec{n}}\Delta\rho_{\vec{n}\vec{n}}|\vec{n}\rangle\langle\vec{n}|+\sum_{\vec{n}}\sum_{\vec{m}\ne\vec{n}}\frac{1}{2(2^{M}-1)}|\rho_{\vec{n}\vec{m}}|^{2}\hat{P}_{\vec{n}\vec{m}}(\rho) , \label{eq:newex}
\end{align}
where 
\begin{align}
\hat{P}_{\vec{n}\vec{m}}(\rho)&=|\vec{n}\rangle\langle\vec{n}|+|\vec{m}\rangle\langle\vec{m}|+\frac{2(2^{M}-1)}{\rho_{\vec{n}\vec{m}}^{*}}|\vec{n}\rangle\langle\vec{m}| \\
\Delta\rho_{\vec{n}\vec{n}}&=\rho_{\vec{n}\vec{n}}-\frac{1}{2(2^{M}-1)}\!\sum_{\vec{m}\ne\vec{n}}\left(|\rho_{\vec{n}\vec{m}}|^2+|\rho_{\vec{m}\vec{n}}|^2\right)\ge 0 . 
\end{align}
Since all the coefficients of the new expansion (\ref{eq:newex}) are positive, 
it is sufficient to prove that each operator in the expansion (\ref{eq:newex}) can be written 
as a Gaussian or as a positive sum over Gaussians. 

\subsection{Diagonal number-state projector}
First we show that the diagonal number-state projector $|\vec{n}\rangle\langle\vec{n}|$ 
in the new expansion (\ref{eq:newex}) with 
positive coefficients $\Delta\rho_{\vec{n}\vec{n}}$ corresponds to a Gaussian operator. 
For individual ladder operators, one has the well-known identities: 
\begin{align}
\hat{b}_{j}=|0\rangle_{j}\langle 1|_{j},\quad \hat{b}_{j}^{\dag}=|1\rangle_{j}\langle 0|_{j} .
\end{align}
If we set $n_{ij}=n_{j}\delta_{ij}$ in Eq.~(\ref{eq:lamex}), 
then $\mu_{ij}=\delta_{ij}(1-2n_{j})/(1-n_{j})$ 
and the Gaussian operator reduces to 
\begin{align}
\hat{\Lambda}&=:\prod_{i,j=1}^{M}(\hat{1}-\hat{b}_{i}^{\dag}\mu_{ij}\hat{b}_{j}):/\det(2\bm{I}-\bm{\mu}) \notag \\
&=\prod_{j=1}^{M}\left[(1-n_{j})\hat{b}_{j}\hat{b}_{j}^{\dag}+n_{j}\hat{b}_{j}^{\dag}\hat{b}_{j}\right] \notag \\
&=\prod_{j=1}^{M}\left[(1-n_{j})|0\rangle_{j}\langle 0|_{j}+n_{j}|1\rangle_{j}\langle 1|_{j}\right] . 
\end{align}
Thus, if $n_{j}$ is chosen as $0$ or as $1$, 
the Gaussian operator itself can be regarded as a diagonal number-state projector:
\begin{align}
|\vec{n}\rangle\langle\vec{n}|=\hat{\Lambda}(n_{j}\delta_{ij}) . 
\end{align}

\subsection{Off-diagonal number-state projector}
Second we show that the mixed projector $\hat{P}_{\vec{n}\vec{m}}(\rho)$ in the expansion (\ref{eq:newex}) 
corresponds to a positive sum over Gaussians. 
Consider a Gaussian operator with 
\begin{align}
n_{ij}=n_{j}\delta_{ij}+\sum_{k=1}^{N_{k}}\nu_{k}\delta_{i,r(k)}\delta_{j,s(k)} , 
\end{align}
where the $n_{j}$ for each $j$ is each either $0$ or $1$ and the locations $(r(k),s(k))$ of the $N_{k}$ 
nonzero off-diagonal elements $\nu_{k}$ satisfy 
$r(k)\ne r(k'),s(k)\ne s(k')$ and $r(k)\ne s(k')$ for any $k,k'(k\ne k')$. 
In other words, if there is a nonzero element in the off-diagonal location $(i,j)$, 
then there will be no other element in the $i$-th row and $j$-th column 
and none in the $j$-th row and $i$-th column, {\it i.e.},
\begin{align}
 \bm{n} = \left[
 \raisebox{-15.5ex}[15.5ex][15.5ex]{\includegraphics[scale=0.73]{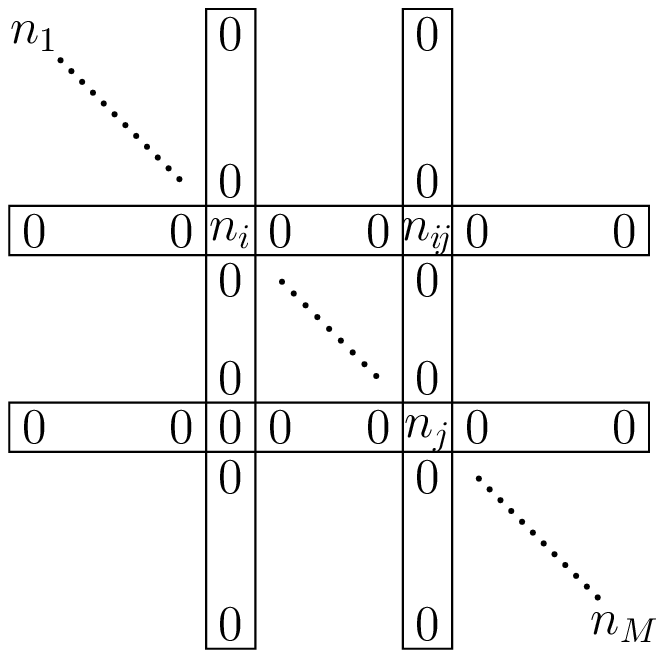}}
 \right] .
\end{align}
This structure means that $\det(\bm{I}-\bm{n})=\prod_{j=1}^{M}(1-n_{j})$ and 
\begin{align}
\mu_{ij}&=\delta_{ij}\frac{1-2n_{j}}{1-n_{j}} \notag \\
&\quad -\sum_{k=1}^{N_{k}}\delta_{j,r(k)}\delta_{i,s(k)}\frac{\nu_{k}}{(1-n_{r(k)})(1-n_{s(k)})} . 
\end{align}
Here again, we take the limit $n_{j}\rightarrow 1$ only in the normalized 
form of the Gaussian to avoid the singularity. 
With these conditions, the Gaussian operator reduces to 
\begin{align}
\hat{\Lambda}&=:\prod_{i,j=1}^{M}\left(1-\hat{b}_{i}^{\dag}\mu_{ij}\hat{b}_{j}\right):/\det(2\bm{I}-\bm{\mu}) \notag \\
&=:\prod_{j=1}^{M}\left[(1-n_{j})\hat{b}_{j}\hat{b}_{j}^{\dag}+n_{j}\hat{b}_{j}^{\dag}\hat{b}_{j}\right]\notag \\
&\qquad\times\prod_{k=1}^{N_{k}}\left[1+\frac{\nu_{k}\hat{b}_{s(k)}^{\dag}\hat{b}_{r(k)}}{(1-n_{r(k)})(1-n_{s(k)})}\right]: \notag \\
&=:\prod_{j=1}^{M}\left[(1-n_{j})|0\rangle_{j}\langle 0|_{j}+n_{j}|1\rangle_{j}\langle 1|_{j}\right]\notag \\
&\qquad\sum_{K}\prod_{k\in K}\frac{\nu_{k}\hat{b}_{s(k)}^{\dag}\hat{b}_{r(k)}}{(1-n_{r(k)})(1-n_{s(k)})}: , 
\end{align}
where the sum over $K$ is the sum over all the possible 
subsets of $\{1,2,\cdots,N_{k}\}$ and has $2^{N_{k}}$ terms. 
For the $j$-th mode, if there is a $k$ such that $s(k)=j$, 
then the diagonal number projector for the $j$-th mode $|n_{j}\rangle\langle n_{j}|$ 
is replaced by the off-diagonal $|1\rangle_{j}\langle0|_{j}$, 
or if $r(k)=j$, the conjugate projector $|0\rangle_{j}\langle 1|_{j}$ is created. 
Thus the Gaussian operator can be represented by a sum over number-state projectors: 
\begin{align}
\hat{\Lambda}=\sum_{K}(\pm)\prod_{k\in K}\nu_{k}\left|\vec{n}_{r(K)}^{s(K)}\right\rangle\left\langle\vec{n}_{s(K)}^{r(K)}\right|\equiv\hat{\Lambda}_{\bm{\nu}}(\vec{n}), \label{eq:sumGau}
\end{align}
where the $j$-th element of the vector $\vec{n}_{s(K)}^{r(K)}$ is defined as 
\begin{align}
\left\{\vec{n}_{s(K)}^{r(K)}\right\}_{j}=
\left\{
  \begin{array}{cl}
     1 & j=r(k),\exists k\in K \\
     0 & j=s(k),\exists k\in K \\
 n_{j} & {\rm otherwise} \\
  \end{array}
\right. . 
\end{align}
A minus sign appears if an odd number of transpositions are required to put 
all the annihilation and the creation operators in a canonical order. 
In this sum over projectors, the diagonal projector 
$|\vec{n}\rangle\langle\vec{n}|$ is contained with coefficient 1. 
In the sum, there also exists the projector that transposes all the $2N_{k}$ 
specified modes with coefficient $\prod_{k=1}^{N_{k}}\nu_{k}$. 
The sum also contains projectors that transpose only subsets of these modes. 
In total, there are $2^{N_{k}}$ terms.

Next by adding other Gaussian operators, 
we eliminate all the intermediate terms from the sum in Eq.~(\ref{eq:sumGau}) 
and leave only $K=N_{k}$ terms. 
First we add the $N_{k}$ Gaussians with one fewer off-diagonal element 
in the $\bm{n}$ matrix, to cancel the projectors that transpose 
$N_{k}-1$ modes. 
Second we add $\null_{N_{k}}C_{2}$ Gaussians 
with two fewer off-diagonal elements, 
to cancel the projectors that transpose $N_{k}-2$ modes. 
This process is repeated until all the
$\sum_{k=1}^{N_{k}-1}\null_{N_{k}}C_{k}=2^{N_{k}}-2$ intermediate terms are removed. 
Finally, we obtain
\begin{align}
\sum_{\bm{\nu}'\subseteq\bm{\nu}}\hat{\Lambda}_{\bm{\nu}'}(\vec{n})
&=(2^{N_{k}}-1)|\vec{n}\rangle\langle\vec{n}|\notag \\
&\qquad\pm\prod_{k=1}^{N_{k}}\nu_{k}\left|\vec{n}_{r(K)}^{s(K)}\right\rangle\left\langle\vec{n}_{s(K)}^{r(K)}\right| , 
\end{align}
where the sum indexed by subsets of $\bm{\nu}$ refers to the sum described above. 

By adding $\hat{\Lambda}_{\bm{\nu}'}(\vec{n})$ and $\hat{\Lambda}_{\bm{\nu}'}(\vec{m})$, 
with different diagonal components $\vec{n}=\vec{n}_{r(K)}^{s(K)}$ and $\vec{m}=\vec{m}_{r(K)}^{s(K)}$, respectively, 
we obtain 
\begin{align}
\hat{P}_{\vec{n}\vec{m}}(\rho)&=|\vec{n}\rangle\langle\vec{n}|+|\vec{m}\rangle\langle\vec{m}|+\frac{2(2^{M}-1)}{\rho_{\vec{n}\vec{m}}^{*}}|\vec{n}\rangle\langle\vec{m}| \notag \\
&=\frac{1}{2^{N_{k}}-1}\sum_{\bm{\nu}'\subseteq\bm{\nu}}\left[\hat{\Lambda}_{\bm{\nu}'}(\vec{n})+\hat{\Lambda}_{\bm{\nu}'}(\vec{m})\right] , 
\end{align}
where 
\begin{align}
\frac{1}{\rho_{\vec{n}\vec{m}}^{*}}=\pm\frac{1}{(2^{M}-1)(2^{N_{k}}-1)}\prod_{k}\nu_{k} .
\end{align}
Thus it is proven that $\hat{P}_{\vec{n}\vec{m}}(\rho)$ can be represented by 
a positive sum over Gaussians and 
hence it is shown that any number-conserving density-matrix operator can be expanded 
by the Gaussian operators with positive coefficients.

\end{document}